\documentclass[a4paper,11pt,reqno]{amsart} 
\usepackage[utf8]{inputenc}
\pdfoutput=1 

\usepackage{graphicx}       
\usepackage{epstopdf}
\usepackage{placeins} 
\usepackage{tikz}
\usetikzlibrary{calc,fadings}

\newcommand\pgfmathsinandcos[3]{%
  \pgfmathsetmacro#1{sin(#3)}%
  \pgfmathsetmacro#2{cos(#3)}%
}

\newcommand\LatitudePlane[3][current plane]{%
  \pgfmathsinandcos\sinEl\cosEl{#2} 
  \pgfmathsinandcos\sint\cost{#3} 
  \pgfmathsetmacro\yshift{\cosEl*\sint}
  \tikzset{#1/.style={cm={\cost,0,0,\cost*\sinEl,(0,\yshift)}}} %
}

\newcommand\DrawLatitudeCircle[2][1]{
  \LatitudePlane{\angEl}{#2}
  \tikzset{current plane/.prefix style={scale=#1}}
  \pgfmathsetmacro\sinVis{sin(#2)/cos(#2)*sin(\angEl)/cos(\angEl)}
  \pgfmathsetmacro\angVis{asin(min(1,max(\sinVis,-1)))}
  \draw[current plane] (\angVis:1) arc (\angVis:-\angVis-180:1);
  \draw[current plane,dashed] (180-\angVis:1) arc (180-\angVis:\angVis:1);
}

\tikzset{%
  >=latex, 
  inner sep=0pt,%
  outer sep=2pt,%
  mark coordinate/.style={inner sep=0pt,outer sep=0pt,minimum size=3pt,
    fill=black,circle}%
}

\usetikzlibrary{arrows,shapes.multipart,backgrounds,fit}
\usepackage{pgfkeys}
\usepackage{lscape}
\usepackage{xcolor}
\usepackage{amsmath}
\usetikzlibrary{decorations.pathreplacing}
\usetikzlibrary{shapes}
\tikzset{%
    every overlay node/.style={%
    draw=gray,fill=white,anchor=north west,},%
}%
\tikzstyle{decision} = [diamond, draw, fill=gray!5, 
    text width=1.6cm, text badly centered, node distance=5em, inner sep=0pt]
\tikzstyle{block} = [rectangle, draw, fill=gray!5, 
    text width=6cm, text centered, rounded corners, minimum height=1.3em]
\tikzstyle{blockS} = [rectangle, draw, fill=gray!30, 
    text width=6cm, text centered, rounded corners, minimum height=1.3em]	    
\tikzstyle{blockEmpty} = [rectangle, draw, 
    text width=6cm, text centered, rounded corners, minimum height=1.3em]
\tikzstyle{line} = [draw, -latex']
\tikzstyle{cloud} = [ellipse, draw, fill=gray!5, 
    text width=1.4cm, text badly centered, node distance=5em, inner sep=0pt, minimum height=1.3em]
\tikzstyle{cloudS} = [ellipse, draw, fill=gray!30, 
    text width=1.4cm, text badly centered, node distance=5em, inner sep=0pt, minimum height=1.3em]  
\tikzstyle{textbf} = [text width=2cm,text centered] 
\tikzstyle{textbfBig} = [text width=6cm,text centered]
\usepackage{subfig} 
\usepackage{caption}
\usepackage{subeqnarray}
\usepackage{multirow}

\usepackage{amsmath} 
\usepackage{siunitx}
\usepackage{amssymb} 
\usepackage{tabularx}
\usepackage{booktabs}
\usepackage{array}
\usepackage{algorithm}
\usepackage{algpseudocode}
\usepackage{bm} 

\input{macros/finiteVolumeMOD.sty} 
\input{macros/tensorNotation.sty} 

\usepackage[margin=0.90in]{geometry}

\usepackage{mathtools}
\usepackage{amsfonts}
\usepackage{amsthm} 

\usepackage{etoolbox}
\makeatletter
\newcommand*{\algrule}[1][\algorithmicindent]{%
  \makebox[#1][l]{%
    \hspace*{.2em}
    \vrule height 0.75\baselineskip depth .25\baselineskip
  }
}
\newcount\ALG@printindent@tempcnta
\def\ALG@printindent{%
    \ifnum \theALG@nested>0
    \ifx\ALG@text\ALG@x@notext
    \else
    \unskip
    \ALG@printindent@tempcnta=1
    \loop
    \algrule[\csname ALG@ind@\the\ALG@printindent@tempcnta\endcsname]%
    \advance \ALG@printindent@tempcnta 1
    \ifnum \ALG@printindent@tempcnta<\numexpr\theALG@nested+1\relax
    \repeat
    \fi
    \fi
}
\patchcmd{\ALG@doentity}{\noindent\hskip\ALG@tlm}{\ALG@printindent}{}{\errmessage{failed to patch}}
\patchcmd{\ALG@doentity}{\item[]\nointerlineskip}{}{}{} 
\makeatother

\title[Computational analysis of single rising bubbles]{Computational analysis of single rising bubbles influenced by soluble surfactant}

\author[C.\ Pesci]{Chiara Pesci}
\address[]{Mathematical Modeling and Analysis, Technische \mbox{Universit{\"a}t} Darmstadt, Germany}
\email[]{pesci@mma.tu-darmstadt.de}
\author[A.\ Weiner]{Andre Weiner}
\address[]{Mathematical Modeling and Analysis, Technische \mbox{Universit{\"a}t} Darmstadt, Germany}
\email[]{weiner@mma.tu-darmstadt.de}
\author[H.\ Marschall]{Holger Marschall}
\address[]{Mathematical Modeling and Analysis, Technische \mbox{Universit{\"a}t} Darmstadt, Germany}
\email[]{marschall@mma.tu-darmstadt.de}
\author[D.\ Bothe]{Dieter Bothe}
\address[]{Mathematical Modeling and Analysis, Technische \mbox{Universit{\"a}t} Darmstadt, Germany}
\email[Corresponding author]{bothe@mma.tu-darmstadt.de}

\date{\today}
\subjclass{flu-dyn}

\begin{document}

\begin{abstract}
This paper presents novel insights about the influence of soluble surfactants on bubble flows obtained by Direct Numerical Simulation (DNS). Surfactants are amphiphilic compounds which accumulate at fluid interfaces and significantly modify the respective interfacial properties, influencing also the overall dynamics of the flow. With the aid of DNS local quantities like the surfactant distribution on the bubble surface can be accessed for a better understanding of the physical phenomena occurring close to the interface. The core part of the physical model consists in the description of the surfactant transport in the bulk and on the deformable interface. The solution procedure is based on an Arbitrary Lagrangian-Eulerian (ALE) Interface-Tracking method. The existing methodology was enhanced to describe a wider range of physical phenomena. A subgrid-scale (SGS) model is employed in the cases where a fully resolved DNS for the species transport is not feasible due to high mesh resolution requirements and, therefore, high computational costs. After an exhaustive validation of the latest numerical developments, the DNS of single rising bubbles in contaminated solutions is compared to experimental results. The full velocity transients of the rising bubbles, especially the contaminated ones, are correctly reproduced by the DNS. The simulation results are then studied to gain a better understanding of the local bubble dynamics under the effect of soluble surfactant. One of the main insights is that the quasi-steady state of the rise velocity is reached without ad- and desorption being necessarily in local equilibrium.
\end{abstract}

\maketitle 

%
%
\section{Introduction}

Surface active agents, so-called surfactants, are present in most multiphase contactors, either as contaminants or added on purpose to change the way how phases interact. In froth flotation, for example, a so-called frother is used to separate hydrophobic from hydrophilic particles. The frother is surface active and renders the particles in question hydrophobic. The particles can then attach to air bubbles, which rise to the surface of the floatation cell and form a froth that can be removed. The efficiency of flotation cells is determined by the probability of bubble-particle collisions, and therefore by the interaction of gas, liquid, particles, and frother. The example of froth flotation demonstrates how complex a system involving surfactants can be. But also systems as simple as a single air bubble rising in tap water may be determined by the presence of surfactant. Experiments have shown that bubbles rising in purified water can reach terminal velocities that are two times higher than in tap water\footnote{see e.g. figure 7.3 on page 172 in the reference}~\cite{clift1978}. This demands that the used substance system must be well determined in order to obtain reliable and reproducible results. The most challenging but also most astonishing property of surfactants is that even traces of it, which are modifying cohesion forces on a molecular level, can cause a tremendous change in the macroscopically observed, sometimes meter-sized, flow patterns.

Levich's \textit{Physicochemical Hydrodynamics}~\cite{levich1962} is one of the first textbooks containing a theoretical treatment of surface forces resulting from an inhomogeneous distribution of a surface active substance on the interface of a rising bubble, and it also describes in much greater detail, for the interested reader, some of the basic concepts outlined hereafter. Bubbles rising in a pure liquid are characterized by a mobile interface, meaning that the fluid elements forming the gas-liquid interface are movable and can be exchanged or displaced. Therefore, the velocity gradients present in the liquid around a rising bubble are smaller than those around a solid body, and less energy is dissipated in the liquid. Consequently, under the same driving force, bubbles rise faster than solid particles. If impurities are present in the surrounding liquid, however, the observed rise velocity varies somewhere between the one of particles with a fully mobile and fully immobile or rigid interface. This observation gave rise to the idea of a partially immobilized interface, which is useful to derive simplified models to account for the influence of surfactants, but which can be misleading sometimes. It is important to clarify that the inhomogeneous surfactant distribution causes additional surface specific forces which in turn change the flow pattern around a rising bubble. The surfactant itself can not render a fluid particle (partially) rigid.

In this work a substance is called surface active if its molecules, present in the liquid bulk phase, accumulate at the gas-liquid interface and lower the surface tension. The process of accumulation is characterized by two steps (see~\cite{chang1995}, section 4 and the reference therein): (1) the exchange of molecules between a surface and a subsurface layer, which is only a few molecule diameters in width, and (2) the transfer of molecules from the bulk liquid into the subsurface layer. The first step is called adsorption and the latter (bulk) mass transfer. In this work we consider only cases of diffusion-controlled adsorption, meaning that the diffusive transport of surfactant molecules from the bulk into the subsurface layer is much slower than their adsorption such that the surfactant concentrations in surface and subsurface layer are always locally  in equilibrium. Because the interface of a rising bubble is mobile and constantly entrained by the surrounding bulk liquid, the adsorbed surfactant is transported to the rear of the bubble, where it accumulates. As a consequence there is a region in the rear part with high surfactant concentration and lowered surface tension, while the upper part stays almost uncontaminated and the surface tension is unchanged. In the transition zone between contaminated and uncontaminated interface segments, strong gradients of surfactant concentration and surface tension result. These surface tension gradients lead to additional, so-called Marangoni forces, acting from points of low towards points of high surface tension. These tangential interface forces have to be balanced by shear forces in the liquid phase. The arising viscous forces act against the Marangoni forces from the top to the bottom and, hence, add to the overall drag force.

The described mechanisms and experimental observations led Davis and Acrivos~\cite{davis1966} to propose a mathematical model which incorporates the idea of a ``stagnant cap''. The interface is divided at a certain polar angle in two rotationally symmetric segments, one fully covered with surfactant and one completely clean. The contaminated cap is stagnant, meaning that the velocity at the interface is zero in a reference frame moving with the bubble center, and the shear stress at the cap is equal to the surface tension gradient. The clean bubble front instead is characterized by zero shear stress. The dividing angle is often referred to as stagnant cap angle. Such a clear separating circle is a strong idealization, assuming that the transition zone from fully contaminated to uncontaminated surface is small compared to the bubble size. A variety of theoretical and numerical studies based on the stagnant cap concept have appeared in the last decades, e.g.~\cite{he1991,fdhilaDuineveld1996,liao2000,zhangFinch2001,dukhin2015,dukhin2016}. One drawback of stagnant cap based models is that dynamic effects cannot be easily included, especially when the assumption of rotational symmetry is violated, as it occurs in most applications. In fact experiments show that the bubble motion is highly transient, especially after the bubble release. Sam et al.~\cite{sam1996} describe the typical transient rise of single bubbles under the influence of different surface active agents (frothers) as a three stage process that has been then observed several times in experiments. After releasing the bubble it accelerates until a maximum terminal velocity is reached; in the second stage the rise velocity starts to reduce until, given sufficient time, a plateau is reached. The constant plateau velocity defines the third stage. Interestingly, the first and second stage depend on the liquid bulk concentration of the surfactant, but the plateau velocity in the third stage seems to be fully determined by the surfactant type alone. Furthermore, the authors observed in their experiments that all investigated bubbles (bubble diameter $d_b < 3mm$), after an initial deformation to an ellipsoidal shape, were almost spherical at the top of the column. Also an influence of the frother concentration on the bubble path was reported: for bubbles showing path instability, the oscillation frequency decreased from the bottom to the top of the column with increasing frother concentration. Even in the case of large bubbles, the path at the column top was rectilinear. Since the work of Mougin and Magnaudet~\cite{mougin2002a} it is known that helical and zig-zagging trajectories of bubbles in the spherical and ellipsoidal regime are associated with pairs of rotating or symmetric vortices in the bubble wake. Sometimes during the initial acceleration, a transition from zig-zag to helical paths can be observed. The reverse transition, from helical to zig-zag, was only reported recently by Tagawa et al.~\cite{tagawaTakagi2014} for contaminated systems. The authors infer that a similar transition between different wake structures may happen. A strong surfactant influence on wake structure, path and shape was also visualized and comprehensively studied by Huang and Saito~\cite{huang2017_1,huang2017_2}. The possible impact of Marangoni forces on lift and drag was deduced from the bubble motion.
All previously mentioned experimental results contribute to partially understand and describe processes occurring on the reactor scale, for instance why the gas hold-up in flotation cells increases from the bottom to the top. However, to fully understand the transient behaviour of contaminated systems, complementary local field information of surfactant concentration, velocity and pressure at the interface and in the liquid bulk is necessary, which is currently only accessible via Direct Numerical Simulations (DNS). Early numerical studies assuming rotational symmetry~\cite{fdhilaDuineveld1996,liao2000} were only able to find a qualitative agreement with the previously described experimental observations, presumably because of too many limiting assumptions in the mathematical model. But also more sophisticated, fully three-dimensional DNS solving the coupled problems of two phase hydrodynamics, and surfactant transport in the bulk and on the interface~\cite{tasoglu2008,tukJas2008} could only partially reproduce and explain the typical three stage process. As we will show in the following chapters this is mainly due to the studied parameter range. The authors study P\'eclet numbers ($\mathrm{Pe}$) below $10^3$ (calculated with the kinematic viscosity of the bulk liquid and the molecular diffusivity of the dissolved surfactant in the bulk). For typical systems instead, $\mathrm{Pe}$ ranges from $10^4$ to $10^7$. The P\'eclet number is a measure for the ratio of convective to diffusive transport of a diluted species. High values of $\mathrm{Pe}$ are associated with thin boundary layers forming along the bubble surface, which determine the surfactant transfer, and hence, the ad- and desorption. The boundary layer width is approximately three to four orders of magnitude smaller than the bubble size~\cite{weiner2017}, which is why it is extremely demanding to resolve them in a DNS.

In this work we use an ALE interface tracking approach~\cite{muzPer1997,jasTuk2006,tukJas2008,tukJas2012} combined with a recently introduced subgrid-scale model methodology~\cite{weiner2017} for the surfactant transfer, which allows us to study realistic systems and to find a good agreement with experimental results. The results for a single rising bubble influenced by different amounts of soluble surfactant are discussed. We present local and global quantities which explain how the surfactant distribution in the bulk and on the interface is related to the macroscopically observed bubble motion, and examine thoroughly different contributions to the overall drag and lift forces. It is the author's intention to provide detailed information which could lead to better scale-reduced models accounting for the influence of contamination in bubbly flows.

%
%
\section{Mathematical Model}
\label{sec:mathMod}

The mathematical model for two-phase flows employs a sharp interface representation, meaning that the interface is represented as a surface of zero thickness with unknown time-dependent shape and location. Consider a fluid domain $\Omega$ containing two immiscible fluids, separated by a deformable interface. The interface, $\Sigma(t)$, separates the domain into two sub-domains, $\Omega^{+}(t)$ and $\Omega^{-}(t)$, corresponding to the two bulk phases. The presence of surfactant in the denser phase and on the interface is taken into account. Under the hypothesis of incompressible Newtonian fluids, isothermal conditions and absence of phase change and chemical reactions, the governing equations are based on the conservation of mass, momentum and surfactant molar mass. For the latter, the additional assumption of negligible inertia of the adsorbed surfactant on the interface is fundamental.

\subsection{Hydrodynamics}
\label{subsec:mathMod_hydro}
The velocity and the pressure field are obtained from the standard two-phase Navier-Stokes equations for incompressible Newtonian fluids. In local formulation, the continuity equation and the momentum balance in the bulk phases $\Omega^{\pm}(t)$ read
\begin{eqnarray}
  \nabla \cdot \vec{v} = 0, \label{eq:m1}\\
  \partial_{t} (\rho \vec{v}) + \nabla \cdot (\rho \vec{v} \otimes \vec{v}) = - \grad p + \nabla \cdot \mathbf{S}^{\mathrm{visc}} + \rho\ \grav,  \label{eq:m2}
\end{eqnarray}
where $\vec{v}$ is the barycentric velocity, $p$ the pressure, $\rho$ the density, $\mathbf{S}^{\mathrm{visc}} = \mu \left(\grad \vec{v} + (\grad \vec{v})^{\sf T}\right)$ the viscous stress tensor and $\mathbf{g}$ the acceleration due to gravity. 
The two bulk phases, separated by the moving interface $\Sigma(t)$, are coupled via transmission (or jump) conditions at the interface:
\begin{eqnarray}
    \dleftsq \vec{v}\drightsq = 0, \label{eq:m3}\\
    \vec{v} \cdot \vec{n}_{\Sigma} = \vec{v}^{\Sigma} \cdot \vec{n}_{\Sigma}, \label{eq:m4}\\
    \dleftsq p\mathrm{\mathbf{I}} - \mathbf{S}^{\mathrm{visc}} \drightsq \cdot \n_{\Sigma} = \sigma \kappa \n_{\Sigma} + \grad_{\Sigma} \sigma, \label{eq:m5}
\end{eqnarray}
where $\vec{v}^{\Sigma}$ is the interface velocity with $\vec{v}^{\Sigma} = \vec{v}_{|\Sigma}$\footnote{The notation $\cdot_{|\Sigma}$ denotes the trace of a quantity defined in $\Omega^{\pm}$ on the interface.} and $\kappa$ the surface curvature defined as $\kappa = - \nabla_{\Sigma} \cdot \vec{n}_{\Sigma}$, with $\grad_{\Sigma} \cdot$ representing the surface divergence\footnote{The surface gradient of a quantity $\phi(\vec{x})$ is defined as: $\grad_{\Sigma} \phi(\vec{x}) = \grad \phi(\vec{x}) - \n_{\Sigma}(\vec{x}) (\grad \phi(\vec{x}) \cdot \n_{\Sigma}(\vec{x}))$ at $\vec{x} \in \Sigma$, where $\phi$ is extended to a neighbourhood of $\Sigma$ as a differentiable function. Then, the surface divergence of a vector \vec{f} is defined as $(\grad_{\Sigma} \cdot \vec{f})(\vec{x}) = \mathrm{tr} (\grad_{\Sigma} \vec{f}) (\vec{x}).$}. The symbol $\sigma$ denotes the surface tension coefficient. In contaminated systems, the surface tension coefficient depends on the local concentration of surfactant on the interface $\sigma = \sigma(c^{\Sigma})$. The notation $\dleftsq \cdot \drightsq$ stands for the jump of a physical quantity across the interface\footnote{The jump of $\phi$ is defined as
$
\dleftsq \phi \drightsq (t,\vec{x}) = \lim_{h \rightarrow 0+} \left( \phi(t, \vec{x}+h\n_{\Sigma}) - \phi(t, \vec{x}-h\n_{\Sigma}) \right),$ $\vec{x} \in \Sigma(t).$}. %
The system of equations governing the hydrodynamic problem is completed by appropriate initial and boundary condition.

\subsection{Surfactant Transport}
\label{subsec:mathMod_st}
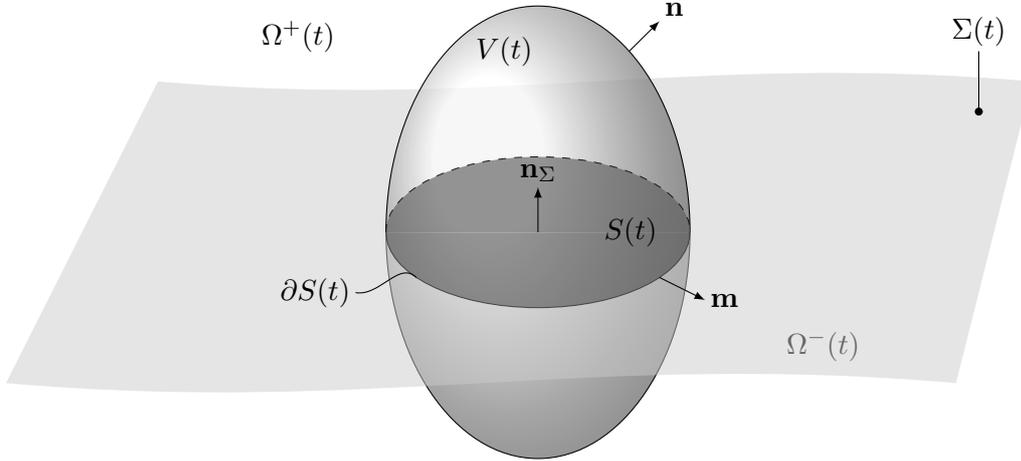
\begin{figure}[ht]
\centering
\begin{tikzpicture}
\def\R{2.0}
\def\RR{3.0} 
\def\angEl{30}

    \fill[white!60!gray, opacity=0.5] (-6,0) -- (-2,0) arc (180:360:2 and -1.0) -- (6,0) -- (6.5,2) to[out=175,in=-5] (-5,2) -- cycle;
    
\draw (3.2,-1.5) node[right]{$\Omega^-(t)$};

    \shade[ball color=white,opacity=0.7] (0,0) circle[x radius=\R, y radius=\RR];
    \draw (0,0) circle[x radius=\R, y radius=\RR];
    \DrawLatitudeCircle[\R]{0} 
    \fill[gray!70!black, opacity=0.65] (-2,0) arc (180:360:2 and -1.0) -- (-2,0) arc (180:360:2 and 1.0) -- cycle;
    \fill[white!60!gray, opacity=0.5] (-6,0) -- (-2,0) arc (180:360:2 and 1.0) -- (6,0) -- (5.5,-2) to[out=175,in=-5] (-7,-2) -- cycle;
 
\draw[->] (1.2,2.4) -- (1.6,2.8) node[above right] {$\mathbf{n}$};
\draw[->] (0,0) -- (0,0.6) node[above] {$\mathbf{n}_{\Sigma}$};
\coordinate[mark coordinate] (sigma) at (5.8,1.6);
\draw (5.8,2.4) node[above]{$\Sigma(t)$};
\draw (5.8,1.6) -- (5.8,2.4);
\draw (-2.6,2.6) node[left]{$\Omega^+(t)$};

\draw (0.8,0) node[right]{$S(t)$}; 
\draw (-2.4,-0.8) node[left]{$\partial S(t)$}; 
\draw (-2.4,-0.8) to[out=-10] (-1.6,-0.6);
\draw[->] (1.6,-0.6) -- (2.2,-0.9) node[right] {$\mathbf{m}$};
\draw (0,2.4) node[left]{$V(t)$};
    
\end{tikzpicture}


%
%
%
%
%
%
%
%
%
\caption{Domain representation for two-phase flows system.}
\label{fig:1}
\end{figure}

The core part of the mathematical model consists of the surfactant transport equations in the liquid phase and on the interface for moving domains.  
Let $V(t)$ be a control volume moving with velocity $\vec{w}$ inside the fluid domain $\Omega$. The boundary of the control volume is denoted by $\partial V(t)$, with $\vec{n}$ being the outer unit normal to $V(t)$. The intersection between the interface and the control volume $\Sigma(t) \cap V(t)$ is denoted as $S(t)$, with the boundary curve $\partial S(t)$ and the outer unit normal $\vec{m} \perp \vec{n}_{\Sigma}$ to $\partial S(t)$; see figure~\ref{fig:1}.
\FloatBarrier
The integral balance of surfactant molar mass for a moving control volume $V(t)$ in absence of chemical reactions (or any other source term)~\cite{bothe2005} reads
\begin{align}
\frac{d}{dt} \Biggl[ \int_{V(t)} c\ dV + \int_{S(t)} c^{\Sigma}\ dS \Biggr] =&
- \int_{\partial V(t)} c \left(\vec{v} - \vec{w} \right) \cdot \n\ dS - \int_{\partial V(t)} \vec{j} \cdot \n\ dS\nonumber\ + \\[1ex]
&- \int_{\partial S(t)} c^{\Sigma} \left(\vec{v}^{\Sigma} - \vec{w} \right) \cdot \vec{m}\ dl - \int_{\partial S(t)} \vec{j}^{\Sigma} \cdot \vec{m}\ dl,
\label{eq:m9}
\end{align}
where $c$ is the molar concentration of surfactant in the bulk ($\mathrm{mol/m^3}$), $c^{\Sigma}$ is the surface molar concentration of surfactant on the interface ($\mathrm{mol/m^2}$), and $\vec{j}$ and $\vec{j}^{\Sigma}$ are the diffusive fluxes in the bulk phase and on the interface, respectively. 
In local formulation the equations for surfactant transport in the bulk phase and on the interface read
\begin{align}
\partial_{t} c + \grad \cdot \left(c \vec{v} + \vec{j} \right) = 0&\quad \mathrm{in}\ \Omega \setminus \Sigma(t), \label{eq:m18}\\
\partial_{t}^{\Sigma} c^{\Sigma} + \grad_{\Sigma} \cdot \left( c^{\Sigma} \vec{v}^{\Sigma} + \vec{j}^{\Sigma} \right) = s^{\Sigma}&\quad \mathrm{on}\ \Sigma(t), \label{eq:m19}
\end{align}
where the sorption term $s^{\Sigma}$ satisfies
\begin{equation}
s^{\Sigma} + \dleftsq \vec{j} \cdot \n_{\Sigma} \drightsq = 0 \quad \mathrm{on}\ \Sigma(t),
\label{eq:m17}
\end{equation}
and equation~\eqref{eq:m19} is a dynamic boundary condition for equation~\eqref{eq:m18}.
The transport equations in the bulk and on the interface are completed by appropriate initial and boundary conditions.

The system of equations~\eqref{eq:m18} -~\eqref{eq:m17} is not closed, i.e.\ additional relations are needed to determine the diffusive fluxes and the source terms as functions of the primitive variables. The derivation of the transport equations can be found, for instance, in~\cite{bothe2005,alke2009,stone1990}.

\subsubsection{Diffusive fluxes}
\label{sec:math_1}
Under the assumption of dilute species concentrations both in the liquid phase and on the interface, the diffusive fluxes are modelled via Fick's law, i.e.
\begin{align}
  \vec{j} = - D\ \grad c&\quad \mathrm{in}\ \Omega^+(t), \label{eq:m24}\\
  \vec{j}^{\Sigma} = - D^{\Sigma}\ \grad_{\Sigma} c^{\Sigma}&\quad  \mathrm{on}\ \Sigma(t). \label{eq:m25}
\end{align}
Furthermore, homogeneous Neumann conditions~\eqref{eq:m26} for the diffusive fluxes at the outer domain boundary are assumed, i.e.
\begin{equation}
  \vec{j} \cdot \vec{n} = 0\quad \mathrm{on}\ \partial \Omega(t). \label{eq:m26}
\end{equation}

\subsubsection{Sorption Process}
\label{sec:math_2}

To model the sorption process, two limiting situations can be considered: diffusion-controlled sorption (fast) and kinetically controlled sorption (slow)~\cite{miller2014}. In the first case the sorption process is much faster than the diffusive transport, while in the latter case the sorption process is slower than the diffusive transport, typically due to the presence of a kinetic barrier. Thus, the transfer rate $s^{\Sigma}$ will be determined in two different ways, while the transmission condition~\eqref{eq:m17} always holds.
In both cases, the effect of surfactant on the interfacial surface tension is described by the surface equation of state which in a general form reads 
\begin{equation}
  \sigma - \sigma_0 = \Pi(c^{\Sigma}).
  \label{eq:m27}
\end{equation}
The function $\Pi(c^{\Sigma})$ in equation~\eqref{eq:m27} assumes a specific expression with respect to the sorption model employed; see~\cite{pesci2015} for more details on the derivation of equation~\eqref{eq:m27} and the full set of sorption models available in our library. For instance, in the Langmuir model, the surface tension equation of state reads
\begin{equation}
  \sigma = \sigma_0 + RT c^{\Sigma}_{\infty} \mathrm{ln} \left( 1 - \frac{c^{\Sigma}}{c^{\Sigma}_{\infty}} \right),
  \label{eq:m28}
\end{equation}
where $R$ is the universal gas constant equal to 8.3144 J/(mol K), $T$ is the absolute system temperature in Kelvin and $c^{\Sigma}_{\infty}$ is the saturated surfactant concentration, i.e. the maximum number of adsorbed molecules per area.
For the application case presented in section~\ref{sec:results} it has been proved that a \emph{fast} model is adequate to describe the sorption mechanism~\cite{kovalchuk2004,aksenenko1998}. For completeness we report in the following section~\ref{sec:math_2_2} the equations for the slow sorption mechanism as they are implemented in our solver, too.

\subsubsection{Diffusion-Controlled Sorption}
\label{sec:math_2_1}

In the case of fast (as opposed to kinetically-controlled transport) sorption, the ad- and desorption rates are locally in equilibrium, i.e.
\begin{equation}
  s^{\mathrm{ads}} \left( c_{|\Sigma}, c^{\Sigma} \right) = s^{\mathrm{des}} \left( c^{\Sigma} \right).
  \label{eq:m29}
\end{equation}
This equality leads to an additional local relationship between $c^{\Sigma}$ and $c_{|\Sigma}$, the so-called \textit{adsorption isotherm}, which needs to be accounted for in the numerical solution. For instance, the Langmuir adsorption isotherm relates the surface and bulk surfactant concentrations by means of the Langmuir equilibrium constant $a$, expressed in mol/m$^3$, and the saturated surface concentration:
\begin{equation}
  c^{\Sigma} = c^{\Sigma}_{\infty} \frac{c/a}{1 + c/a}.
  \label{eq:m30}
\end{equation}
%

\subsubsection{Kinetically-Controlled Sorption}
\label{sec:math_2_2}

In the case of kinetically-controlled sorption, the source term at the interface is computed as
\begin{equation}
  s^{\Sigma} = s^{\mathrm{ads}} \left( c_{|\Sigma}, c^{\Sigma} \right) - s^{\mathrm{des}} \left( c^{\Sigma} \right),
  \label{eq:m31}
\end{equation}
where $s^{\mathrm{ads}} \left( c_{|\Sigma}, c^{\Sigma} \right)$ and $s^{\mathrm{des}} \left( c^{\Sigma} \right)$ describe the rate of ad- and desorption, respectively. Note that the rate of adsorption is a function of the bulk concentration near the interface and the concentration of the adsorbed species, while the desorption rate is usually assumed to be a function of the adsorbed species only. 
From equation~\eqref{eq:m17} and the diffusive fluxes according to equation~\eqref{eq:m24}, a Neumann boundary condition for the bulk species equation is derived, namely
\begin{equation}
  \left(\grad c\right)_{|\Sigma} \cdot \n_{\Sigma} = - s^{\Sigma}/D.
  \label{eq:m32}
\end{equation}
%

%
%
\section{Numerical Model}
\label{sec:numMod}

The solution procedure is based on the arbitrary Lagrangian Eulerian (ALE) Interface Tracking method, originally presented by Hirt et al.~\cite{hirt1974}, later further developed by Muzaferija and Peri\'c~\cite{muzPer1997} and extended by Tukovi\'c and Jasak~\cite{tukJas2012}. Collocated Finite Volume / Finite Area methods are applied to solve the transport equations on unstructured meshes of general topology with moving mesh support. The interface is represented by a computational surface mesh (boundary mesh) advected in a semi-Lagrangian manner under the enforcement of jump conditions at the interface, whereas the volume mesh is updated through automatic mesh motion with Laplacian smoothing in order to preserve a high mesh quality. The interface divides the computational domain in two  disconnected sub-domains. The coupling between the two is enforced by the boundary conditions for pressure and velocity at $\Sigma(t)$ derived from the jump conditions~\eqref{eq:m3}~to~\eqref{eq:m5}. The governing equations are discretized in time using a second-order backward scheme known also as Gear's method~\cite{ferzigerPeric1996}. 
The two fluid domains $\Omega^{\pm}(t)$ are discretized by a finite number of convex polyhedral control volumes $V_P$. 
The centroid of the control volume is denoted by $P$, and the one of the neighbouring cell by $N$. The cell faces $f$ are of polygonal shape with area $S_f$ and area normal vector $\vec{S}_f$. In analogy to the volume discretization, the interface $\Sigma(t)$ is subdivided into polygonal control areas\footnote{The computational surface mesh can be seen as the boundary of the volume mesh, that is the faces approximating the interface belong to the boundary cells of the volume mesh.}. The center of a control area is again denoted by $P$ and the neighbouring one by $N$. 
The two control areas are separated by the edge $e$, characterized by the edge vector $\mathrm{\vec{e}}$, length $L_{e}$ and bi-normal $\mathrm{\vec{m}}_{e}$ (perpendicular to both $\mathrm{\mathbf{e}}$ and the edge normal vector $\mathrm{\n}_{e} = \left( \mathrm{\n}_{1} + \mathrm{\n}_{2} \right)/2$).

\subsection{Hydrodynamics and mesh motion}
\label{subsec:num_1}

The pressure-velocity coupling is solved applying the iterative pressure implicit with splitting of operators (PISO) algorithm~\cite{issa1986}. A modified version of the Rhie-Chow interpolation suggested in~\cite{tukJas2012} is employed to prevent a decoupling of pressure and velocity. 
A detailed description of the flow field solution and the mesh motion can be found in~\cite{tukJas2012,pesci2015}. 

\subsection{Surfactant transport}
\label{subsec:num_2}

In our system only one surfactant species is considered, while in~\cite{dieterKissling2015A,dieterKissling2015B} the methodology and the results for multicomponent surfactant systems in free-surface flows were presented. For cases where a fully resolved DNS for the species transport is not feasible due to high computational costs and numerical stability issues, a subgrid-scale model is employed. 

\subsubsection{Equation Discretization}
\label{subsubsec:num_2_1}

A Finite Volume method is applied to discretize the species transport equation in the liquid phase. In this case the transported quantity is the surfactant molar concentration $c$. The transport equation in integral form can be derived from~\eqref{eq:m9}. Applying Fick's law~\eqref{eq:m24} to describe the diffusive fluxes it reads
\begin{equation}
  \frac{d}{dt} \int_{V(t)} c\ dV + \int_{\partial V(t)} \left( c (\vec{v} -\vec{w}) - D \grad c \right) \cdot \n \ dS = 0.
  \label{eq:n34}
\end{equation}
The fully discretized transport equation for the control volume $V_P$ then reads
\begin{equation}
  \frac{ 3 c_{P}^{n} V_{P}^{n} - 4 c_{P}^{o} V_{P}^{o} + c_{P}^{oo} V_{P}^{oo}}{\Delta t} + \sum_f \bm{\phi}_f c_f^{n} = \sum_f D_f \left( \grad c\right)_f^{n} \cdot \mathbf{S}_f,
  \label{eq:n35}
\end{equation}
where $\bm{\phi}_f = \mathrm{\mathbf{S}}_{f} \cdot \left(\vec{u} - \vec{w} \right)_f$ is the face  flux. We denote the discrete velocity as $\vec{u}$ to distinguish between the discrete and the continuous quantity. The superscripts $n,\ o$ and $oo$ represent values evaluated at the new time instance $t^{n}$ and the two previous time instance $t^{o} = t^{n}-\Delta t$ and $t^{oo} = t^{o}-\Delta t$. The discretized concentration field is defined in the cell centres $P$ as $c_P$. Then, as required by the discretization of the diffusive and convective terms, the quantities $\left( \grad c\right)_f$ and $c_f$ have to be approximated at the faces centres.

The surfactant transport on the interface can also be derived from~\eqref{eq:m9}. Applying the Finite Area method, the local discretized form of the equation is obtained for the control surface $S_P$,
\begin{equation}
  \begin{split}
  \frac{ 3 (c_{P}^{S})^{n} (S_{P})^{n} - 4 (c_{P}^{S})^{o} (S_{P})^{o}\ +\ (c_{P}^{S})^{oo} (S_{P})^{oo}}{\Delta t} + \sum_{e} (\bm{\phi}_{e}^{S} c_{e}^{S,n}) =\\ \sum_{e} D_{e}^{\Sigma} \left( \grad_{S} c^{S}\right)_{e}^{n} \cdot (\mathrm{\mathbf{m}}_{e} L_{e})\ +\ s_{P}^{S} S_{P}
  \end{split}
  \label{eq:n36}
\end{equation}
with the relative edge flux $\bm{\phi}_{e}^{S} = (\mathrm{\mathbf{m}}_{e} L_{e}) \cdot (\vec{u} - \mathbf{w})_{||}$ and $\grad_{S}$ representing the discrete counterpart of the surface gradient operator $\grad_{\Sigma}$. The quantity ${c}^{S}$ denotes the discretized counterpart of the continuous quantity $c^{\Sigma}$ in the face center $c_{P}^{S}$ or interpolated on the edge center $c_{e}^{S}$. The terms $s_{P}^{S}$ is the discrete source term\footnote{The source term can be split in explicit $s_{P,\mathrm{exp}}^{S}$ and implicit $s_{P,\mathrm{imp}}^{S}$ parts, respectively. In case of \emph{fast} sorption processes the source term appears only in an explicit form, thus the splitting is not necessary.}.

The diffusion terms (bulk and surface transport) can be decomposed into orthogonal and non-orthogonal contributions, treating the first one implicitly and the second one explicitly; see~\cite{tukJas2008}.

Equations in the bulk and on the interface are solved with a Preconditioned Bi-conjugate Gradient (PBiCG) linear solver, with a Diagonal Incomplete-LU preconditioner (DILU) and tolerance $1 \cdot 10^{-12}$.

\subsubsection{Sorption Process}
\label{subsubsec:num_2_2}

The coupling between bulk and interfacial surfactant transport is achieved applying a Dirichlet (\emph{fast} sorption) or a Neumann (\emph{slow} sorption) boundary condition to the diffusive term in~\eqref{eq:n35} and the respective constitutive equation for the source term in~\eqref{eq:n36} derived from the sorption model. In our code, a \emph{sorption model library} is available~\cite{pesci2015}, where multiple sets of models, both \emph{fast} and \emph{slow} sorption models, are implemented. Depending on the chosen model, the solver will automatically use the respective boundary conditions and source term. As in this work we use a fast sorption model, slow sorption is not treated in this section, but its numerical treatment can be found in~\cite{pesci2015} and~\cite{pesciSPP2017}.

For diffusion controlled (fast) sorption processes, the source term for the surface concentration equation is computed from the transmission condition~\eqref{eq:m17} as
\begin{equation}
  s^{\Sigma} = \vec{j} \cdot \n_{\Sigma} = - D (\grad c \cdot \n )_{\Sigma} =: s_{\mathrm{fast}}^{\Sigma}.
  \label{eq:n38}
\end{equation}
Then the discretized surface transport equation~\eqref{eq:n36} is solved to obtain the new surface concentration field of the surfactant species. Since the adsorption isotherm $c^{\Sigma} = f (c_{|\Sigma})$ is known, e.g.~equation~\eqref{eq:m30}, the value of
\begin{equation}
  c_{|\Sigma} = f^{-1} (c^{\Sigma})\quad \mathrm{at}\ \Sigma
  \label{eq:n37}
\end{equation}
is taken as a Dirichlet boundary condition for the discretized surfactant bulk equation~\eqref{eq:n35}. 
After solving the interfacial and bulk surfactant transport equations, the surface tension $\sigma = \sigma(c^{\Sigma})$ is updated according to the chosen sorption model.

\subsubsection{Implicit SGS Model}
\label{subsubsec:num_2_3}

Consider the species, in this case surfactant, transport problem in the liquid phase. The species transport along the bubble interface is mainly governed by two transport processes, namely advection in streamwise direction and diffusion in interface normal direction. For the bubbles under investigation (Reynolds number $\operatorname{Re} \approx 10^{2}$ and P\'{e}clet\footnote{The Reynolds number is defined as $\operatorname{Re} = uL/\nu$, the P\'{e}clet number as $\operatorname{Pe} = \operatorname{Re} \cdot \operatorname{Sc} = uL/D $ and the Schmidt number as $\operatorname{Sc} = \nu/D$, where $u$ is the velocity, $L$ the characteristic length, $\nu$ the kinematic viscosity and $D$ the molecular diffusivity.} number $\operatorname{Pe} \approx 10^7$) the species transport is dominated by advection, leading to a very thin concentration boundary layer around the bubble ($\delta_s \approx 10^{-6}\ \mathrm{m}$). Thus, a fully resolved 3D DNS for the species transport is not feasible due to the high computational costs. In previous studies, for instance in~\cite{cuenot1997}, this issue was faced using a very fine grid on the axisymmetric case with bubbles at steady state, i.e. with a non deformable interface. Moreover, the hydrodynamics is solved only in the liquid phase. This approach is not suitable for the study of the initial transient of the bubble rise and the effect of surfactant on it. An effective solution to the thin species boundary layer problem is the use of a subgrid-scale (SGS) model, a by now standard approach in mass transfer problems~\cite{botheFL2013}, to approximate the surfactant boundary layer in the vicinity of the bubble. The main idea behind the SGS model is to employ an appropriate model-function to compute the numerical (SGS) fluxes on all cell faces of an interface cell. These SGS fluxes are used to correct the numerical fluxes to accurately predict the species transport close to the interface, even if the concentration boundary layer is fully embedded in a single cell layer. 
Our approach is based on the latest development of the SGS model presented in~\cite{weiner2017}, although here the transport equation is coupled to the sorption process at the interface and solved implicitly to improve the numerical stability and to allow for larger time steps.
In~\cite{weiner2017} the transport equations are solved explicitly with a direct modification of diffusive fluxes and concentration values at the required faces. Since our solution is implicit, i.e.~the fluxes contain the unknown variable $(c_f)^n,\ (\grad c)_f^n$, we modify the diffusion coefficient and the advective term as described in the following subsections.
It has been shown~\cite{weiner2017} that the SGS model can reduce the mesh resolution requirements near the interface by a factor of ten or more.

Applying the SGS model to the bulk surfactant transport results in the following discretized transport equation (from~\eqref{eq:n35}) solved with locally modified diffusion coefficients and advection flux-field:
\begin{equation}
    \label{eq:n54}
    \frac{ 3 c_{P}^{n} V_{P}^{n} - 4 c_{P}^{o} V_{P}^{o} + c_{P}^{oo} V_{P}^{oo}}{\Delta t} + \sum_f \bm{\phi}^{\mathrm{SGS}}_f c_f = \sum_f D^{\mathrm{SGS}}_f \mathbf{S}_f \cdot \left( \grad c\right)_f.
\end{equation}
The derivation of $\bm{\phi}^{\mathrm{SGS}}$ and $D^{\mathrm{SGS}}$ is reported in Appendix~\ref{subsec:app_0}.

\subsubsection{SGS Model and Fast Sorption}
\label{subsubsec:num_2_4}
The inverse expression of the adsorption isotherm~\eqref{eq:n43} serves as a Dirichlet boundary condition for the bulk transport. The bulk transport is coupled to the surface balance via the source term~\eqref{eq:n38}. Also for computing the source term, we apply the locally corrected SGS diffusion coefficients, i.e.
\begin{equation}
  s^{\Sigma}_{f_i^{\Sigma}} = - D^{\mathrm{SGS}}_{f_i^{\Sigma}} \left( \partial_n c \right)^{\mathrm{num}}_{f_i^{\Sigma}}.
  \label{eq:n55}
\end{equation}
%

\subsection{Validation}
\label{subsec:num_3}
The validation of the pure hydrodynamics has been conducted comparing with the experimental data by Duineveld~\cite{duineveld1995} for single bubbles rising in pure water and can be found in~\cite{pesciSPP2017}. There, rise velocity and aspect ratio for bubbles with radii ranging between 0.75 and 1.0 mm were considered and found in very good agreement with the experimental data; see figure 15.13, page 418 in~\cite{pesciSPP2017}. The validation for the sorption source term for fast and slow sorption processes can be found in~\cite{pesciSPP2017}, too. The validation of the implicit SGS model can be found in appendix~\ref{subsec:app_3_a}.

%

%
%
\section{Results and Discussion}
\label{sec:results}

A single air bubble rising in aqueous solution contaminated by surfactant is considered. For this prototypical problem a direct comparison with experimental results is possible. The experimental data and a short description of the corresponding set-up can be found in~\cite{pesciSPP2017}. More details on the experimental set-up are presented in~\cite{krzan2002,krzan2007}. Briefly, a digital camera was used to record the bubble motion at various distances from the orifice. Four to eight images of the bubble were obtained for each camera position illuminating the region of interest with a strobe frequency from 100 to 200 Hz. The higher frequency was used for the initial acceleration stage. From the distances between the subsequent positions of the bubble and knowing the strobe frequency, the local bubble velocity is computed. The measurement at each camera position was performed at least three times and mean local velocity values were calculated.

%
\subsection{Simulations set-up}
\label{subsec:5_1}

\begin{figure}[ht]
\centering
\subfloat[][\emph{Full domain}.]
{\includegraphics[width=.45\textwidth]{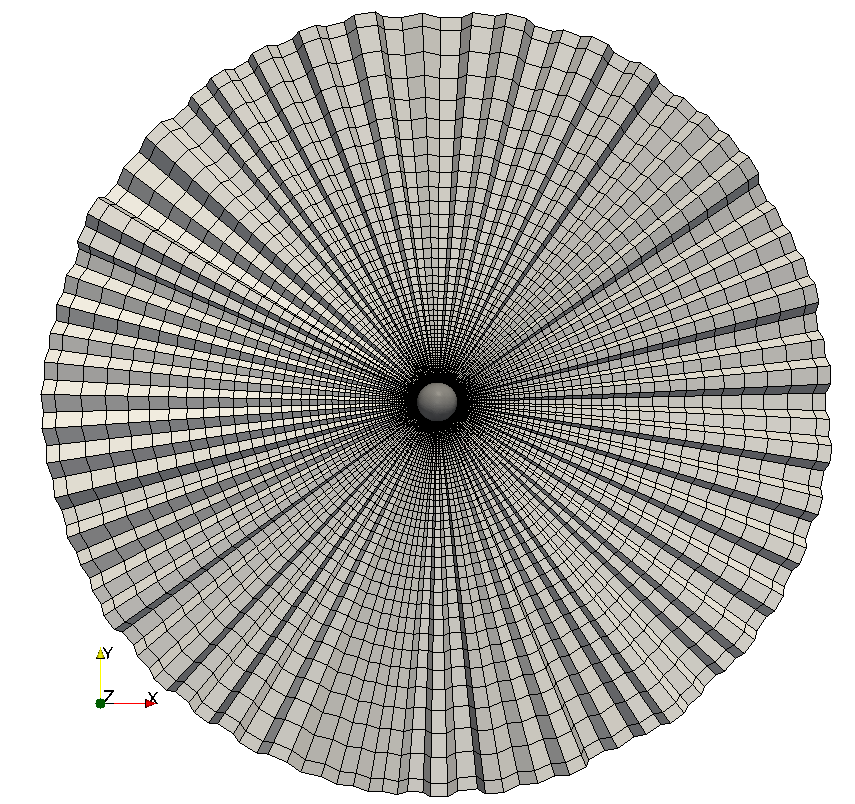}\label{fig:v8a}} \quad
\subfloat[][\emph{Enlarged view of the bubble region}.]
{\includegraphics[width=.45\textwidth]{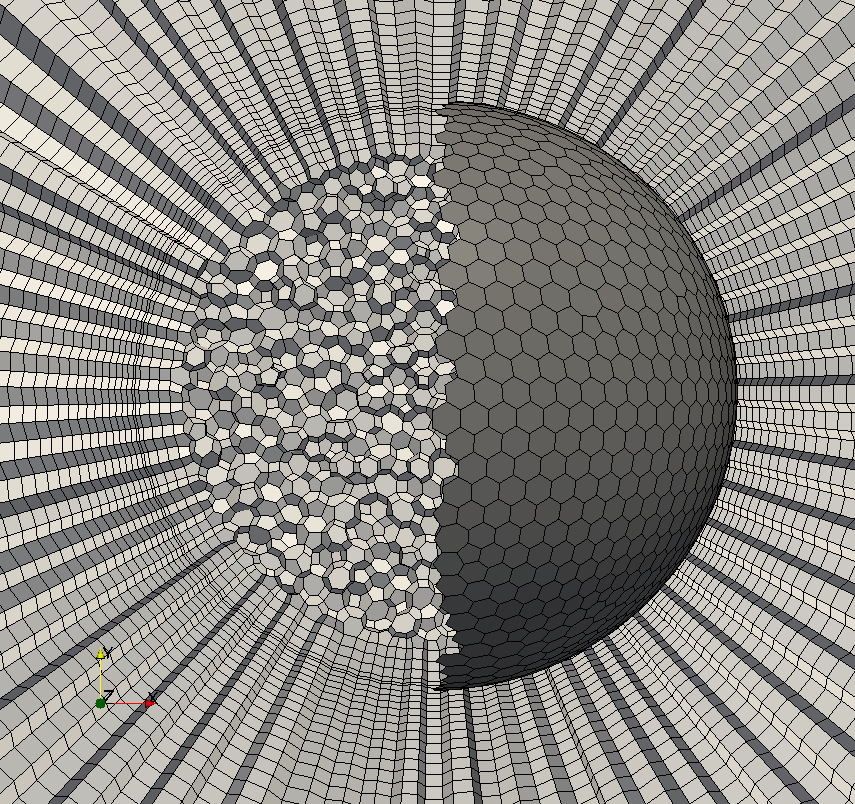}\label{fig:v8b}}
\caption{3D computational domain for a rising bubble. Inner, outer and surface (dark grey on the right) meshes.}
\label{fig:v8}
\end{figure}

The material properties used in the simulations are reported in tables~\ref{tab:tr2} and~\ref{tab:tr3}. The bubble diameter is $d_B = 1.45\ \mathrm{mm}$. The initial shape of the bubble is a sphere positioned in the center of a spherical domain with radius twenty times the bubble radius. 
The computational domain is divided into two sub-domains, one representing the gas phase and the other one representing the liquid phase. The meshes used for the simulations consist of polyhedral cells in the gas phase and prismatic cells with polyhedral base in the liquid phase, as can be seen in figure~\ref{fig:v8}. The calculation is performed in a moving reference frame (MRF) that follows the bubble center during its rise, while the interface is deformable. The presence of a non-inertial reference frame located in the center of the bubble is taken into account via a correction in the momentum equation ($\rho \mathrm{\vec{a}_{MRF}}$ added to the momentum equation) and the velocity boundary condition at the outer domain boundary, $\vec{v}_{\mathrm{out}} = - \vec{v}_{\mathrm{MRF}}$\footnote{The boundary condition \emph{inletOutlet} available in OpenFOAM is used. The inlet velocity is set to $- \vec{v}_{\mathrm{MRF}}$, at the outlet a \emph{zeroGradient} condition is set.}.
A (constant) time step $\Delta t \approx \mathcal{O}\left(10^{-6}\right) \mathrm{s}$ is chosen to fulfil the criterion for the interface numerical stability~\cite{tukJas2012}, i.e. $\Delta t < \sqrt{\rho^+\ \mathrm{min}(l_{PN})^3/(2 \pi \sigma)}$ with $\mathrm{min}(l_{PN})$ being the minimum distance between two face centres on the interface. The surfactant used in the experiments is the non-ionic dodecyl-dimethyl-phosphine-oxide (C$_{12}$DMPO). Its sorption process is modelled via the fast Langmuir sorption model. For the simulations the bubble shape is initialized as a sphere with zero initial velocity. To model the surfactant transport in the bulk phase in the vicinity of the interface, the SGS model described in section~\ref{subsubsec:num_2_3} and Appendix~\ref{subsec:app_0} is used. From the available experimental data we consider the clean case and other three different initial surfactant concentrations as a reference.
\FloatBarrier
\begin{table}[ht]
\caption{Fluid properties.}
\label{tab:tr2}
\centering
\begin{tabular}{ccccc}
\toprule
 $\rho^+\ \mathrm{kg/m^3}$ & $\mu^+\ \mathrm{kg/(ms)}$ & $\rho^-\ \mathrm{kg/m^3}$ & $\mu^-\ \mathrm{kg/(ms)}$ & $\sigma_0\ \mathrm{N/m}$\\
\midrule
 $997.3$ & $9.3 \cdot 10^{-4}$ & $1.1965$ & $1.83 \cdot 10^{-5}$ & $0.0724$\\
\bottomrule
\end{tabular}
\end{table}
\FloatBarrier
\begin{table}[ht]
\caption{Surfactant (C$_{12}$DMPO) properties, fast Langmuir adsorption model parameters.}
\label{tab:tr3}
\centering
\begin{tabular}{ccccc}
\toprule
 $c^{\Sigma}_{\infty}\ \mathrm{mol/m^2}$ & $a_L\ \mathrm{mol/m^3}$ & $D\ \mathrm{m^2/s}$ & $D^{\Sigma}\ \mathrm{m^2/s}$ & $T\ \mathrm{K}$\\
\midrule
 $4.17 \cdot 10^{-6}$ & $4.85 \cdot 10^{-3}$ & $5 \cdot 10^{-10}$ & $5 \cdot 10^{-7}$ & $296$\\
\bottomrule
\end{tabular}
\end{table}
\FloatBarrier
The surface diffusivity $D^{\Sigma}$ is only an estimate, since it is not possible to accurately measure it. Nevertheless, a parameter study with $D^{\Sigma}$ varying in the range of $[10^{-6}\ ...\ 10^{-9}]\ \mathrm{m^2/s}$ confirmed that its variation has only a minor effect on the sorption dynamics and rise velocity, because the transport is advection dominated.

\begin{figure}[ht]
\centering
\includegraphics[width=1.0\textwidth]{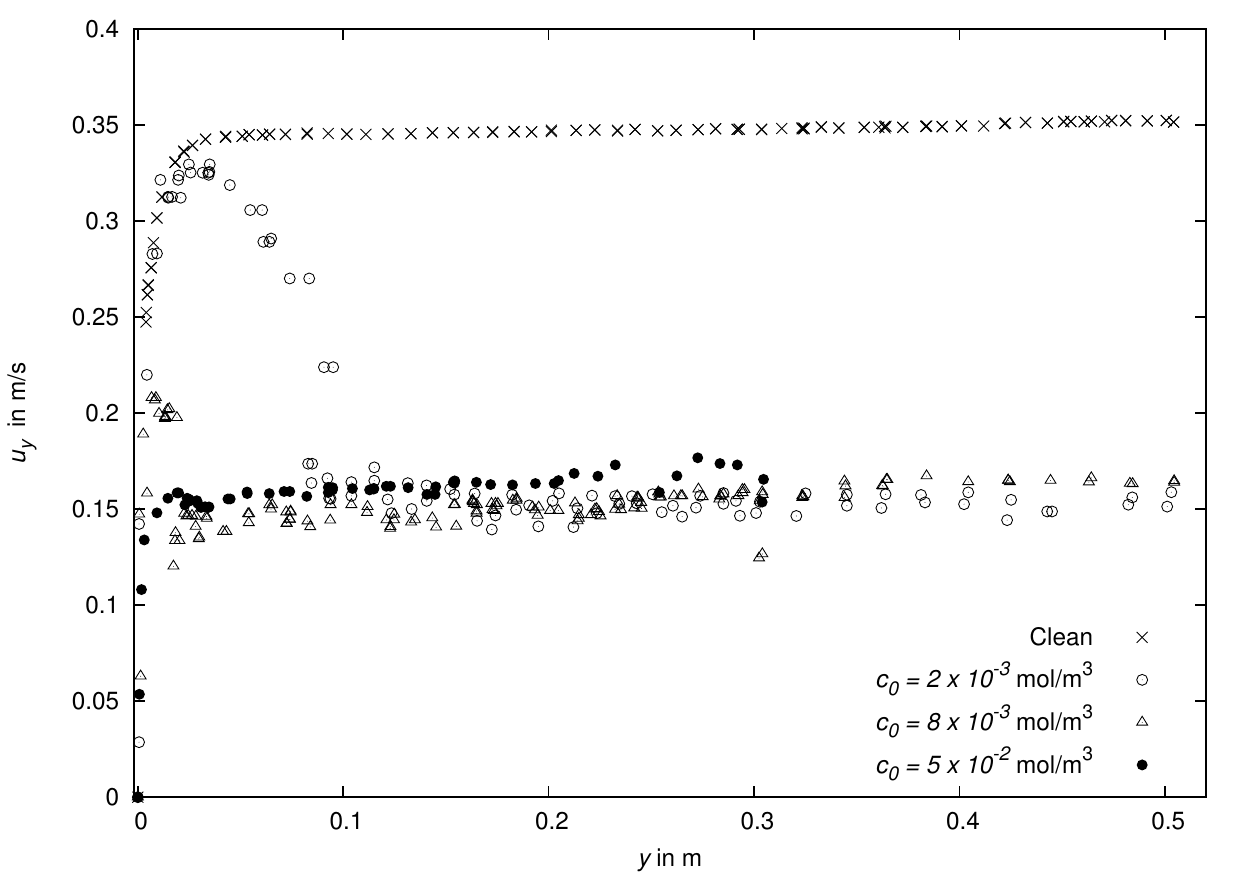}
\caption{Experimental bubble center rise velocities in rise direction $y$. Data from~\cite{pesciSPP2017}.}
\label{fig:expRef3}
\end{figure}

The selected experimental results from~\cite{pesciSPP2017} are given in figure~\ref{fig:expRef3} and they will be the base for our discussion of the simulation results. According to~\cite{pesciSPP2017}, the average accuracy of the experimental data is $\pm 5 \%$. Three different initial concentrations in the liquid phase\footnote{The surfactant concentration in the gas phase is set to zero.} are considered, a relatively small one, $c_{0,1} = 2 \cdot 10^{-3}\ \mathrm{mol/m^3}$, an intermediate one, $c_{0,2} = 8 \cdot 10^{-3}\ \mathrm{mol/m^3}$, and a relatively high one, $c_{0,3} = 5 \cdot 10^{-2}\ \mathrm{mol/m^3}$. To this three initial surfactant concentrations correspond Marangoni numbers $\operatorname{Ma}$ of 34, 49 and 70, which express the ratio between surface tension and viscous forces. $\operatorname{Ma}$ is computed as
\begin{equation}
\operatorname{Ma} = \frac{RT c^{\Sigma}_{\infty}}{\mu^+ U_{max}},
\label{eq:Ma}
\end{equation}

where $U_{max}$ is the peak rise velocity reached by the bubble. Moreover, the respective surface equilibrium concentrations computed from the Langmuir isotherm~\eqref{eq:m30} are $c^{\Sigma}_{\mathrm{eq},1} = 1.2175 \cdot 10^{-6}\ \mathrm{mol/m^2}$, $c^{\Sigma}_{\mathrm{eq},2} = 2.596 \cdot 10^{-6}\ \mathrm{mol/m^2}$ and $c^{\Sigma}_{\mathrm{eq},3} = 3.801 \cdot 10^{-6}\ \mathrm{mol/m^2}$. 
\FloatBarrier
In figure~\ref{fig:expRef3}, the well-known velocity profile of rising bubbles under the effects of surfactants can be observed. The bubble rising in clean water (crosses), thus with a fully mobile surface, after an initial acceleration reaches a constant velocity that is the terminal velocity. The same can be observed for bubbles rising in highly contaminated solutions (filled circles). After an initial acceleration, the bubble velocity reaches a constant value, although it is much lower than the velocity for a mobile surface. At intermediate concentrations (empty circles, triangles) there is still an acceleration phase, but after reaching the peak velocity the bubble decelerates. The bubbles keep decelerating until they reach a quasi-steady terminal velocity which is similar to the case with very high contamination.

In applications involving bubbly flows it is fundamental to correctly reproduce the initial transient stage of the bubble rise, because it determines the position of the bubble and perhaps also how it will interact with other bubbles. Thus in sections~\ref{subsec:5_2},~\ref{subsec:5_3} and~\ref{subsec:5_4} the attention is focused on correctly reproducing the transient velocity profiles.

%
\subsection{Discussion on under-resolved species boundary layers}
\label{subsec:5_2}

The simulation results for the clean case have already been compared to the experimental ones in~\cite{pesciSPP2017} showing a very good agreement. These results are reproduced in section~\ref{subsubsec:5_5_1} with additional information about the bubble path.
\begin{figure}[ht]
\centering
\includegraphics[width=1.0\textwidth]{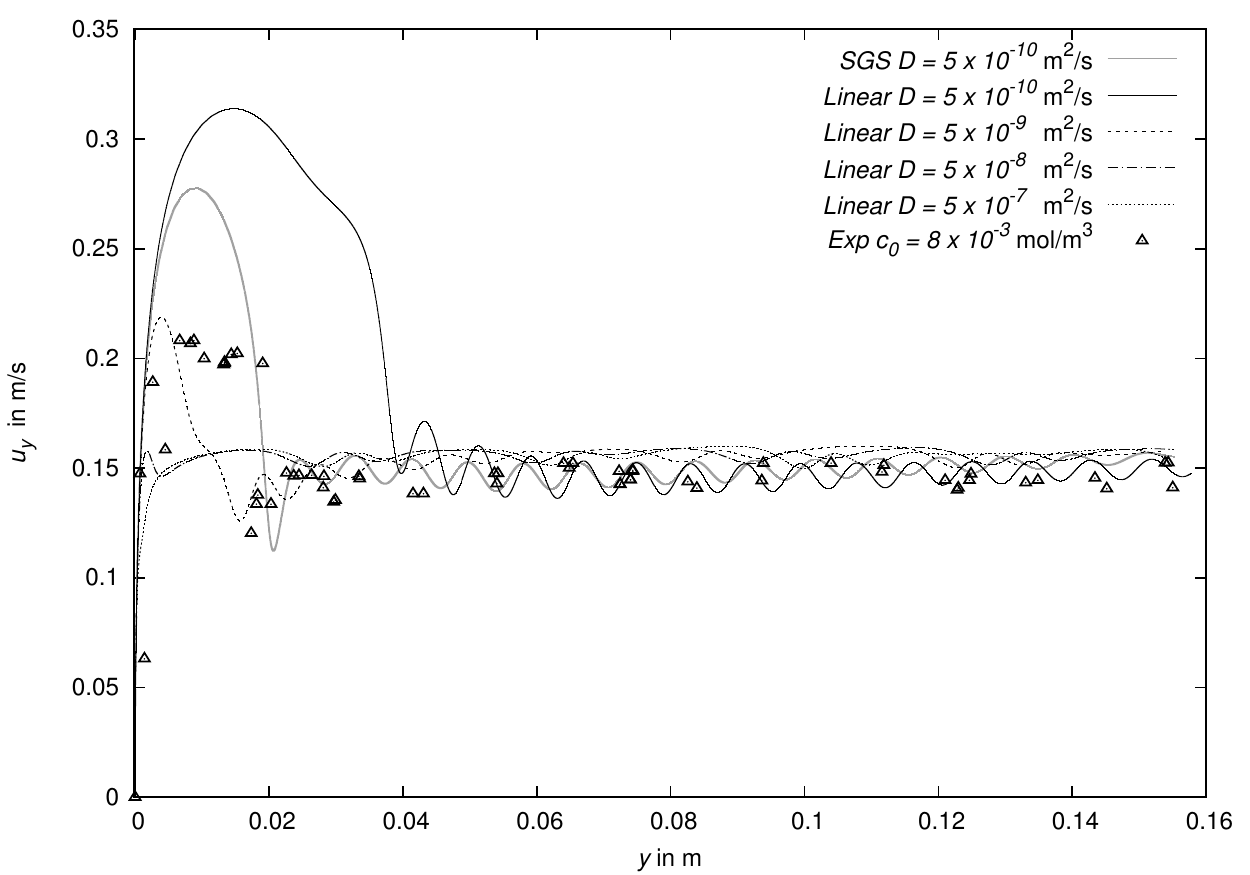}
\caption{Comparison between simulations without (black lines) and with SGS model (grey line); $c_0 = 8 \cdot 10^{-3}\ \mathrm{mol/m^3}$; simulated time $t = 1\ \mathrm{s}$.}
\label{fig:LINvsSGS}
\end{figure}
\FloatBarrier
The surfactant transport problem is a typical case with highly non-linear concentration profiles at the interface in a very thin boundary layer. Thus a standard linear interpolation from the cell centres to the face centres leads to over- or underestimated diffusive and convective fluxes normal to the interface, resulting in an unphysically thick boundary layer. Only thanks to the application of the SGS model described in section~\ref{subsubsec:num_2_3} it becomes possible to study cases with real diffusion coefficients for the surfactant in the liquid phase. The usage of physical diffusivities is imperative to get the correct transient velocity, since it is not only affecting the surfactant bulk transport but also the sorption mechanism itself, as described in section~\ref{sec:numMod} and in~\cite{pesciSPP2017}. A comparison between the standard interpolation and the flux correction by the SGS model is given in figure~\ref{fig:LINvsSGS}. The results there refer to the intermediate surfactant bulk concentration $c_{0,2} = 8 \cdot 10^{-3}\ \mathrm{mol/m^3}$. A first set of simulations is run without SGS modelling to test the sensitivity to different diffusivities with a fixed mesh resolution (first cell thickness $l \approx$ 16 $\mu$m). For a realistic diffusivity, the rise velocity is overpredicted; see figure~\ref{fig:LINvsSGS}. On the other hand, increased diffusion coefficients result in thicker species boundary layers that can be resolved by this mesh, but at the same time they speed-up the adsorption process and, consequently, the rise velocity approaches the steady state value too quickly. 

Figure~\ref{fig:LINvsSGS} depicts also the velocity profile obtained with the SGS approach and the physical diffusivity. The initial transient velocity is reproduced much better, but the velocity peak is still overestimated. This difference can be explained considering the bubble formation and detachment time in the experiments. As it is known from experimental works, e.g.~\cite{krzan2007,malysa2011,ulaganathan2016}, the initial transient velocity depends strongly on the time of bubble formation and release. During the bubble formation process, the newly generated bubble surface is exposed to the contaminated solution. Thus, when the bubble detaches from the capillary, its interface holds already a certain amount of surfactant. This relatively small (not above $10 \%$ of $c^{\Sigma}_{\mathrm{eq}}$) initial surface contamination influences the peak rise velocity. From the experiments, the adsorption time for detaching bubble is known to be about 1.6 s, hence, during this time there would be a diffusion of surfactant towards the growing bubble surface. The surface coverage at release is a function of time and bulk surfactant concentration, and it can be estimated as
\begin{equation}
c^{\Sigma}_0(t) =  \frac{1}{3} \left( 2 c_0 \sqrt{\frac{3 D t}{7 \pi}} \right),
\label{eq:estCs0}
\end{equation}
a formula taken from~\cite{miller1995ch}~(pages 118-119). A summary of the estimated surface coverages at detachment is reported in table~\ref{tab:estCS}. Within our simulation set-up, different detachment times can be investigated varying the initial surfactant surface concentration.
\begin{table}[ht]
\caption{Initial surface coverage estimates at release time $t_{rel} = 1.6\ \mathrm{s}$ with \\$D = 5 \cdot 10^{-10}\ \mathrm{m^2/s}$.}
\label{tab:estCS}
\centering
\begin{tabular}{ccccc}
\toprule
 $c_{0}\ \mathrm{mol/m^3}$ & $2 \cdot 10^{-3}$ & $8 \cdot 10^{-3}$ & $5 \cdot 10^{-2}$\\
\midrule
$c^{\Sigma}_0(t_{rel})\ \mathrm{mol/m^2}$ & $1.39 \cdot 10^{-8}$ & $5.57 \cdot 10^{-8}$ & $3.48 \cdot 10^{-7}$\\
 & or & or & or \\
 & $1.14 \%\ c^{\Sigma}_{\mathrm{eq},1}$ & $2.14 \%\ c^{\Sigma}_{\mathrm{eq},2}$ & $9.16 \%\ c^{\Sigma}_{\mathrm{eq},3}$\\
\bottomrule
\end{tabular}
\end{table}
\FloatBarrier
Before presenting these results, a mesh sensitivity study of the full problem with SGS modelling is necessary. Note that for the simulations corresponding to figures~\ref{fig:LINvsSGS} and~\ref{fig:meshDep} the initial surface concentration was set to zero, $c^{\Sigma}(t = 0) = 0\ \mathrm{mol/m^2}$.

%
\subsection{Mesh sensitivity study}
\label{subsec:5_3}

\begin{figure}[ht]
\centering
\includegraphics[width=1.0\textwidth]{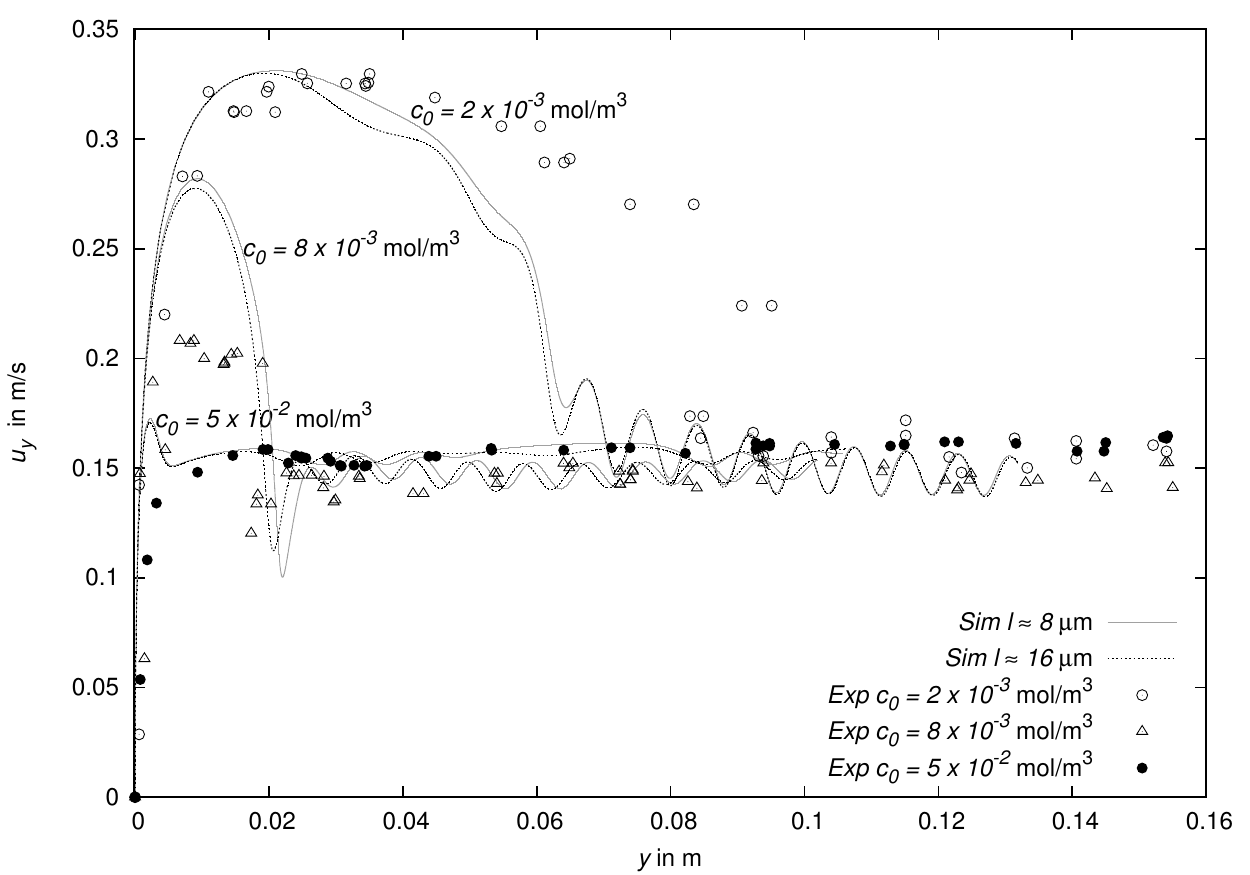}
\caption{Rise velocity for three initial surfactant bulk concentrations. Results for two mesh resolutions (continuous lines - fine mesh; dotted lines - coarse mesh); simulated time up to $t = 0.6\ \mathrm{s}$.}
\label{fig:meshDep}
\end{figure}

To study the dependency of the numerical results with respect to the mesh resolution, simulations with different initial bulk concentrations and zero initial surface coverage are performed on two different meshes, a fine one ($\approx$ 320000 cells) with a first layer thickness of $l \approx 8\ \mu$m and 3700 faces on the interface, and a coarser one ($\approx$ 160000 cells) with a first layer thickness of $l \approx 16\ \mu$m and 2400 faces on the interface. 
As can be noticed from figure~\ref{fig:meshDep} the biggest difference between fine and coarse mesh is encountered in the decelerating phase for the smallest initial bulk concentration. In fact, for higher $c_0$, the bubble rises slower, thus the Reynolds number is smaller and consequently the hydrodynamic boundary layer thicker. A thicker hydrodynamic boundary layer is then well resolved by a coarser mesh, too. Even though there is a small difference between the coarse and the fine mesh results, for the simulations that are reported below we decided to use the coarser mesh because of the required computational time. Only for the least contaminated bubble, the bubble path is reported both for the coarse and the fine meshes, since the helical path was more pronounced in the latter case. The fine and coarse cases ran in parallel (MPI) on three and five cores, respectively, with the interface (liquid side) and its counterpart (gas side) on the same processor. The computations took between thirty and sixty days to reach 1 s of simulated physical time.
\FloatBarrier
%

%
\subsection{Bubble shape and path under the influence of surfactant}
\label{subsec:5_4}

\subsubsection{Initial surface coverage}
\label{subsubsec:5_4_1}

We vary the detachment time via pre-contaminating the bubble surface, while the initial shape deformation at detachment is neglected. 
Since equation~\eqref{eq:estCs0} provides only an estimate of the initial surface coverage at release, we found it appropriate to conduct a parameter study varying $c^{\Sigma}_0$ for the different bulk concentrations to obtain a more precise value of the initial surface contamination. 

\begin{figure}[ht]
\centering
\subfloat[][\emph{$c_0 = 2 \cdot 10^{-3}\ \mathrm{mol/m^3}$}.]
{\includegraphics[width=.48\textwidth]{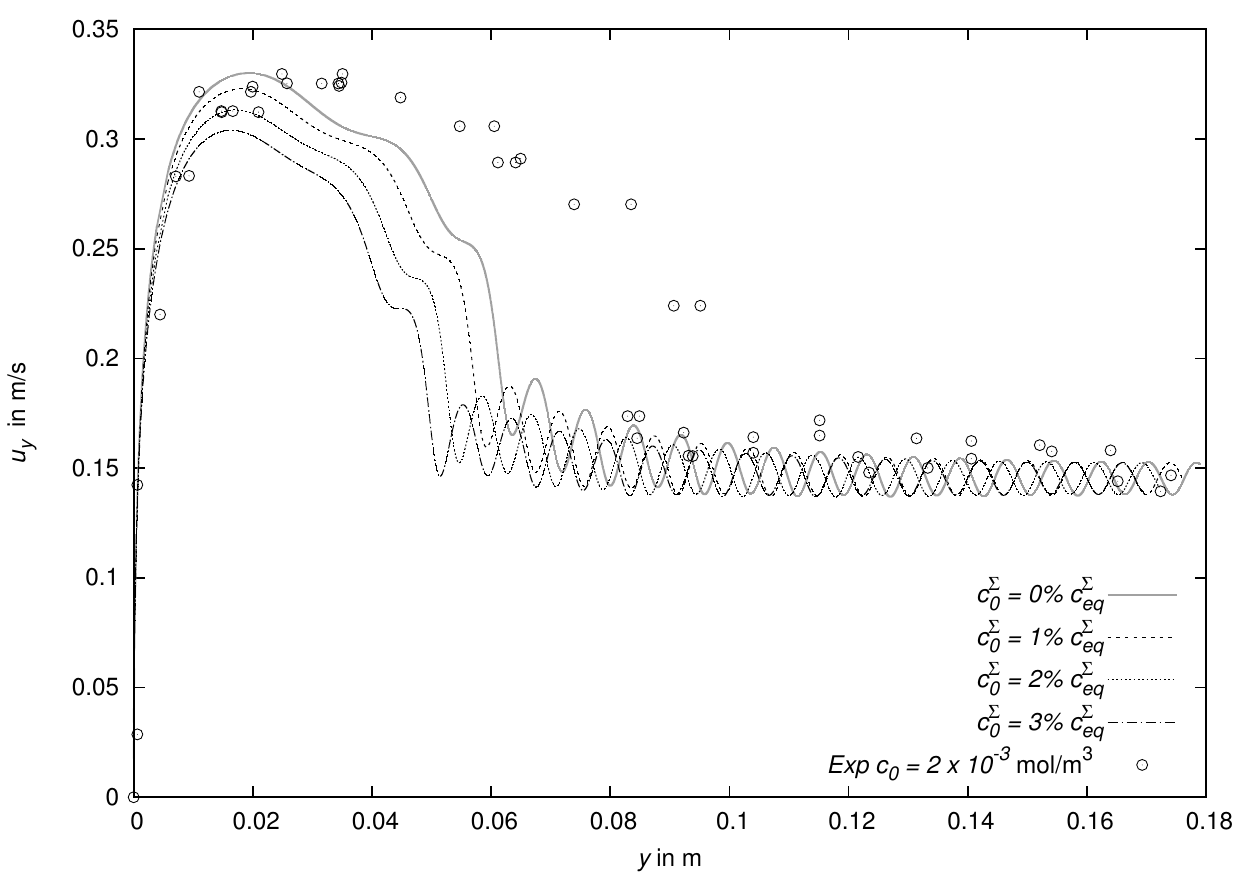}\label{fig:b3cS0}} \quad
\subfloat[][\emph{$c_0 = 8 \cdot 10^{-3}\ \mathrm{mol/m^3}$}.]
{\includegraphics[width=.48\textwidth]{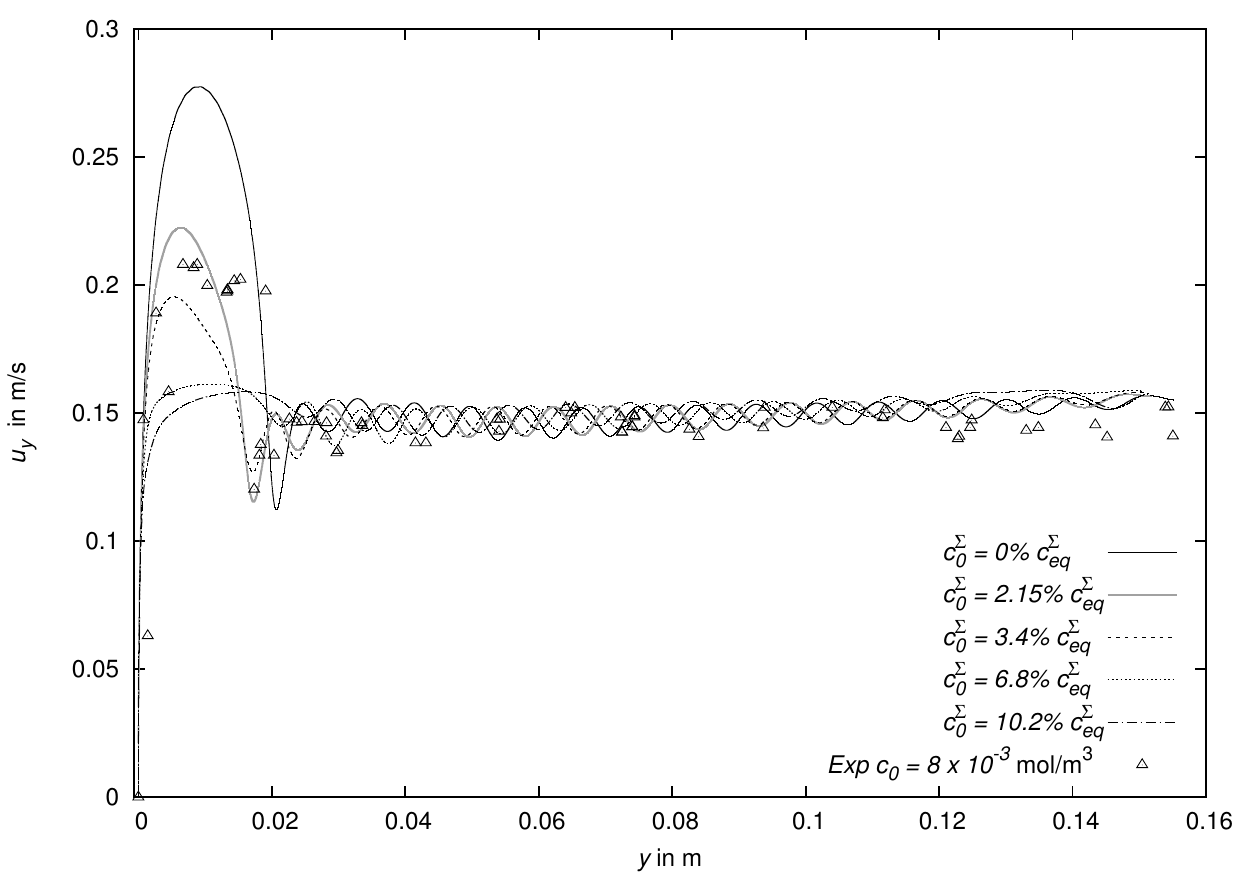}\label{fig:b5cS0}} \\
\subfloat[][\emph{$c_0 = 5 \cdot 10^{-2}\ \mathrm{mol/m^3}$}.]
{\includegraphics[width=.48\textwidth]{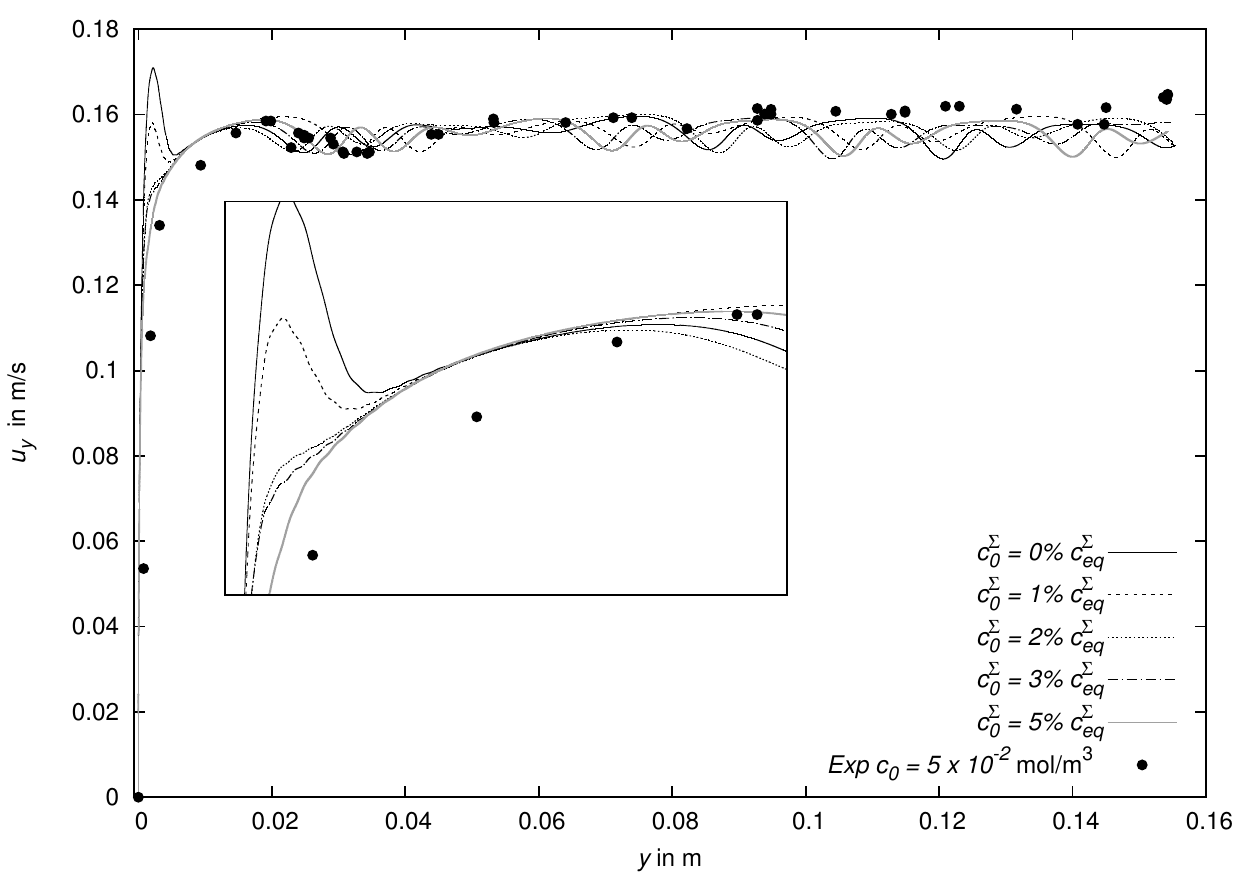}\label{fig:b8cS0}}
\caption{Study on the effects of the initial surface coverage on the rise velocity; simulated time $t = 1\ \mathrm{s}$.}
\label{fig:simTrio}
\end{figure}

Figure~\ref{fig:b3cS0} shows that for a small initial bulk surfactant concentration the surface coverage at detachment must have been almost zero (estimated value $\approx 1\%\ c^{\Sigma}_{\mathrm{eq},1}$), since the simulation results for $c^{\Sigma}_0 = 0\ \mathrm{mol/m^2}$ are the closest to the experimental ones. After reaching the peak velocity, the bubble starts to decelerate until the rise velocity oscillates around its steady state value. The most noticeable difference between the experimental and the numerical results for the case in figure~\ref{fig:b3cS0} is that in the simulation the bubble decelerates sooner than in the experiments. This discrepancy can result from small perturbations occurring at different times for simulations and experiments. In fact, the case studied is strongly sensitive to the onset of path instability. Perturbations triggering path instabilities are caused by different mechanisms in experiments and simulations. In experiments, perturbations could derive for instance from initial shape deformations. In numerical simulations, such perturbations can be numerical errors which are highly dependent on the mesh topology. Moreover, the discrepancy between experiments and simulations is only more pronounced for the least contaminated case which is also the case with the highest oscillations in the experimental data; see figure~\ref{fig:b358pathXZ} and table~\ref{tab:oscFreq}.
For intermediate and high initial surfactant bulk concentrations, the presence of initial surface contamination is evident; see figures~\ref{fig:b5cS0} and~\ref{fig:b8cS0}. The higher the initial bulk concentration, the more contaminated the bubble surface at release and the lower the velocity peaks. Figure~\ref{fig:b5cS0} shows that the best agreement between numerical and experimental results is obtained with an initial contamination of approximately $2\%\ c^{\Sigma}_{\mathrm{eq},2}$ which is in agreement with the estimated value in table~\ref{tab:estCS}. For the highest initial bulk concentration, see figure~\ref{fig:b8cS0}, a very good agreement with the experimental results is found already for $c^{\Sigma}_0 \approx 5\%\ c^{\Sigma}_{\mathrm{eq},3}$, that is a smaller value than the predicted one by equation~\eqref{eq:estCs0}. In fact, with a further increase of the initial surface contamination above the $5 \%\ c^{\Sigma}_{\mathrm{eq},3}$, the rise velocity profile does almost not change any more.

It is also interesting to note from figure~\ref{fig:simTrio} that after the initial transition period, all the bubble rise velocity values present small amplitude oscillations around a similar mean velocity value.
\FloatBarrier
%

\subsubsection{Effects of the initial surface coverage on bubble shape and path}
\label{subsubsec:5_4_2}

\begin{figure}[ht]
\centering
\includegraphics[width=1.0\textwidth]{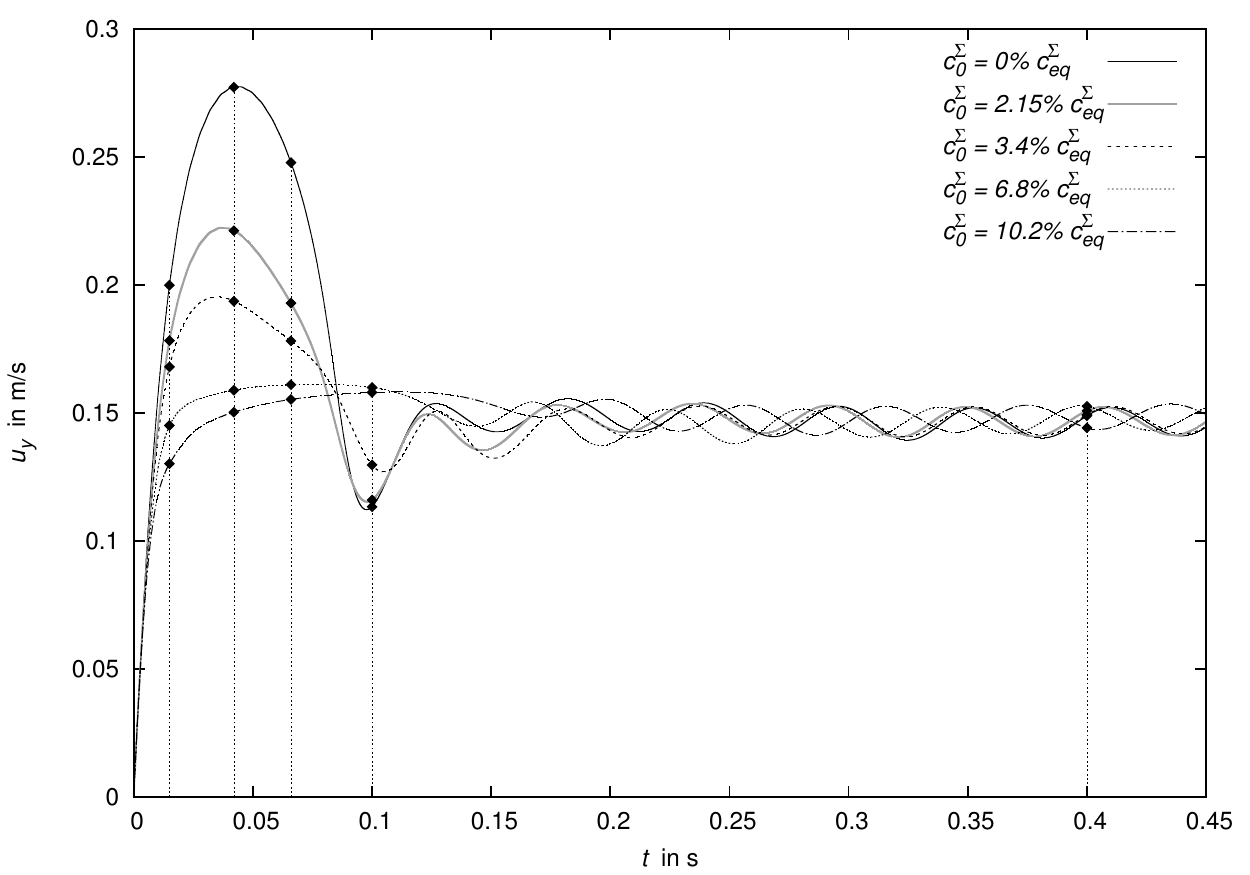}
\caption{Influence of the detachment time for the initial bulk concentration, $c_0 = 8 \cdot 10^{-3}\ \mathrm{mol/m^3}$.}
\label{fig:b5TU}
\end{figure}

\begin{figure}[ht]
\centering
\includegraphics[width=1.0\textwidth]{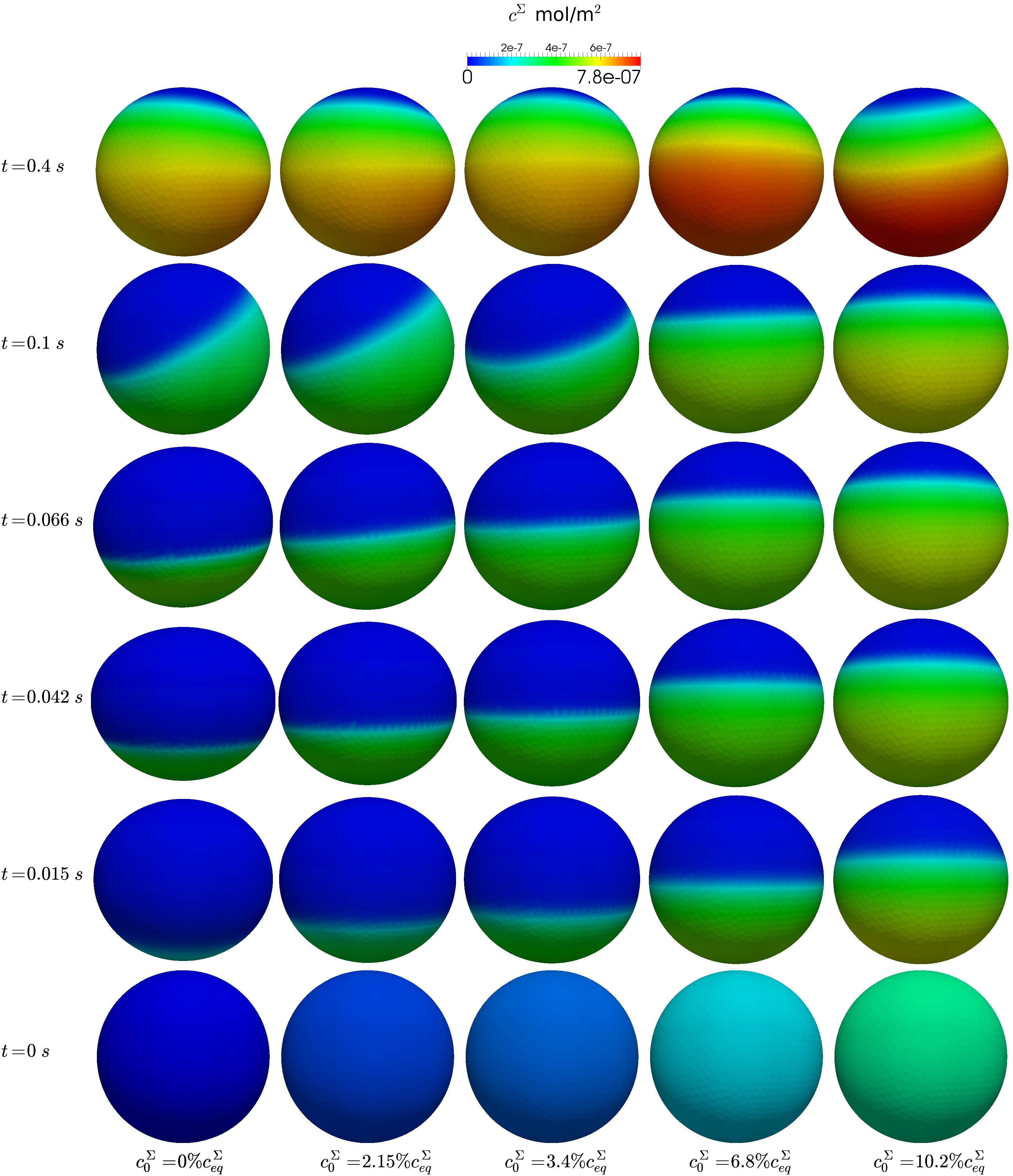}
\caption{Influence of the detachment time on the bubble shape and local surface coverage for the initial bulk concentration $c_0 = 8 \cdot 10^{-3}\ \mathrm{mol/m^3}$.}
\label{fig:b5_shapes}
\end{figure}

In this section the effects of the detachment time, or better the initial surface coverage for the simulations, are investigated in terms of bubble shapes and paths. Consider the intermediate initial bulk concentration, \mbox{$c_0 = 8 \cdot 10^{-3}\ \mathrm{mol/m^3}$}, that is the case shown in figure~\ref{fig:b5cS0}. The velocity profiles for the different initial surface coverages are plotted again in figure~\ref{fig:b5TU} but over time. In figure~\ref{fig:b5TU}, five time instances are marked where the bubble shape and the surface coverage are then compared and studied in figure~\ref{fig:b5_shapes}. In figure~\ref{fig:b5_shapes}, from the bottom to the top, the five bubbles are shown in their rise at the selected time instances (every column shows one of the bubbles rising), while from left to right the initial surface concentration increases (see the surface coverage at $t = 0\ \mathrm{s}$). The bubble surfaces are coloured by the local surfactant surface concentration. From figure~\ref{fig:b5TU} and~\ref{fig:b5_shapes} it is clearly visible that increasing $c^{\Sigma}_0$ results in a less deformed interface and a slower bubble. In fact, for $c^{\Sigma}_0 = 0\% c^{\Sigma}_{eq},\ 2.15\% c^{\Sigma}_{eq}$ and $3.6\% c^{\Sigma}_{eq}$, respectively, the bubble surface is still deforming and reaches its maximum aspect ratio ($AR = 1.27,\ 1.1,\ 1.06$, respectively) with the peak velocity. During the deceleration phase the bubbles are going back to a more spherical shape; see $t = 0.066$ s. For the two cases on the right of figure~\ref{fig:b5_shapes} with the highest initial surface coverage, the amount of surfactant on the interface is high enough to result in an almost not deformed interface ($AR = 1.04$). These bubbles accelerate until reaching the quasi-steady state velocity and their shape remains spherical.

\begin{figure}[ht]
\centering
\includegraphics[width=1.0\textwidth]{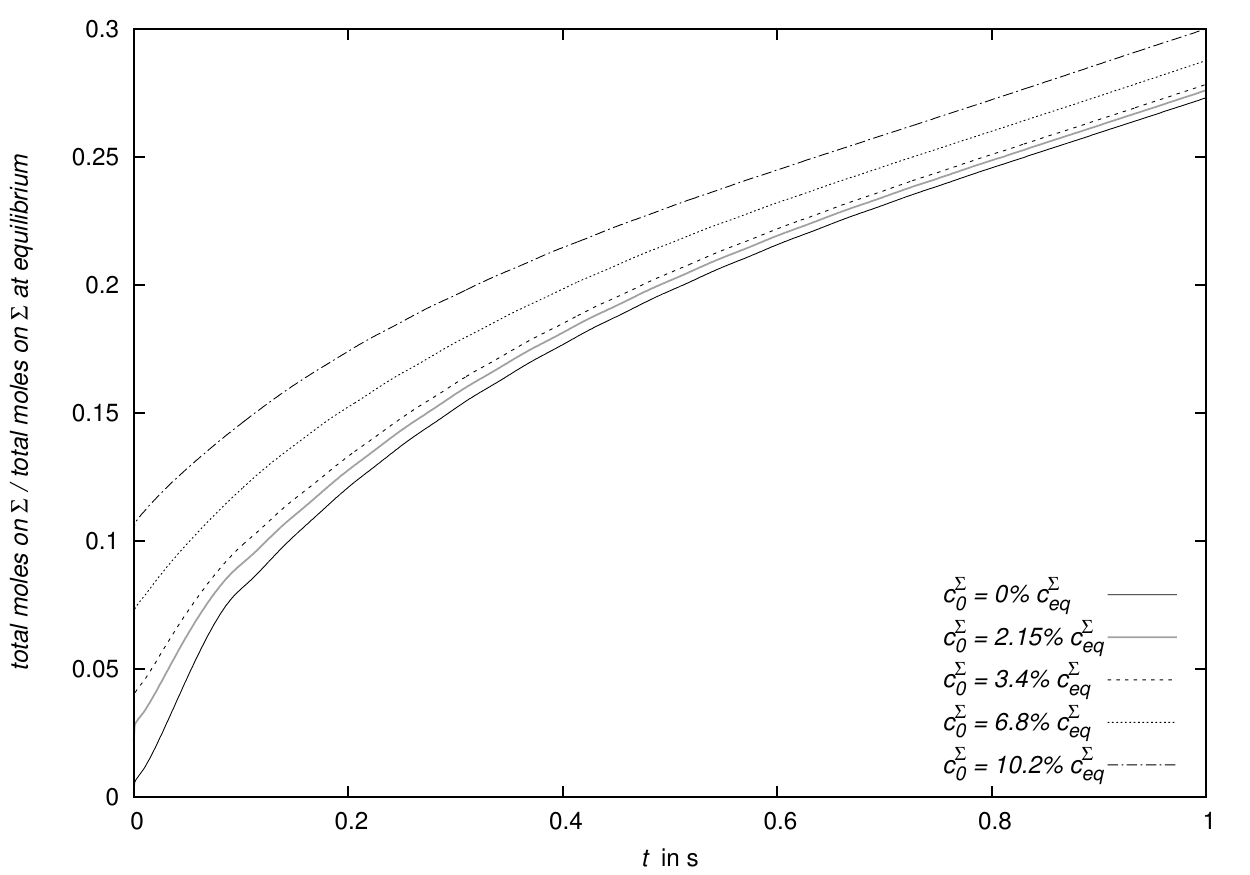}
\caption{Influence of the detachment time on the total amount of surfactant on the interface divided by the respective equilibrium value for the initial bulk concentration $c_0 = 8 \cdot 10^{-3}\ \mathrm{mol/m^3}$.}
\label{fig:b5_totCs}
\end{figure}

If we consider the latest time ($t = 0.4\ \mathrm{s}$) in figure~\ref{fig:b5_shapes}, the bubbles have a similar velocity though, surprisingly, they do not have the same surface coverage. Moreover, with different $c^{\Sigma}_0$ (and/or different initial bulk concentrations $c_0$, see figure~\ref{fig:meshDep}) we obtain similar terminal velocities, but with a different final surface coverage that is not yet the equilibrium value, $c^{\Sigma}_{\mathrm{eq}}$, and not even close to it. To confirm this, we show in figure~\ref{fig:b5_totCs} the total amount of surfactant on the interface with respect to time. Here it can be seen that even at $t = 1$ s the total amount of surfactant on the interface is less than 30$\%$ of the equilibrium value. For the smallest initial surface concentration, the total amount of surfactant on the interface grows more rapidly than in the other cases. This behaviour can be explained by the fact that P\'{e}clet and Reynolds numbers are higher for smaller $c^{\Sigma}_0$. Also the concentration difference between bulk and interface is larger (for a given bulk concentration and varying the initial surface concentration). This results in stronger advective transport, thus thinner concentration boundary layers. 
Instead, from $t \approx 0.6$ s, when the bubbles have approximately the same terminal velocity, the total amount of surfactant on the interface grows similarly for each bubble.

\begin{figure}[ht]
\centering
\subfloat[][\emph{Top view of the bubble path}.]
{\includegraphics[width=.48\textwidth]{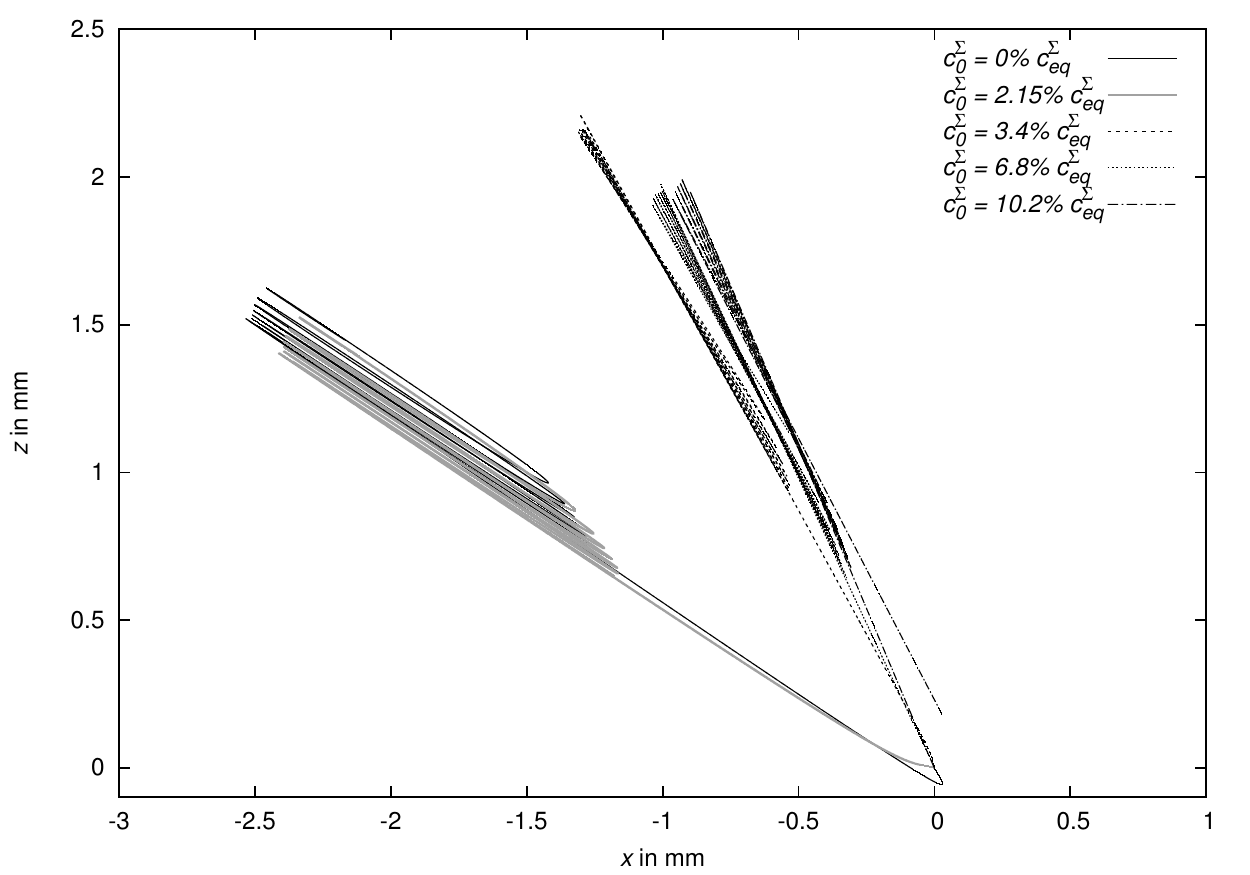}\label{fig:b5pathXZ}} \quad
\subfloat[][\emph{Lateral view of the bubble path, where $x^{\prime} = \sqrt{x^2 + z^2}$}.]
{\includegraphics[width=.48\textwidth]{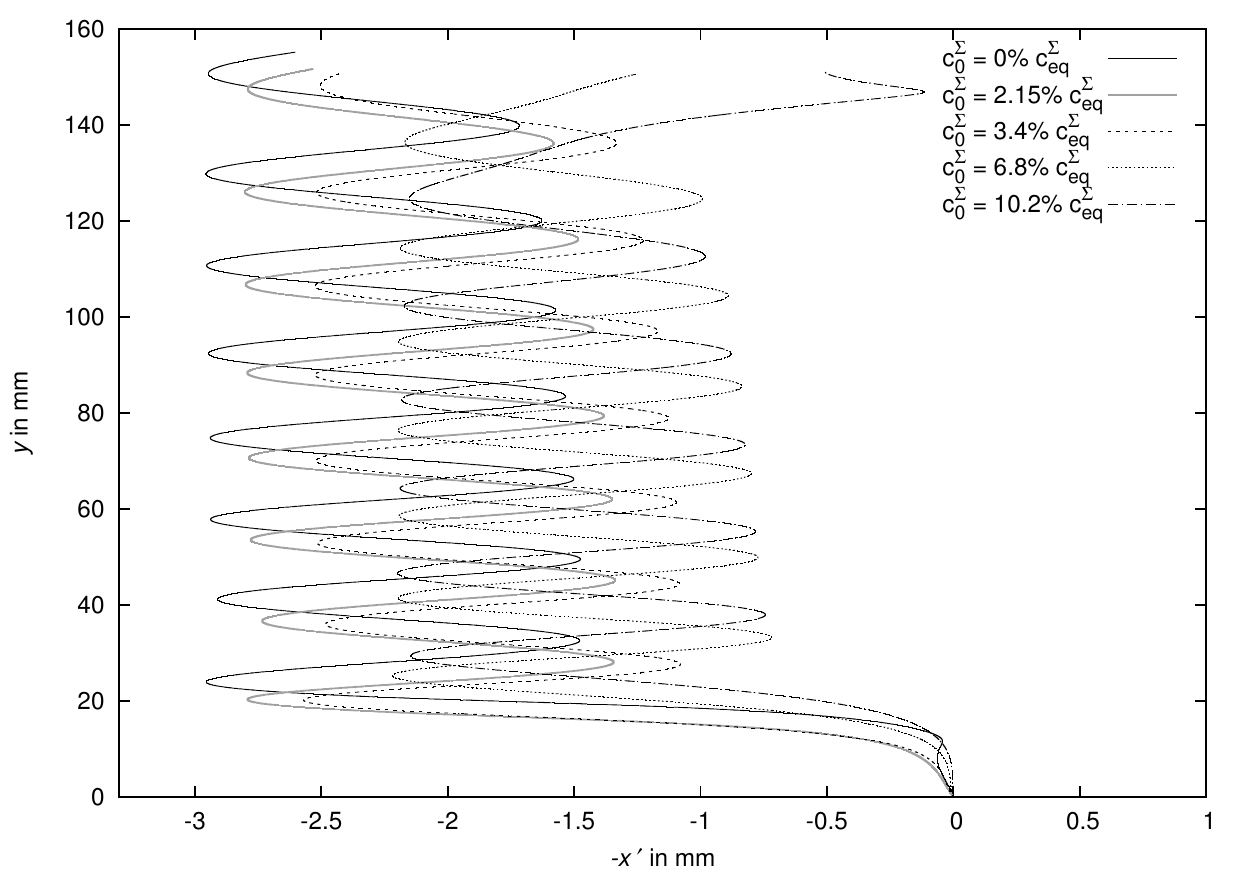}\label{fig:b5pathXY}}
\caption{Effects of the initial surface coverage on the bubble path for the initial bulk concentration $c_0 = 8 \cdot 10^{-3}\ \mathrm{mol/m^3}$.}
\label{fig:b5path}
\end{figure}

In figure~\ref{fig:b5path} the respective bubble paths are depicted. From the top view in figure~\ref{fig:b5pathXZ} it can be observed that all the bubbles follow a zig-zag path, but the onset of path instability occurs later for less contaminated surfaces, as shown by the path front view, figure~\ref{fig:b5pathXY}. 
\FloatBarrier
\FloatBarrier
%
%
\FloatBarrier
%

%
\subsection{Effects of surfactant on the onset of path instability}
\label{subsec:5_5}
\begin{figure}[ht]
\centering
\includegraphics[width=1.0\textwidth]{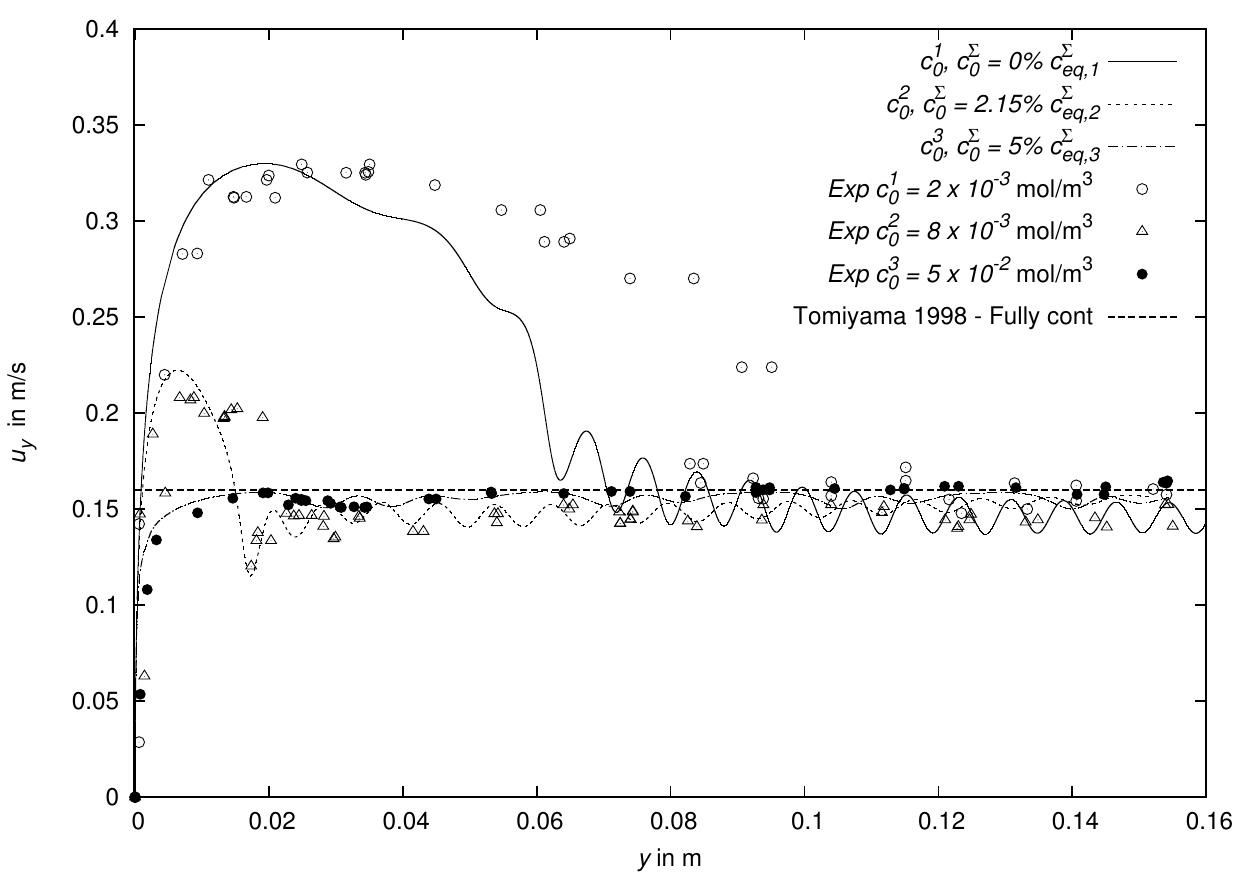}
\caption{Bubble rise velocity, influence of the initial bulk concentration with pre-contaminated surface.}
\label{fig:b358_vel_expBest}
\end{figure}

The simulation results which agree best with the experimental ones from figure~\ref{fig:simTrio} (grey curves) are selected for the rest of the discussion and reported in figure~\ref{fig:b358_vel_expBest}. All the simulations run until reaching $t = 1\ \mathrm{s}$ of physical time.
In figure~\ref{fig:b358_vel_expBest}, also the estimated velocity from the correlation for fully contaminated systems proposed by Tomiyama et al.~\cite{tomiyama1998} (equation (33) in the reference) is plotted. All the simulation results, including the least contaminated case, are in very good agreement with this estimated velocity at quasi-steady state.
\FloatBarrier
A further indicator of agreement between experimental and numerical results at quasi-steady state may be the comparison of the standard deviation of the rise velocities for \mbox{$0.1\ \mathrm{m} < y < 0.16\ \mathrm{m}$}. In fact, the numerical results show pronounced oscillations that are not clearly visible from the experiments. The values for the standard deviation reported in table~\ref{tab:sdtDevVel} show a similar trend, i.e. oscillations decrease with increasing bulk concentration. The magnitude is in agreement between simulations and experiments, too.
\begin{table}[ht]
\caption{Standard deviation from the mean velocity value at quasi-steady state, $0.1\ \mathrm{m} < y < 0.16\ \mathrm{m}$.}
\label{tab:sdtDevVel}
\centering
\begin{tabular}{ccc}
\toprule
  & Experiment & Simulation\\
\midrule
$\operatorname{Ma} = 34$ & 0.00706 & 0.00668\\
$\operatorname{Ma} = 49$ & 0.00430 & 0.00310\\
$\operatorname{Ma} = 70$ & 0.00218 & 0.00247\\
\bottomrule
\end{tabular}
\end{table}
\FloatBarrier
For completeness, also the frequency of the horizontal velocity and the vortex shedding from the rear part of a rigid sphere are computed as reported in~\cite{tagawaTakagi2014} from~\cite{tsugeHibino1971} and~\cite{kim1990}, respectively,\footnote{The frequency of the bubble horizontal velocity is computed according to~\cite{tsugeHibino1971} as $f = \frac{u_b}{d_e} 0.1 c_D^{0.734}$, where $u_b$ is the averaged quasi-steady velocity and $d_e$ is the bubble equivalent diameter. The frequency of the vortex shedding from the rear part of a rigid sphere is $f_v = \frac{\omega \nu_l \operatorname{Re}}{\pi d_e^2}$ (from~\cite{kim1990}), where $\omega$ is taken equal to 0.30 as in~\cite{tagawaTakagi2014,kim1990} and $\operatorname{Re}$ is computed based on $u_b$. The drag coefficient $c_D$ is computed equating the drag to the buoyancy force as in~\cite{tagawaTakagi2014} (equation 2.7 in the reference), thus $c_D = \frac{4 d_e g}{3 u_b^2}$, where $g$ is the gravitational acceleration.} and compared to the simulation results. From table~\ref{tab:oscFreq} it can be seen that the oscillation frequencies of the vertical velocity are approximately twice the horizontal ones, as expected from~\cite{tagawaTakagi2014,mougin2006,deVries2002p}. Moreover, the least and intermediate contaminated cases show a good agreement between the numerical and literature results. The most contaminated case is not relevant for this comparison since the velocity oscillations are not as regular as the other two cases. 
\begin{table}[ht]
\caption{Oscillation frequencies of the velocity components compared to the frequencies $f$ and $f_v$ reported in~\cite{tagawaTakagi2014} from~\cite{tsugeHibino1971} and~\cite{kim1990}, respectively.}
\label{tab:oscFreq}
\centering
\begin{tabular}{ccccccc}
\toprule
  & \multirow{2}*{$c_D$} & \multirow{2}*{$\operatorname{Re}$} & \multirow{2}*{$f$} & \multirow{2}*{$f_v$} & \multicolumn{2}{c}{Simulations}\\
  &       &     &                     &       & $f_{u_x,u_z}$ & $f_{u_y}$\\  
\midrule
$\operatorname{Ma} = 34$ & 0.88 & 228 & 9.228 & 9.646 & 9.33 & 25.00\\
$\operatorname{Ma} = 49$ & 0.82 & 237 & 9.074 & 9.999 & 8.86 & 16.67\\
$\operatorname{Ma} = 70$ & 0.79 & 241 & 8.982 & 10.218 & 4.75 & 10.04\\
\bottomrule
\end{tabular}
\end{table}
\FloatBarrier
\begin{figure}[ht]
\centering
\includegraphics[width=1.0\textwidth]{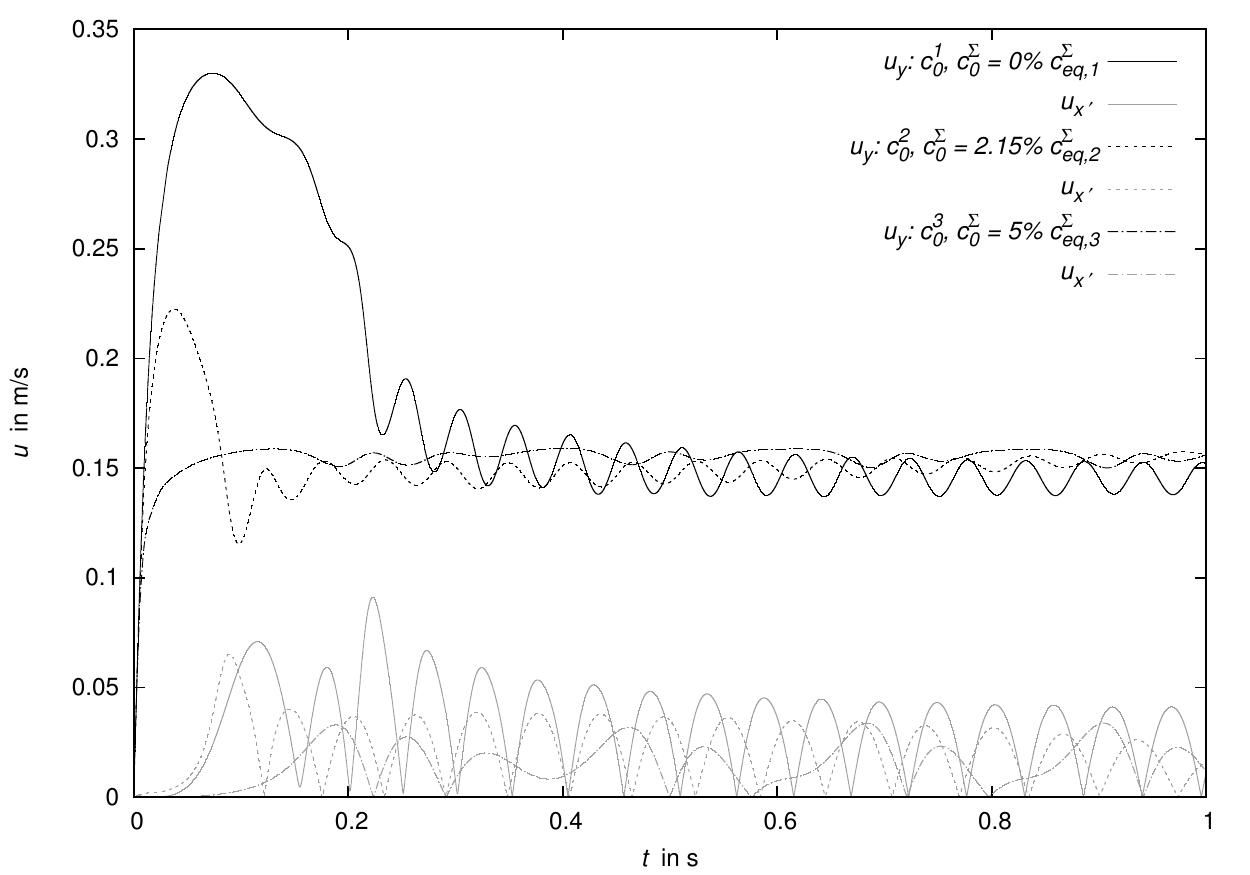}
\caption{Bubble rise velocity, influence of the initial bulk concentration with $u_{x^{\prime}} = \sqrt{u_x^2 + u_z^2}$.}
\label{fig:b358_vel}
\end{figure}
\FloatBarrier
%

\subsubsection{Bulk concentration}
\label{subsubsec:5_5_1}
\begin{figure}[ht]
\centering
\subfloat[][\emph{Top view of the bubble path}.]
{\includegraphics[width=.478\textwidth]{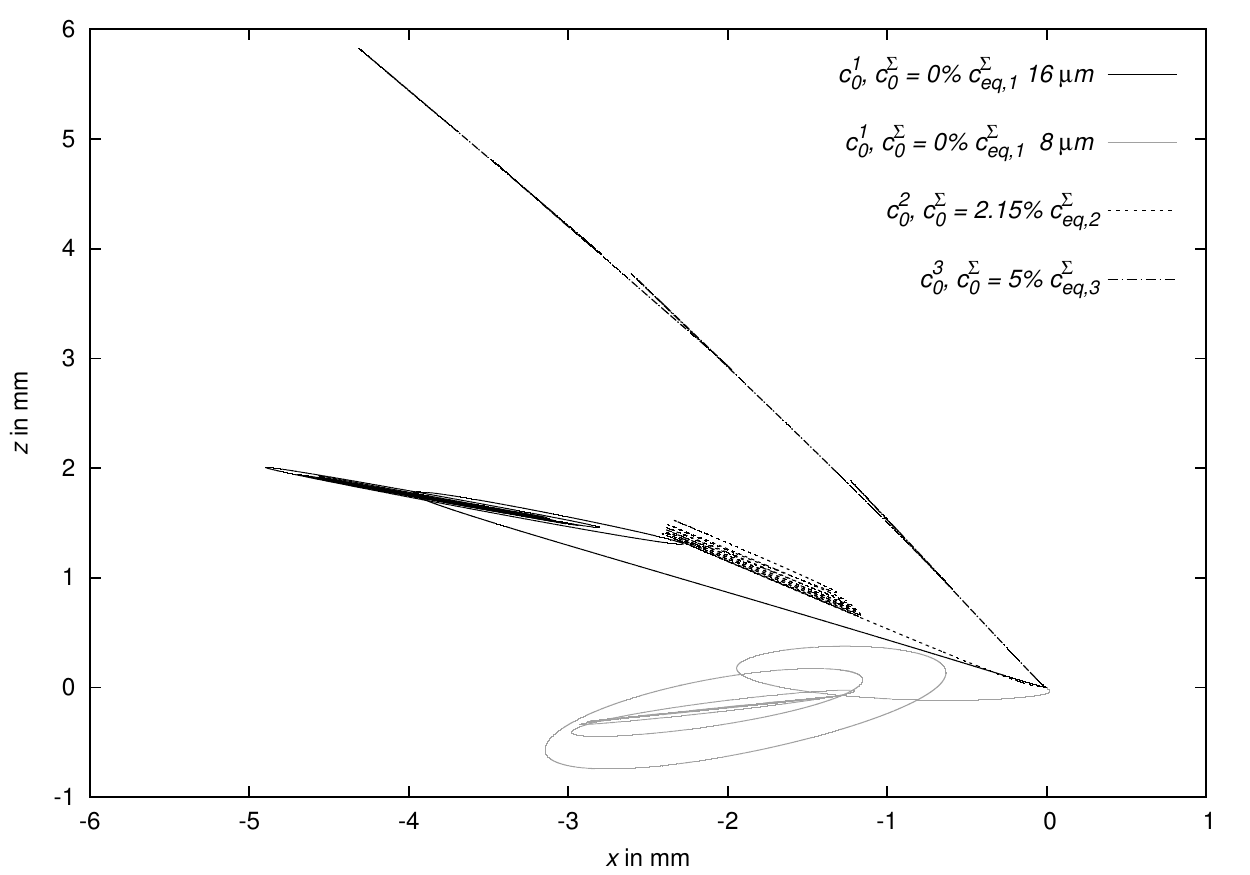}\label{fig:b358pathXZ}} \quad
\subfloat[][\emph{Lateral view of the bubble path, where $x^{\prime} = \sqrt{x^2 + z^2}$}.]
{\includegraphics[width=.478\textwidth]{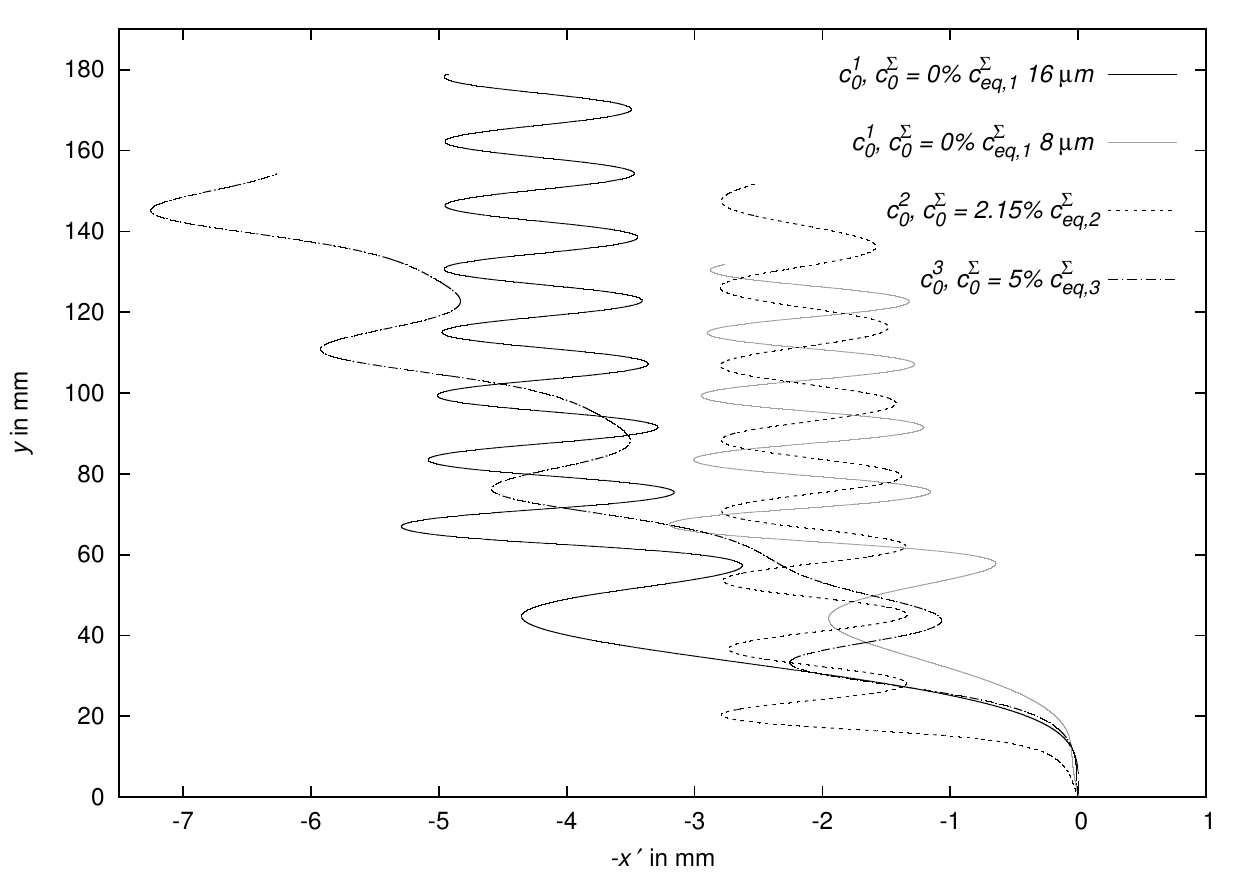}\label{fig:b358pathXY}}
\caption{Study on the effects of the initial bulk concentration on the bubble path.}
\label{fig:b358path}
\end{figure}

The velocity components along the rise direction $y$ and in the $x-z$ plane for the three bubbles under investigation are reported in figure~\ref{fig:b358_vel}. The respective bubble paths are given in figures~\ref{fig:b358path} (top view $x-z$ in~\ref{fig:b358pathXZ} and lateral view $x^{\prime}-y$ in~\ref{fig:b358pathXY}, where $x^{\prime} = \sqrt{x^2 + z^2}$).

Even though the bubbles reach a similar terminal velocity, their lateral velocity components and paths show a significant difference. The bubble rising in the least contaminated aqueous solution ($c_0^1$, $\operatorname{Ma} = 34$) follows first a helical path until it starts to oscillate around its terminal velocity \mbox{(t $\approx$ 0.35 s)} and then turns into a zig-zag path. The amplitude of this zig-zag path is around one bubble diameter. While the shift from zig-zag to helical path was already observed for clean bubbles~\cite{canoLozano2016}, the transition from helical to zig-zag trajectory occurs only in presence of surfactant and was first reported by Tagawa et al.~\cite{tagawaTakagi2014}. Our simulation results can serve as a further confirmation of this phenomenon. 
For the intermediate surfactant bulk concentration ($c_0^2$, $c^{\Sigma}_0 = 2.15\% c^{\Sigma}_{eq}$, $\operatorname{Ma} = 49$), see also figure~\ref{fig:b5path}, after the initial transient stage when the bubble accelerates and then decelerates towards its quasi-steady state, the bubble follows a zig-zag path (starting from $t \approx 0.11$ s) with an amplitude around 0.7 bubble diameters. 

The bubble rising in the most contaminated solution ($c_0^3$, $c^{\Sigma}_0 = 5\% c^{\Sigma}_{eq}$, $\operatorname{Ma} = 70$), after the initial acceleration, at $t \approx 0.22$ s starts to follow a zig-zag path, but with a pronounced drift towards one side. 
\FloatBarrier
\begin{figure}[ht]
\centering
\subfloat[][\emph{Rise velocity components}.]
{\includegraphics[width=.48\textwidth]{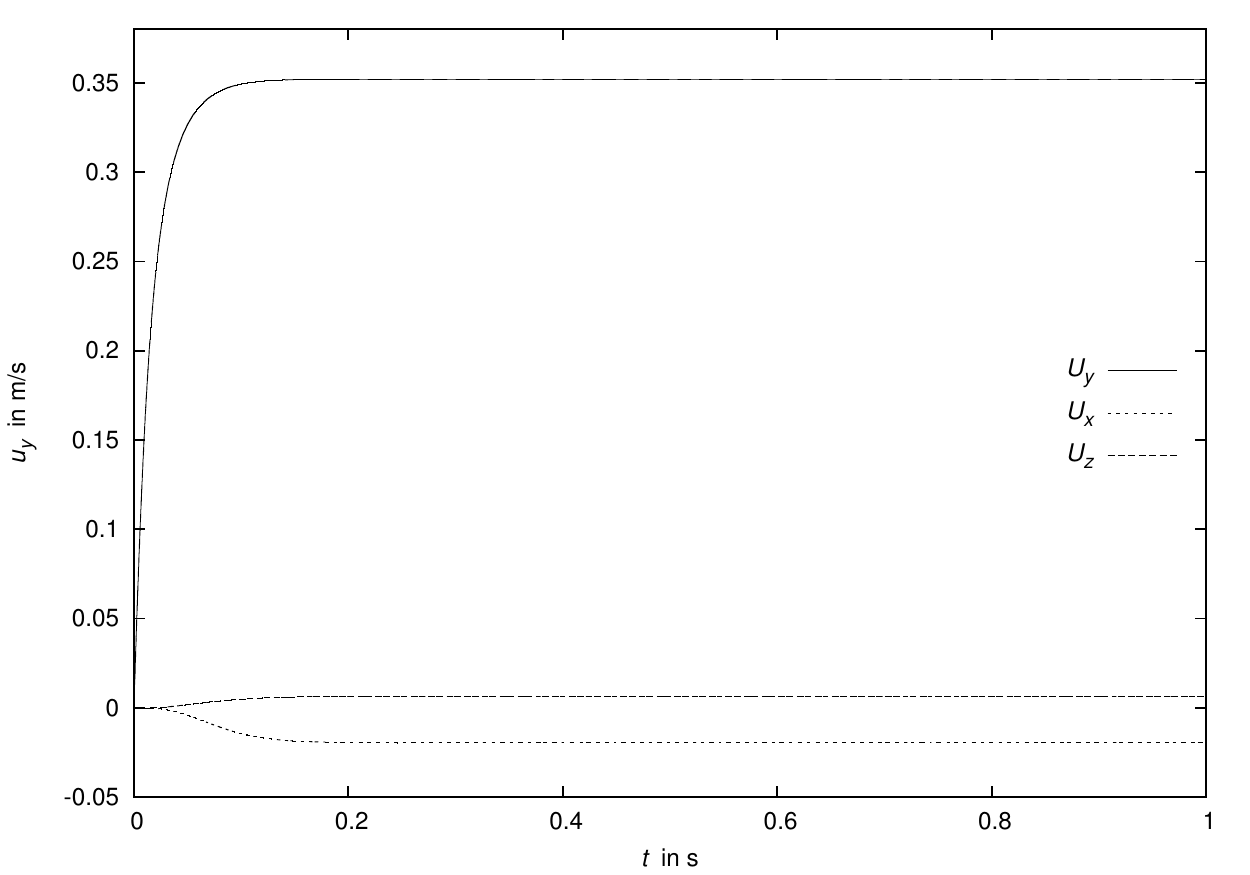}\label{fig:cleanResultsTU}} \quad
\subfloat[][\emph{Top view of the bubble path}.]
{\includegraphics[width=.48\textwidth]{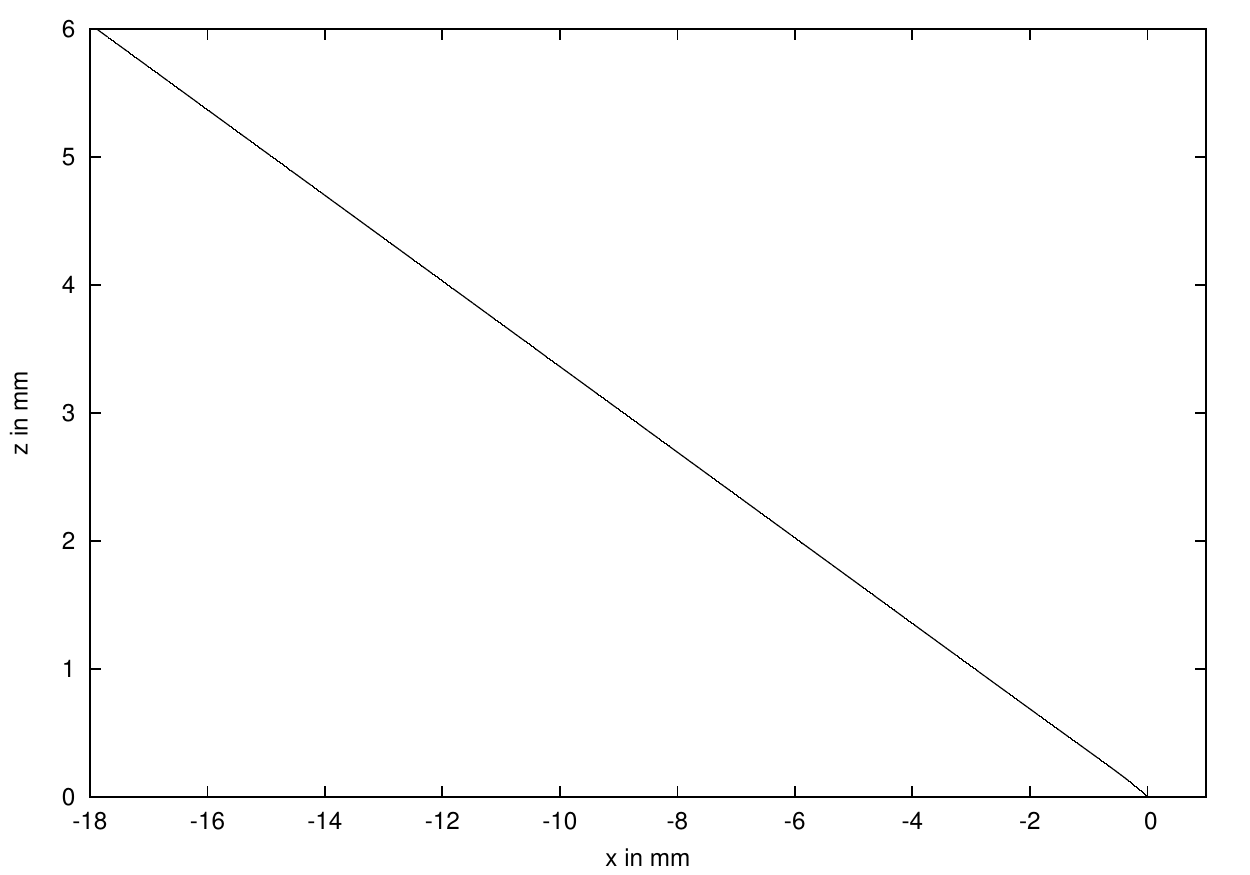}\label{fig:cleanResultsXZ}}
\caption{Bubble rising in clean water, evidence of the later drift.}
\label{fig:cleanResults}
\end{figure}
\FloatBarrier
Lateral migration is a known effect both from experimental and numerical works~\cite{deVries2002p,albert2015} for bubbles close to the path instability regime. Our own studies have confirmed this trend, too. For small bubbles rising in clean systems, the lateral drift is almost zero, while for larger bubbles (but not yet path unstable) a significant migration can be observed. The lateral migration can be observed also for the bubble under investigation ($d = 1.45\ \mathrm{mm}$) rising in clean water, as reported in figure~\ref{fig:cleanResults}. In fact, looking at the lateral components of the rise velocity (figure~\ref{fig:cleanResultsTU}) it can be noticed that they are non-zero. This causes the drift visualized in the top view of the bubble path, see figure~\ref{fig:cleanResultsXZ}. We assume that in our set-up the instabilities are triggered by the unstructured nature of the computational mesh.

\begin{figure}[ht]
\centering
\includegraphics[width=0.94\textwidth]{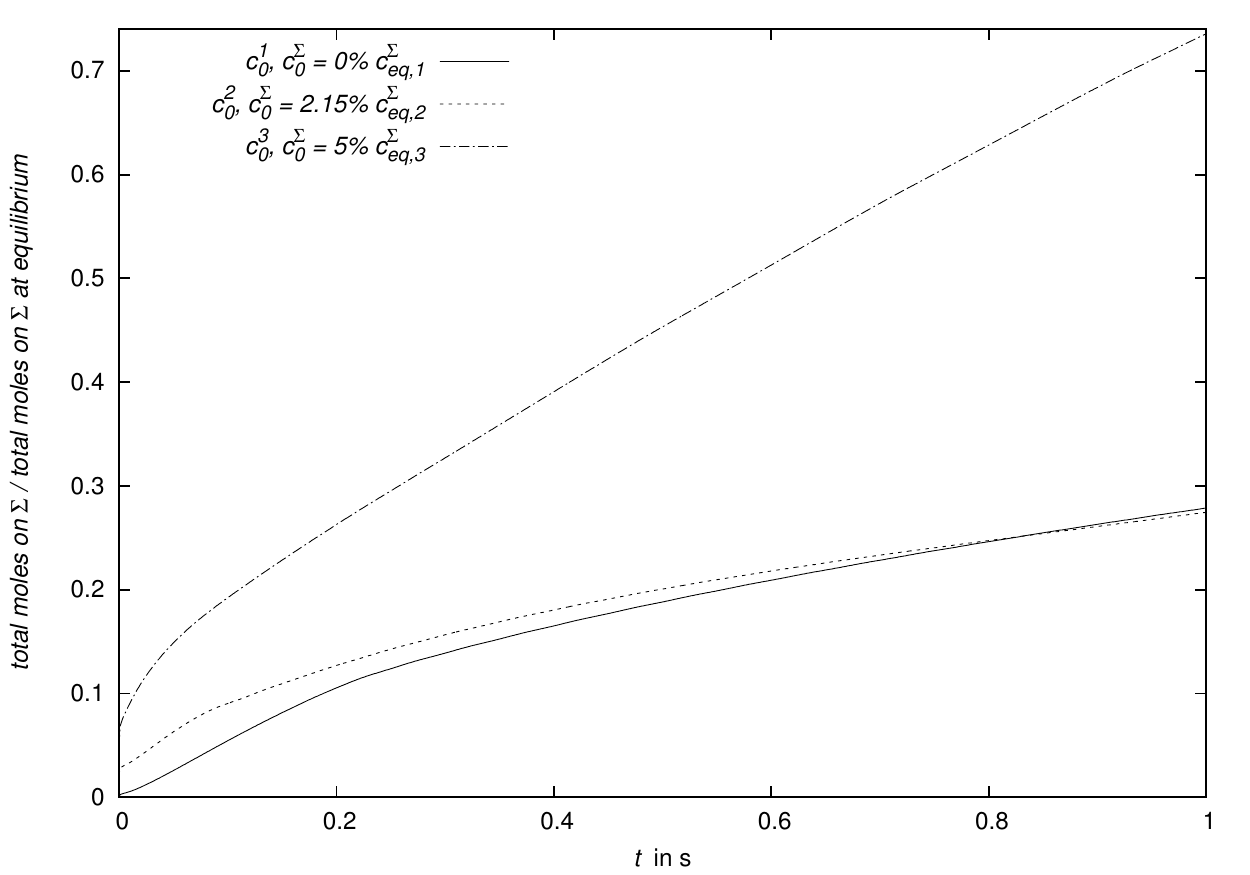}
\caption{Temporal evolution of the total amount of surfactant on the interface divided by the respective equilibrium values for the three selected initial surface and bulk concentrations.}
\label{fig:totCs358}
\end{figure}

\begin{figure}[ht]
\centering
\includegraphics[width=0.94\textwidth]{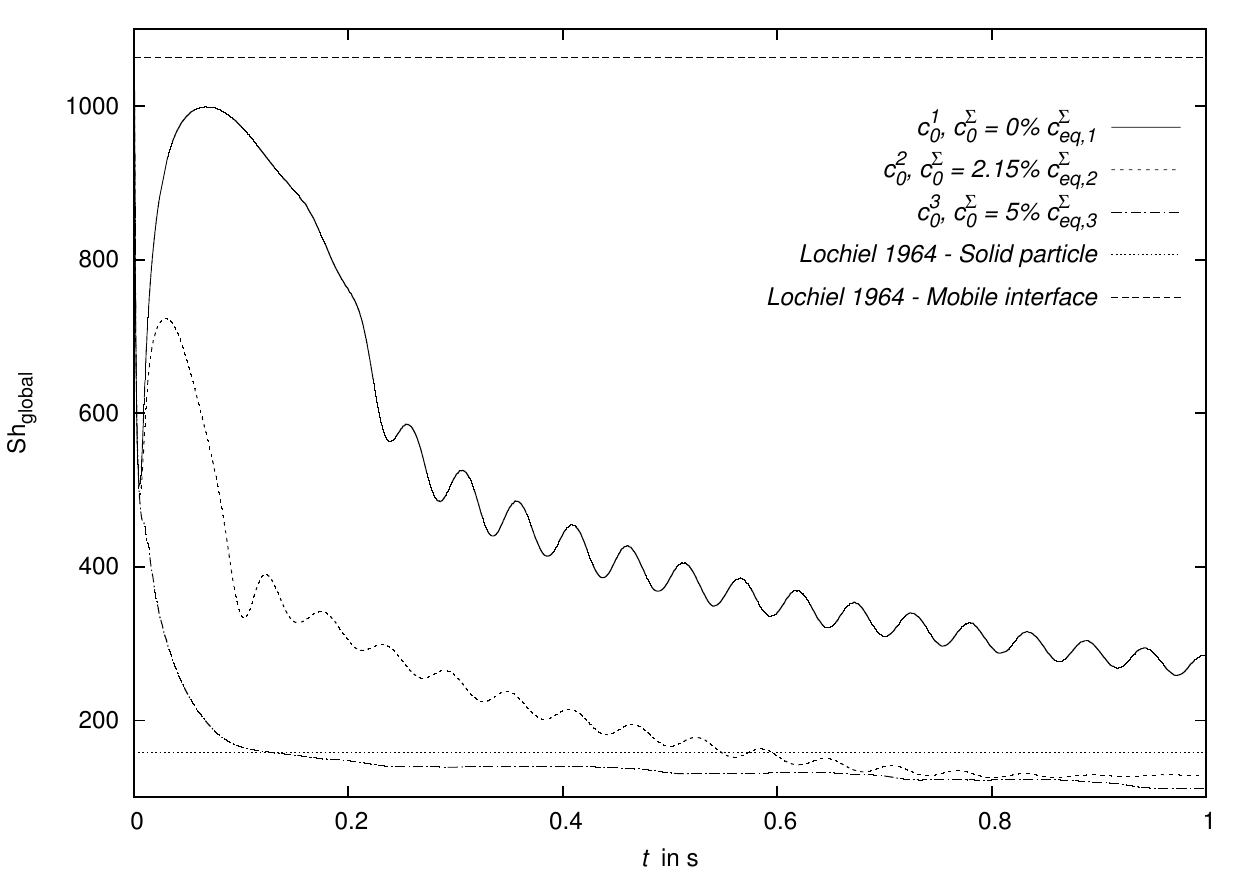}
\caption{Global Sherwood number referred to surfactant transfer. The surface area variation is less than 3$\%$.}
\label{fig:glSh}
\end{figure}

The temporal evolution of the total amount of surfactant on the interface is depicted in figure~\ref{fig:totCs358}. It is remarkable that for all studied cases the surface coverage is much smaller than the respective equilibrium concentration. Nonetheless, the quasi-steady state terminal velocity is reached. This finding is relevant because it shows that the steady state velocity can be reached without an equilibrium between ad- and desorption and without the bubble being ``fully contaminated". This situation will also have a large impact on the mass transfer processes in contaminated systems. From the slopes of the depicted curves in figure~\ref{fig:totCs358}, it becomes visible that the bubble rising in the most contaminated liquid is adsorbing the surfactant much quicker than in the other two cases. The bubbles rising in low and medium contaminated liquid follow a similar trend, even though the bulk concentration in the latter case is about four times higher. There are mainly three effects causing this behaviour: (1) With increasing surfactant bulk concentration the initial concentration difference between interface and bulk increases, and therefore also the driving force for mass transfer is higher. (2) The first effect is mitigated because at the same time the bubble accumulates surfactant quicker. (3) Since the surfactant distribution on the interface is coupled with the bubble hydrodynamics via the Marangoni forces, the shape of the surfactant boundary layer changes. In general, an increasing amount of surfactant will slow down the bubble, and therefore decrease the advective transport which in turn decreases the driving force for mass transfer. The last effect may be expressed as  the dimensionless surfactant gradient at the sub-layer in terms of the global Sherwood number, figure~\ref{fig:glSh}. In the initial state, when the bubble is formed in the experiment, or at the very beginning of the numerical simulation, the bubble is stagnant, and a surfactant boundary layer forms very quickly at the liquid-gas interface, driven by pure diffusion. This process is not depicted in figure~\ref{fig:glSh}, since the concentration difference is the highest and the boundary layer formation happens on a time scale much smaller than the course of the bubble rise from the initial release up to the quasi-steady state, i.e. $\mathcal{O}(1)\ \mathrm{s}$. When the bubble starts to rise it accelerates and the initial boundary layer becomes thinner due to the strong convective transport. The cleaner the system, the higher the maximum rise velocity, and hence the more pronounced this effect will be. After the initial increase in the acceleration phase, the Sherwood number decreases rapidly as the bubble decelerates. When the bubble velocity reaches a quasi-steady state, the Sherwood number for the cases with low and medium contamination keep decreasing, but at a much slower rate. This is because the Marangoni forces are constantly increasing with increasing surface contamination. The Marangoni forces, in turn, influence the shape of the hydrodynamic boundary layer, and therefore also of the surfactant boundary layer. For the most contaminated bubble, a further increase of surfactant on the interface does not lead to an increase of the Marangoni forces. A more detailed view of all forces acting on the bubble will be given in the sections~\ref{subsubsec:5_5_3} and \ref{subsec:5_6}.
In figure~\ref{fig:glSh} the correlations for mass transfer problems based on the boundary layer theory from Lochiel et al.~\cite{lochiel1964} are plotted, too. Two limiting situations are considered, that is fully mobile interface (equation (58) in~\cite{lochiel1964}) and solid particle (equation (86) in~\cite{lochiel1964}). It is very interesting to notice that the global Sherwood number computed for the adsorbed surfactant tends to a value very close to the predicted one for solid particles. Moreover, for the least contaminated case the global Sherwood number at the beginning of the rise is comparable with the one of a clean bubble. Note that there are only two reference lines given, based on the Reynolds number of the clean case, $\operatorname{Re} = 544$, and the average Reynolds number for the contaminated cases, $\operatorname{Re} = 235$. We did not want to put much emphasis on the mass transfer similarity to solid particles since the physical effects leading to a comparable quantitative outcome in both cases are actually very different.

So far, we described what we could observe from the simulation results in terms of rise velocity, surface coverage and path. Nevertheless, to really disclose the bubble dynamics, a study of the local flow field in the proximity of the interface and the forces acting on the bubble surface, in particular the local and global Marangoni forces generated by a non-uniform surface tension and their interplay with deformable interfaces, viscous and pressure forces is performed below.
\FloatBarrier
\FloatBarrier
%

\subsubsection{Vorticity}
\label{subsubsec:5_5_2}
\begin{figure}[ht]
\centering
\subfloat[][\emph{Rise velocity}.]
{\includegraphics[width=.48\textwidth]{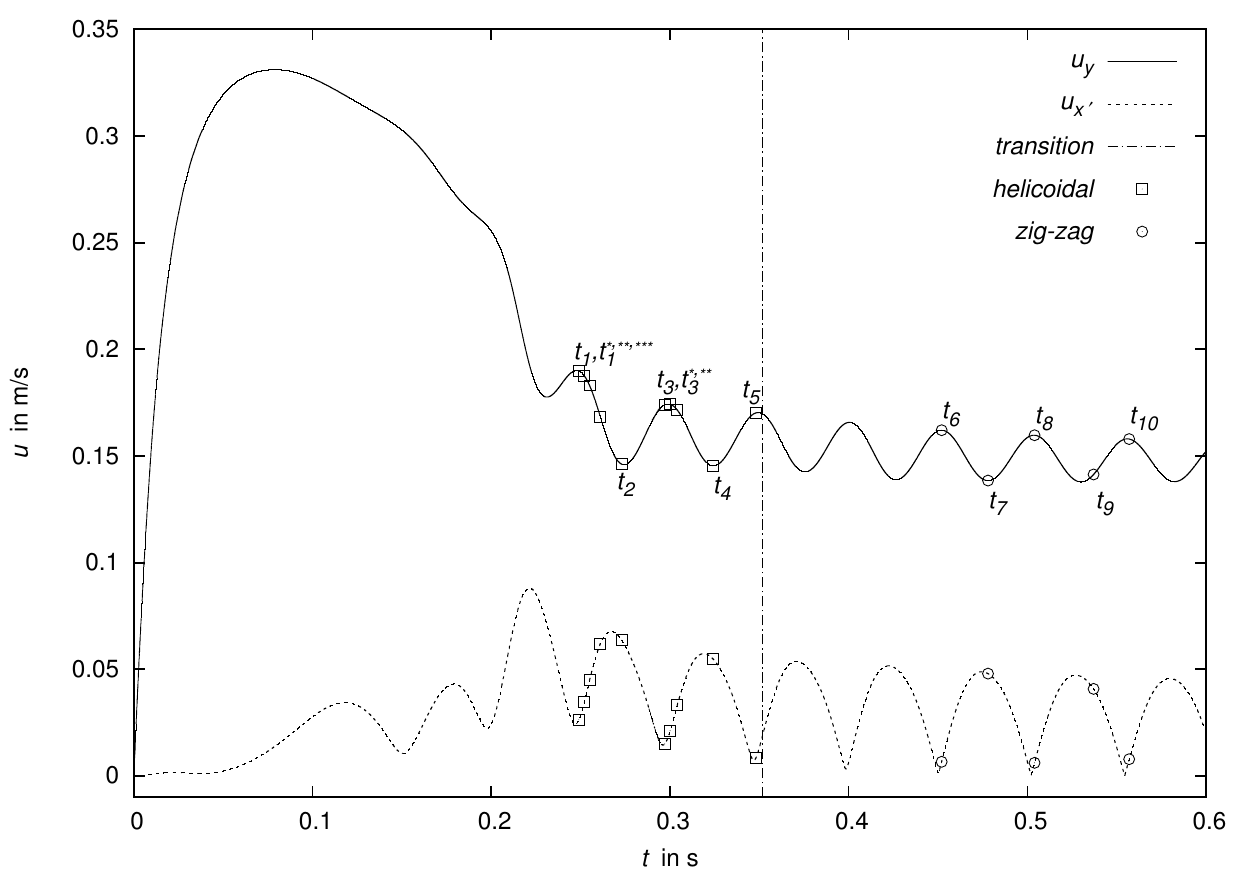}\label{fig:b3v1}} \quad
\subfloat[][\emph{Lateral view of the bubble path}.]
{\includegraphics[width=.48\textwidth]{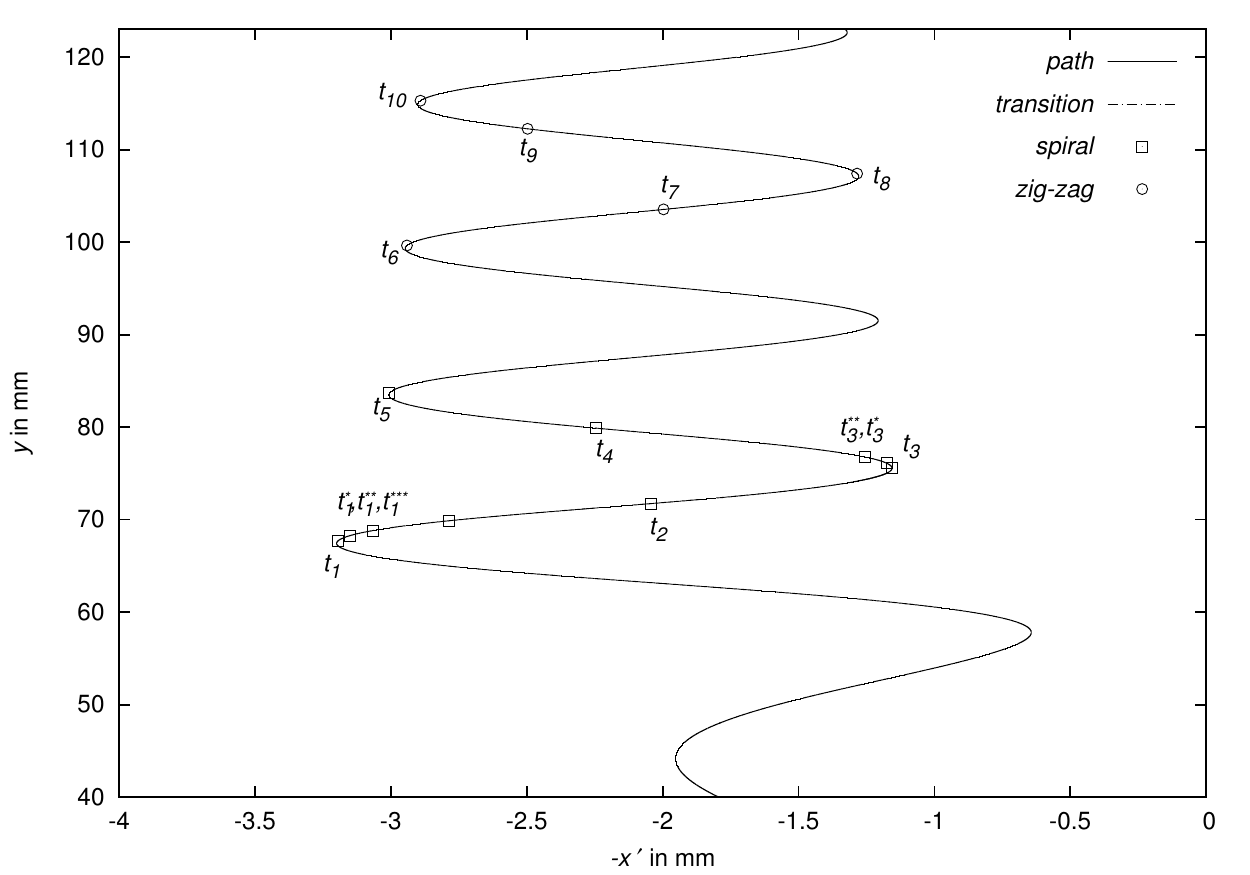}\label{fig:b3v2}}  \\
\subfloat[][\emph{$t_1$}.]
{\includegraphics[width=.11\textwidth]{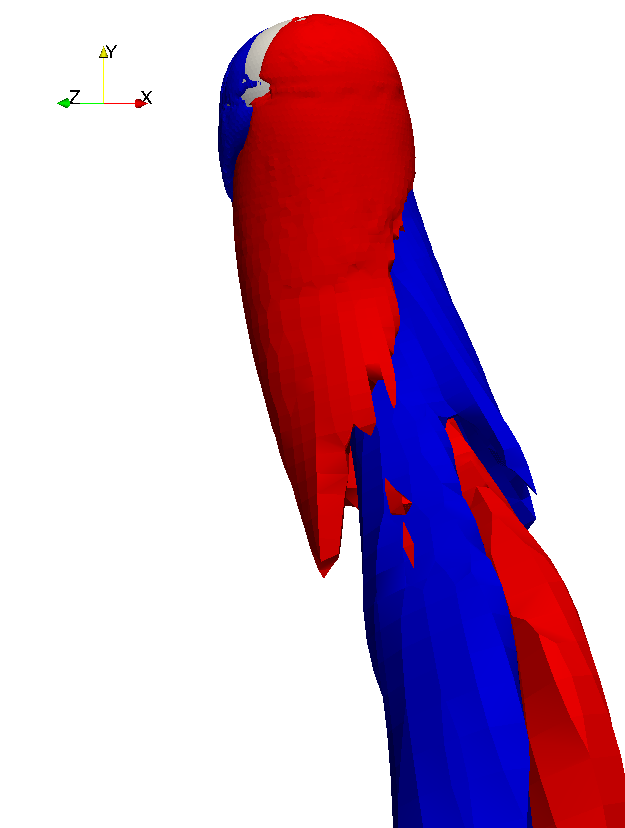}\label{fig:b3va}}$\quad \quad $
\subfloat[][\emph{$t_1^{*}$}.]
{\includegraphics[width=.11\textwidth]{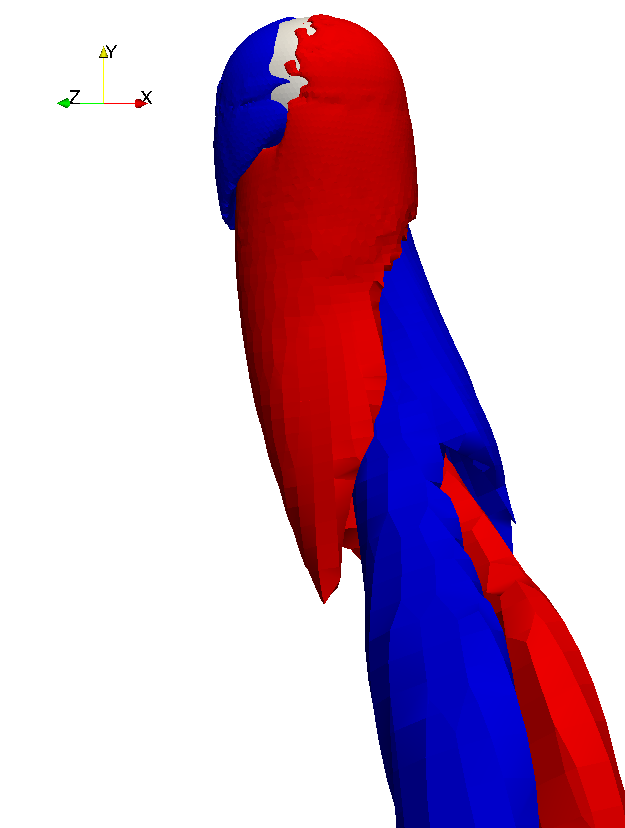}\label{fig:b3va1}}$\quad \quad $
\subfloat[][\emph{$t_1^{**}$}.]
{\includegraphics[width=.11\textwidth]{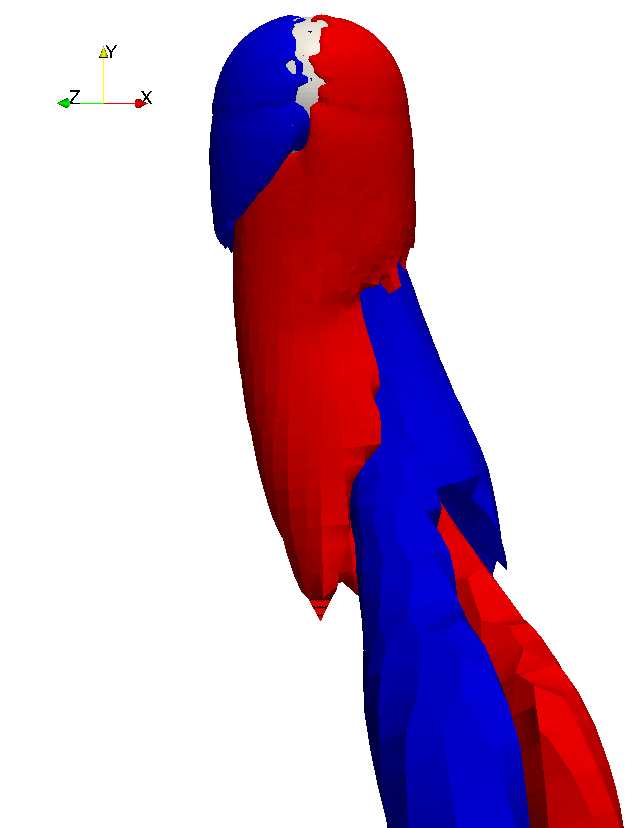}\label{fig:b3va2}}$\quad \quad $
\subfloat[][\emph{$t_1^{***}$}.]
{\includegraphics[width=.11\textwidth]{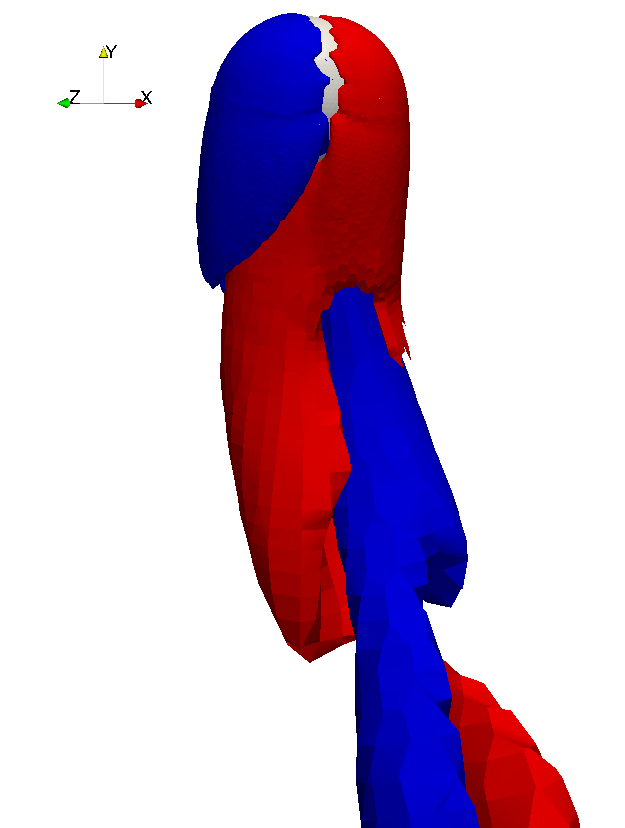}\label{fig:b3va3}}$\quad \quad $
\subfloat[][\emph{$t_2$}.]
{\includegraphics[width=.11\textwidth]{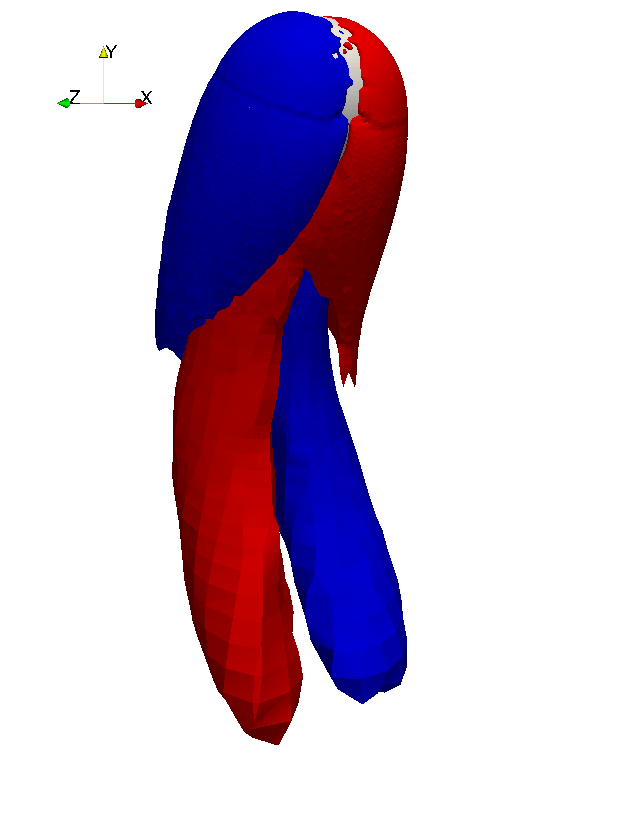}\label{fig:b3vb}}\\
\subfloat[][\emph{$t_3$}.]
{\includegraphics[width=.11\textwidth]{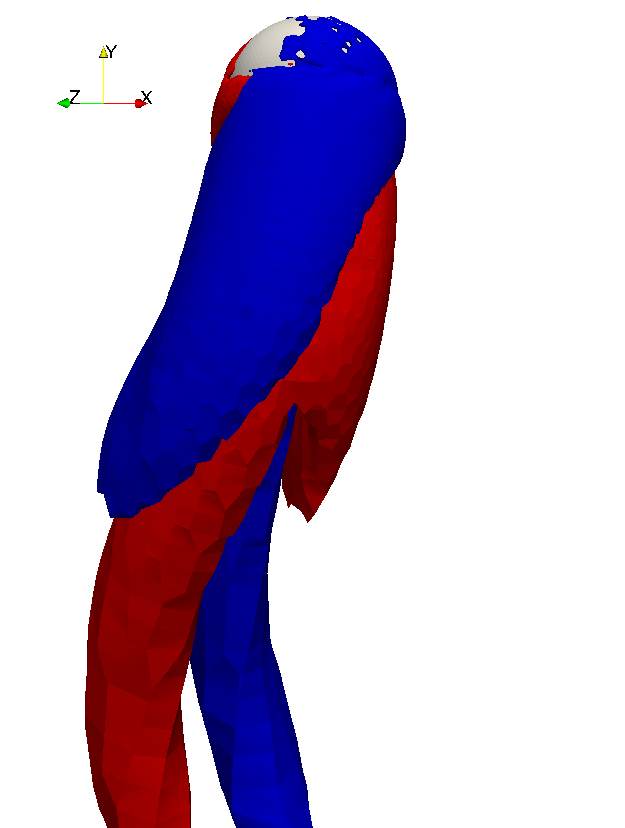}\label{fig:b3vc}}$\quad \quad $
\subfloat[][\emph{$t_3^{*}$}.]
{\includegraphics[width=.11\textwidth]{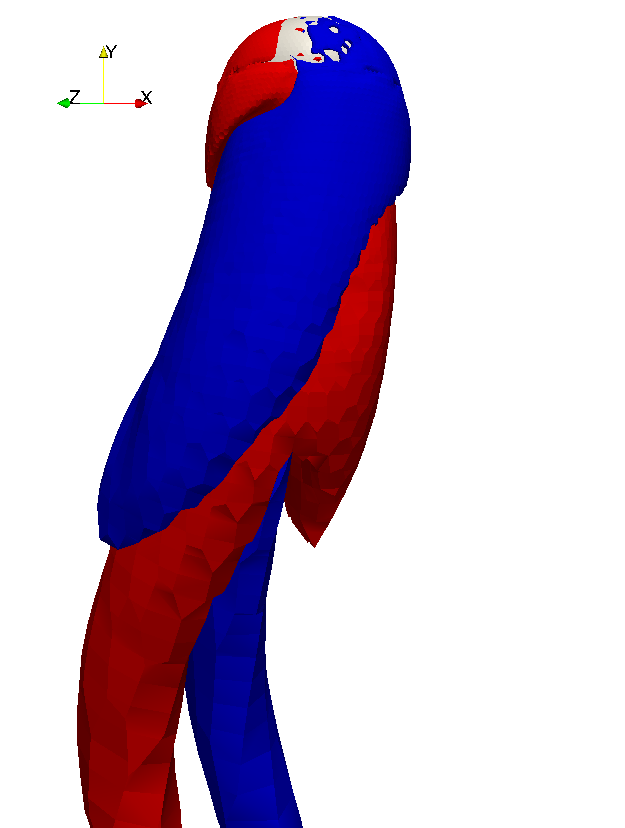}\label{fig:b3vc1}}$\quad \quad $
\subfloat[][\emph{$t_3^{**}$}.]
{\includegraphics[width=.11\textwidth]{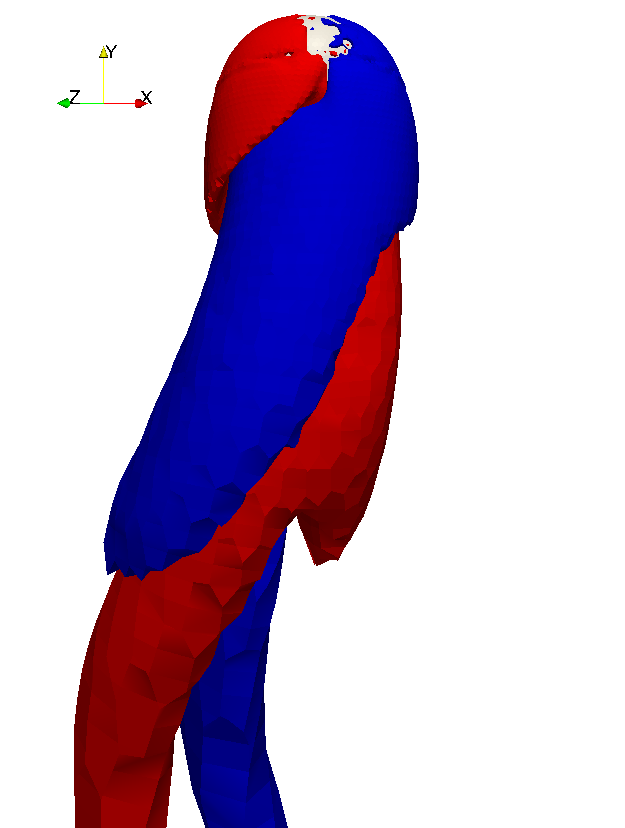}\label{fig:b3vc2}}$\quad \quad $
\subfloat[][\emph{$t_4$}.]
{\includegraphics[width=.11\textwidth]{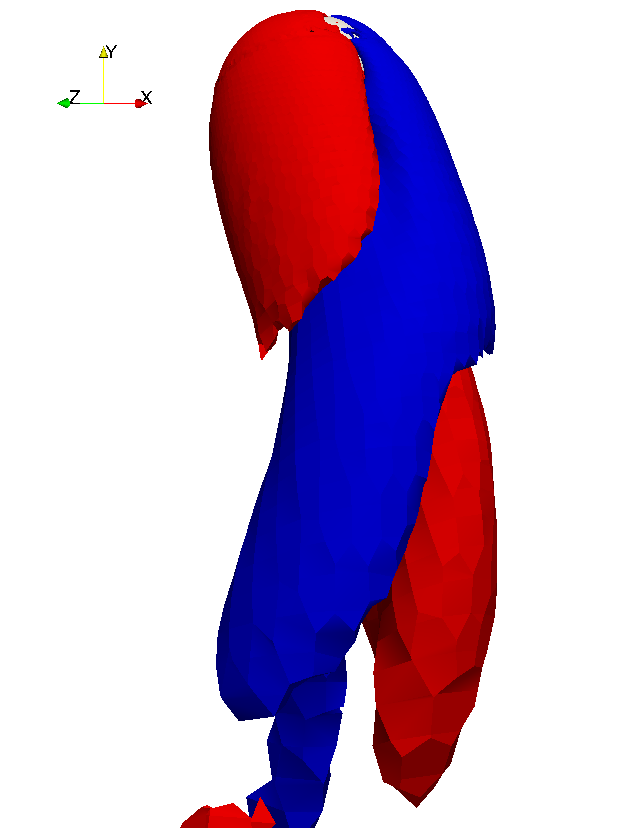}\label{fig:b3vd}}$\quad \quad $
\subfloat[][\emph{$t_5$}.]
{\includegraphics[width=.11\textwidth]{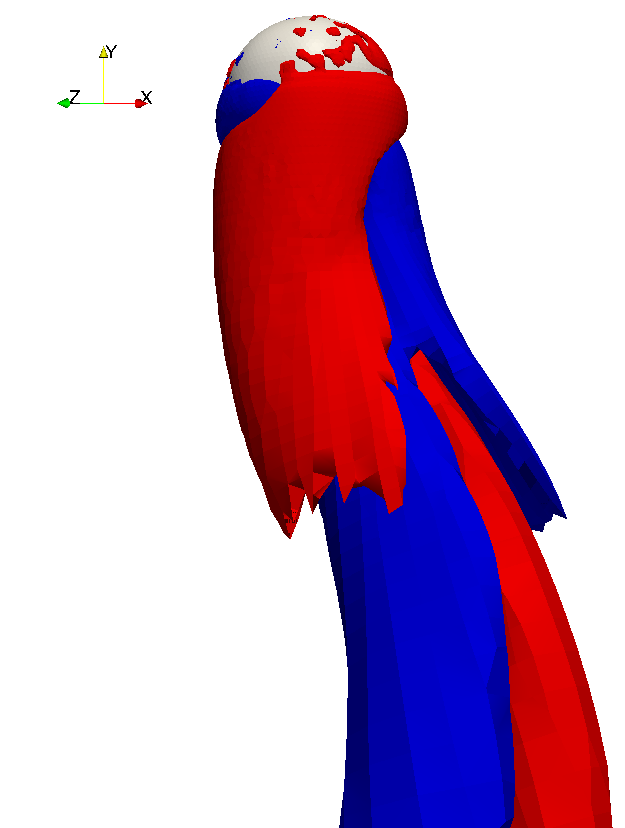}\label{fig:b3ve}}  \\
\subfloat[][\emph{$t_6$}.]
{\includegraphics[width=.11\textwidth]{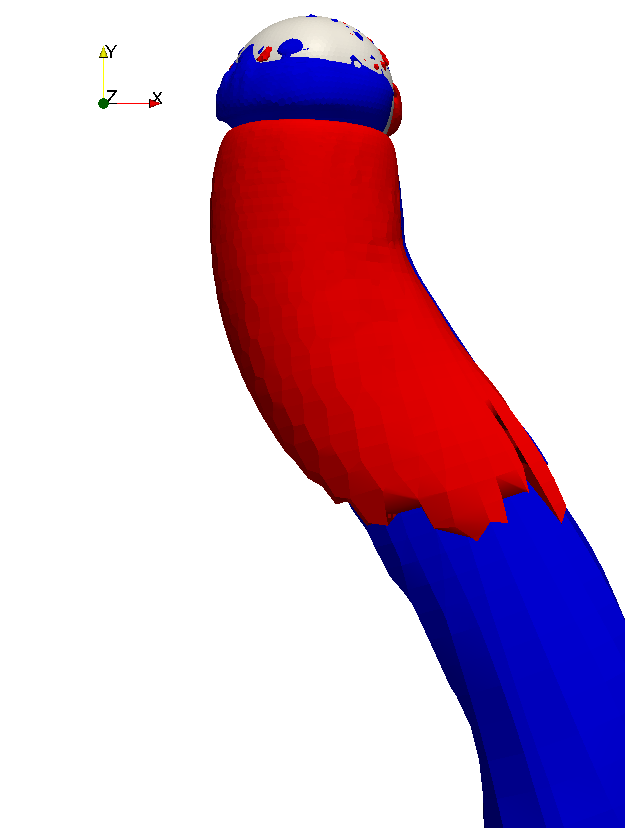}\label{fig:b3vf}}$\quad \quad $
\subfloat[][\emph{$t_7$}.]
{\includegraphics[width=.11\textwidth]{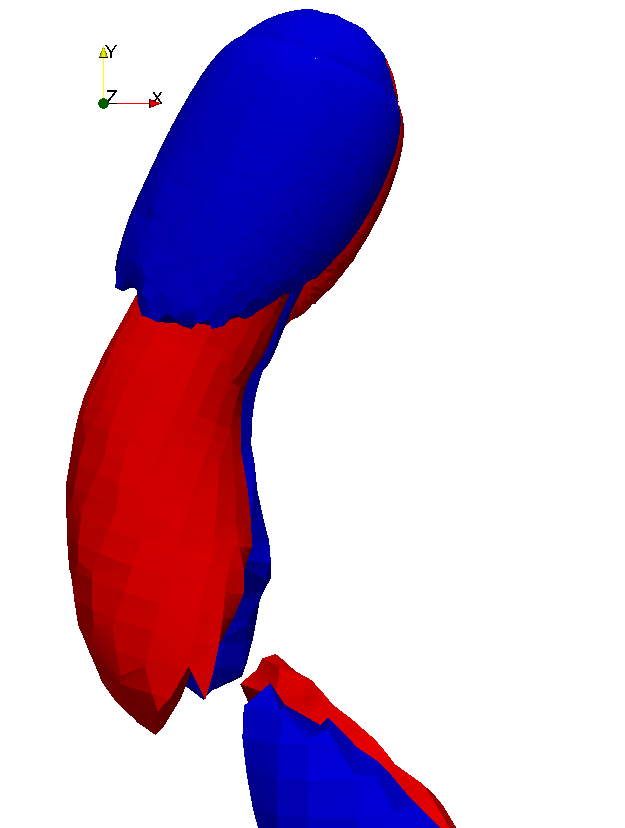}\label{fig:b3vg}}$\quad \quad $
\subfloat[][\emph{$t_8$}.]
{\includegraphics[width=.11\textwidth]{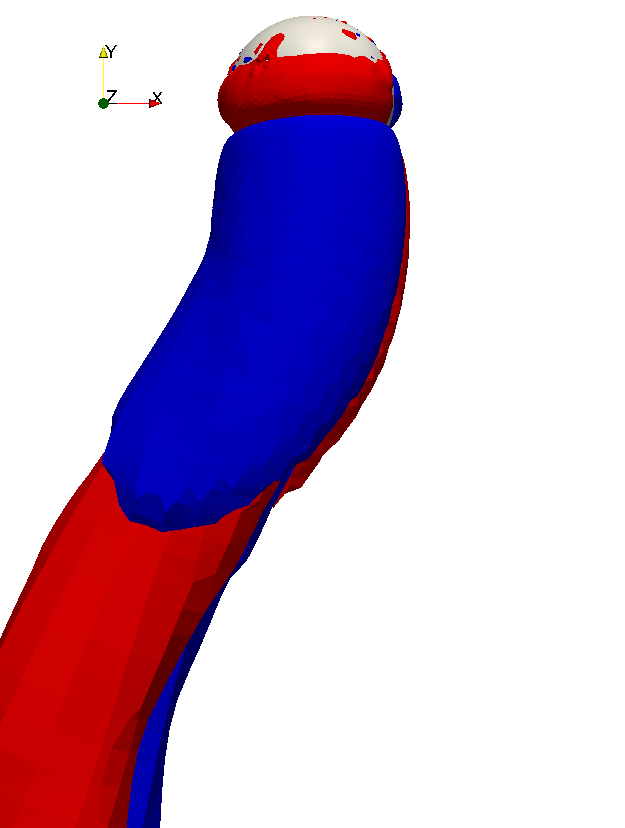}\label{fig:b3vh}}$\quad \quad $
\subfloat[][\emph{$t_9$}.]
{\includegraphics[width=.11\textwidth]{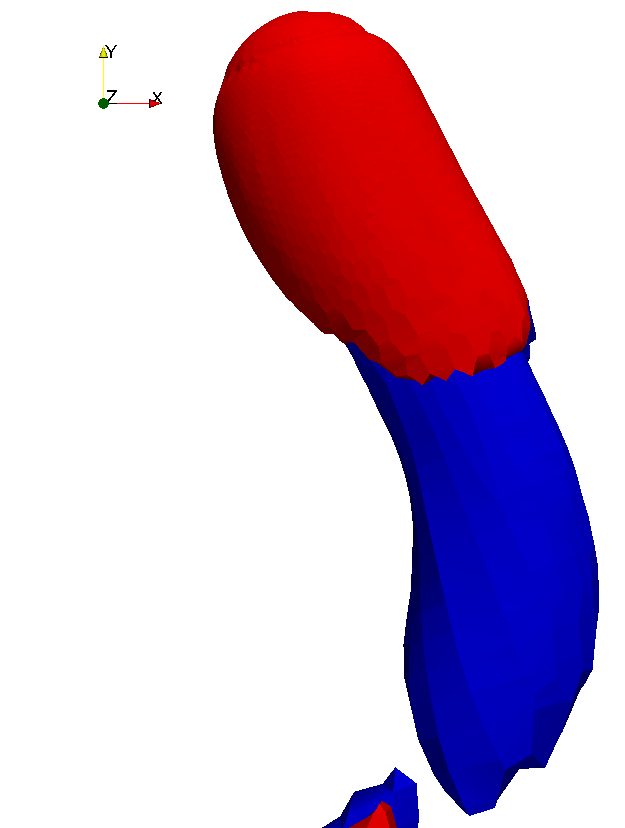}\label{fig:b3vi}}$\quad \quad $
\subfloat[][\emph{$t_{10}$}.]
{\includegraphics[width=.11\textwidth]{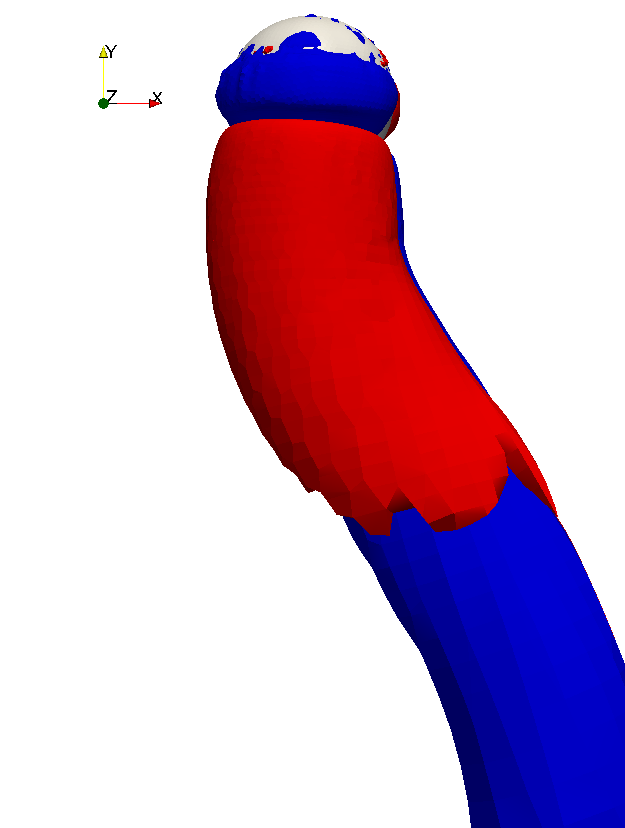}\label{fig:b3vj}}  \\
\caption{Vorticity contour plot ($\omega_y = \pm 20$ 1/s) at different time instances, $c_0 = 2 \cdot 10^{-3}\ \mathrm{mol/m^3}$, $c^{\Sigma}_0 = 0$.}
\label{fig:b3vort}
\end{figure}

\begin{figure}[ht]
\centering
\subfloat[][\emph{Rise velocity}.]
{\includegraphics[width=.48\textwidth]{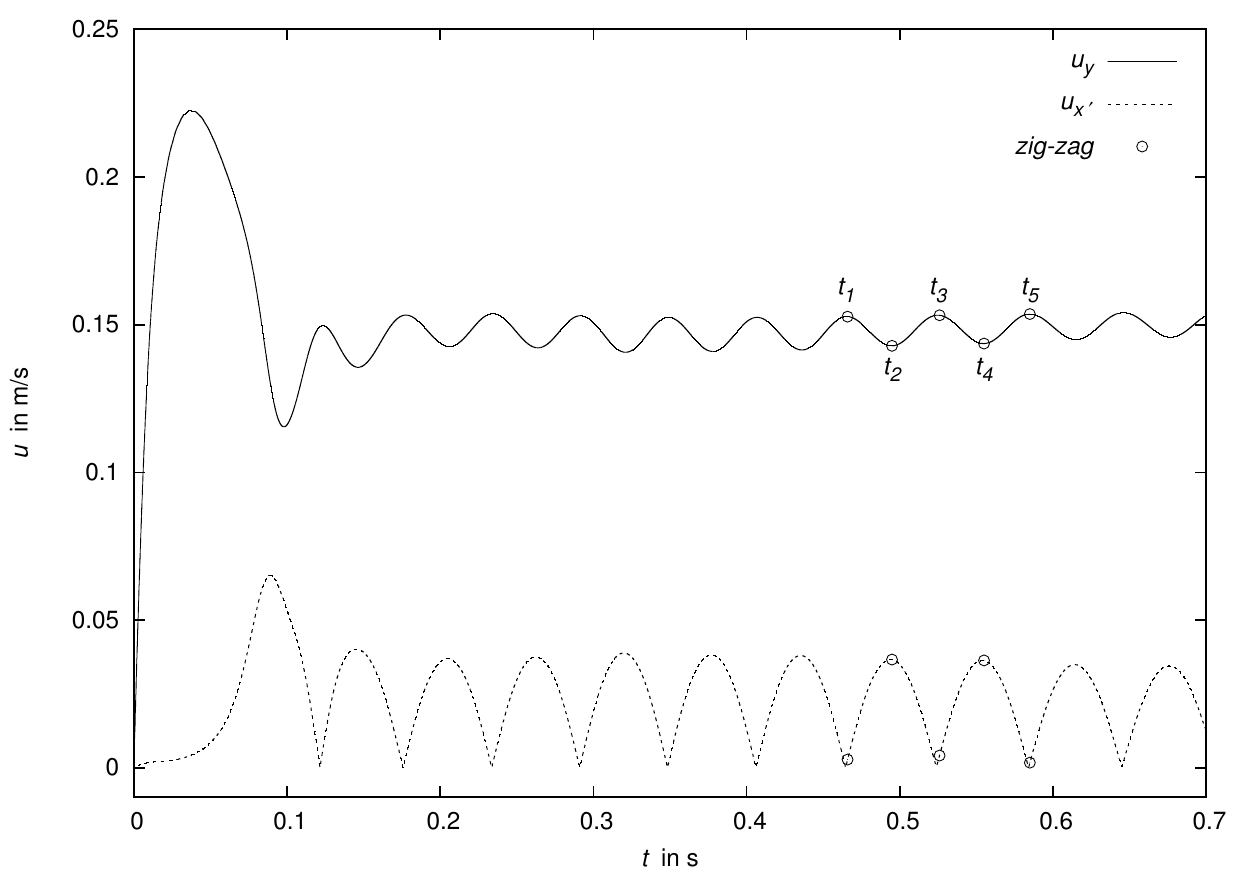}\label{fig:b5v1}} \quad
\subfloat[][\emph{Lateral view of the bubble path}.]
{\includegraphics[width=.48\textwidth]{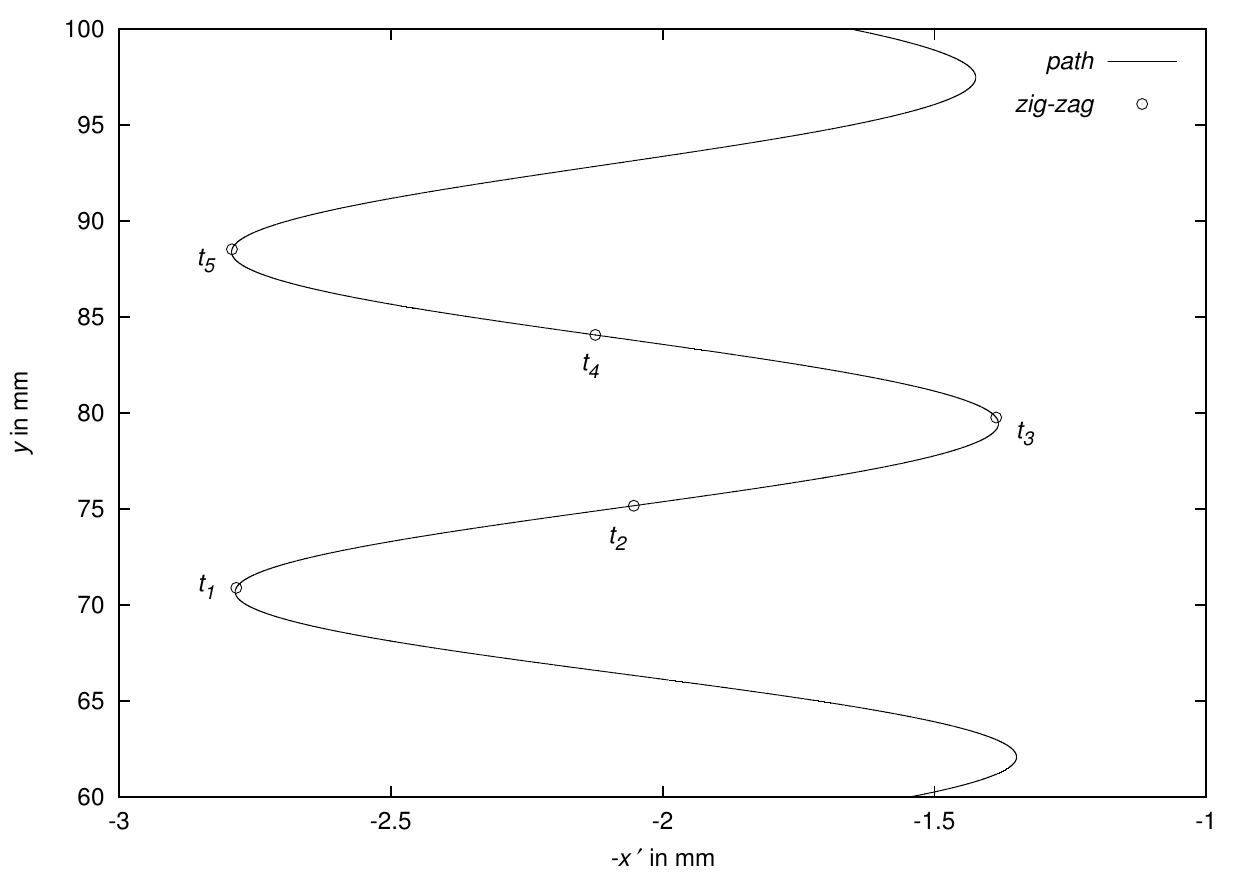}\label{fig:b5v2}}  \\
\subfloat[][\emph{$t_1$}.]
{\includegraphics[width=.15\textwidth]{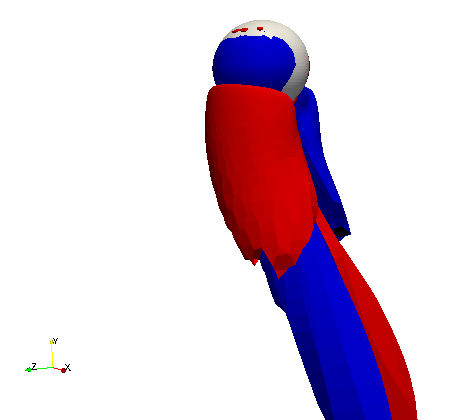}\label{fig:b5va}}$\ $
\subfloat[][\emph{$t_2$}.]
{\includegraphics[width=.15\textwidth]{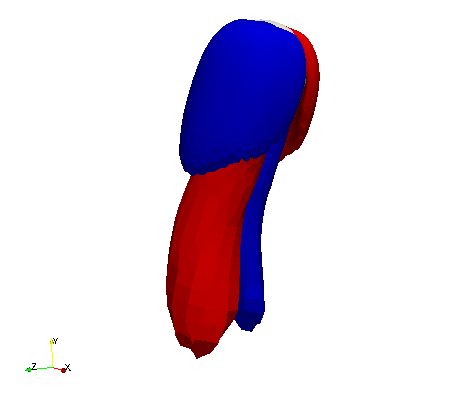}\label{fig:b5vb}}$\ $
\subfloat[][\emph{$t_3$}.]
{\includegraphics[width=.15\textwidth]{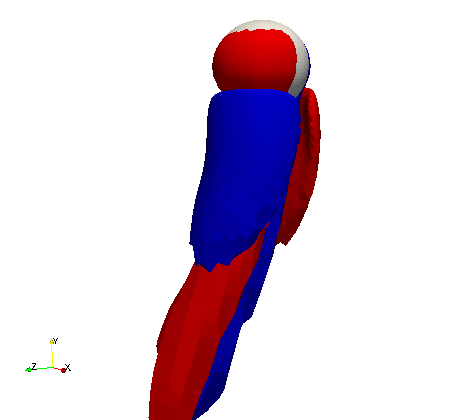}\label{fig:b5vc}}$\ $
\subfloat[][\emph{$t_4$}.]
{\includegraphics[width=.15\textwidth]{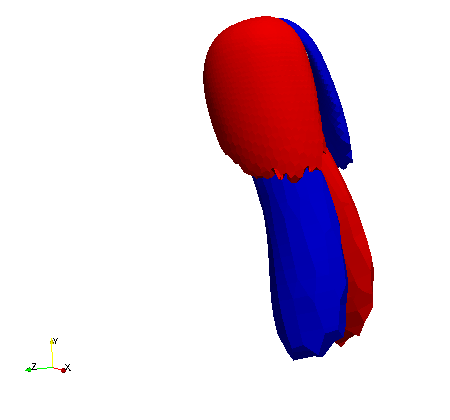}\label{fig:b5vd}}$\ $
\subfloat[][\emph{$t_5$}.]
{\includegraphics[width=.15\textwidth]{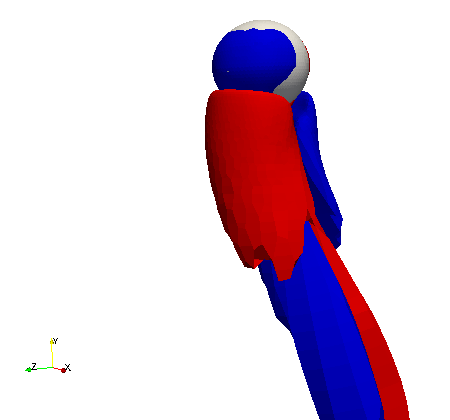}\label{fig:b5ve}}$\ $
\caption{Vorticity contour plot ($\omega_y = \pm 40$ 1/s) at different time instances, $c_0 = 8 \cdot 10^{-3}\ \mathrm{mol/m^3}$, $c^{\Sigma}_0 = 2 \% c^{\Sigma}_{eq,2}$.}
\label{fig:b5vort}
\end{figure}

\begin{figure}[ht]
\centering
\subfloat[][\emph{Rise velocity}.]
{\includegraphics[width=.48\textwidth]{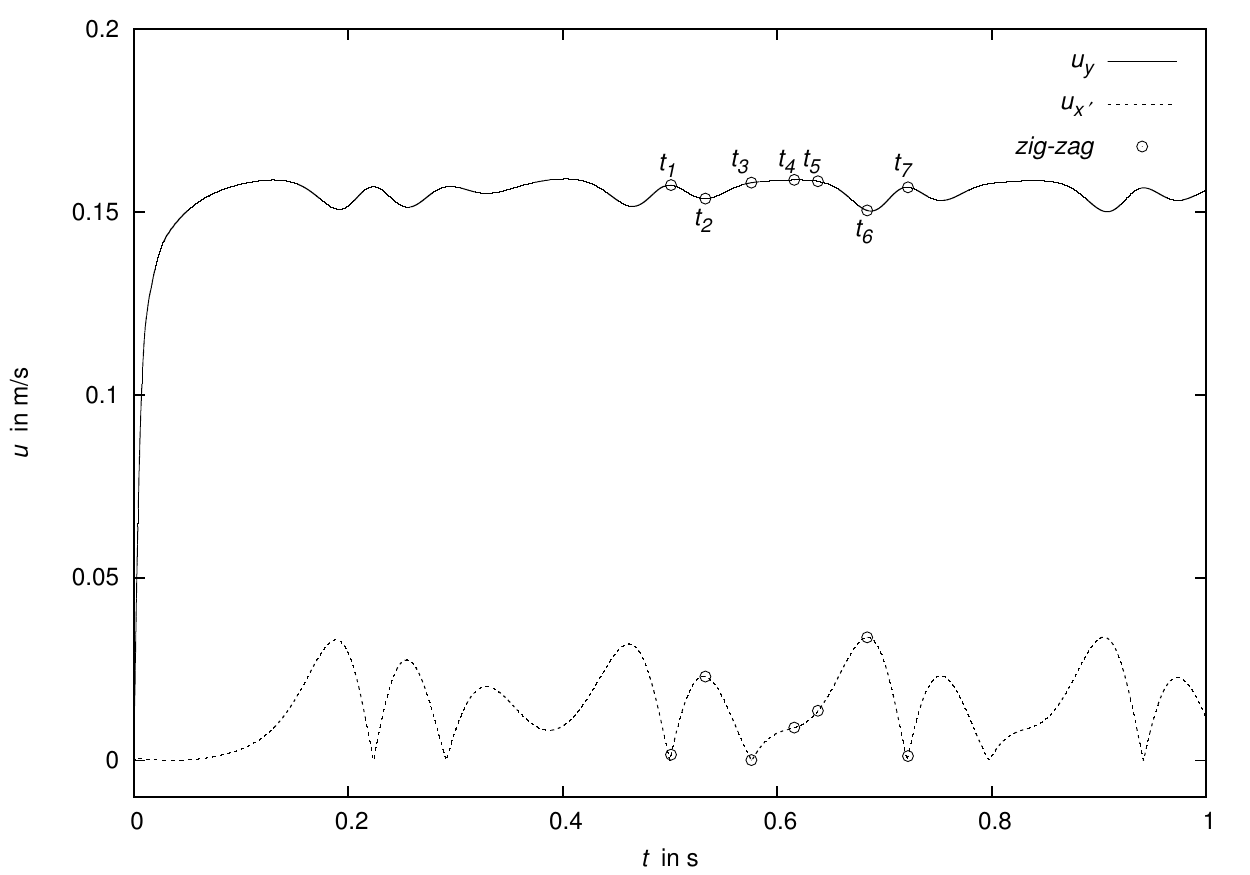}\label{fig:b8v1}} \quad
\subfloat[][\emph{Lateral view of the bubble path}.]
{\includegraphics[width=.48\textwidth]{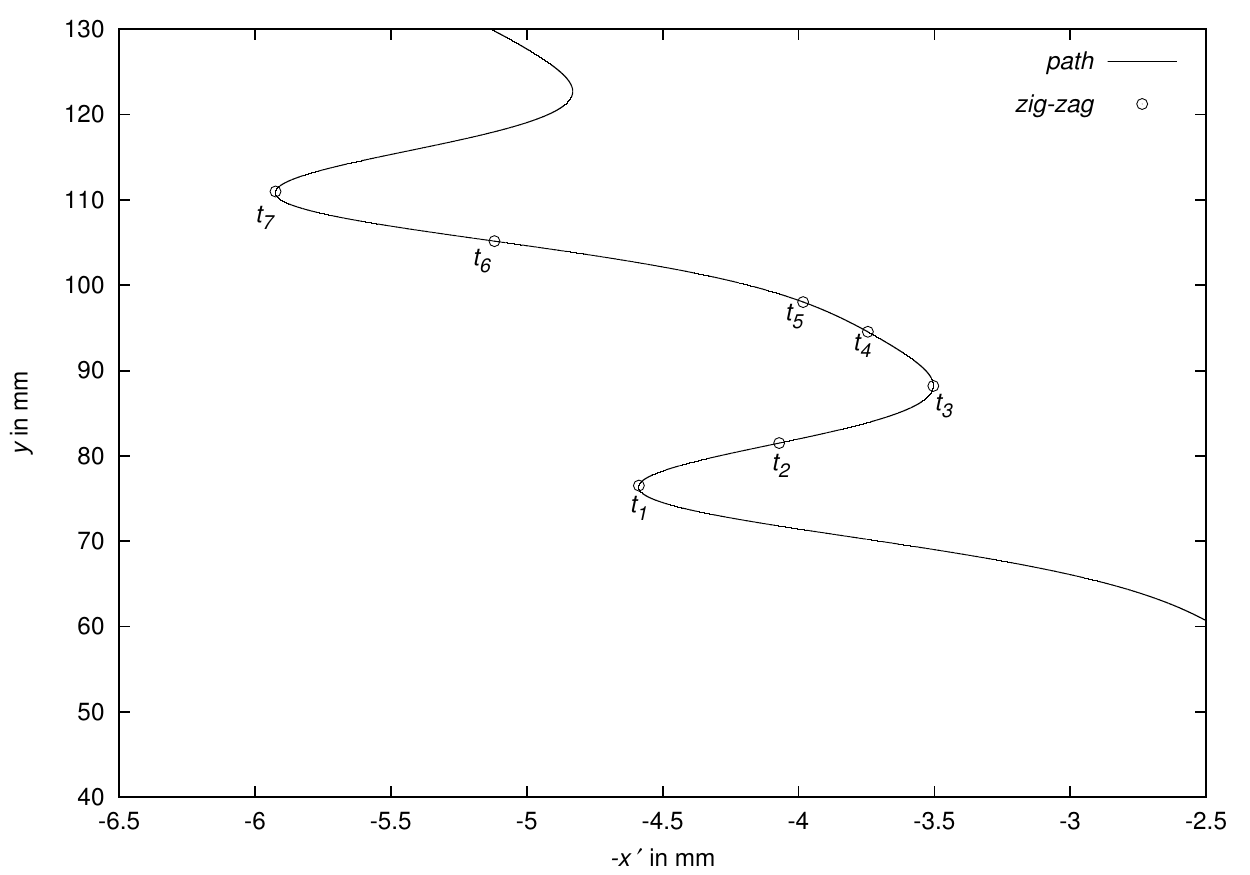}\label{fig:b8v2}}  \\
\subfloat[][\emph{$t_1$}.]
{\includegraphics[width=.15\textwidth]{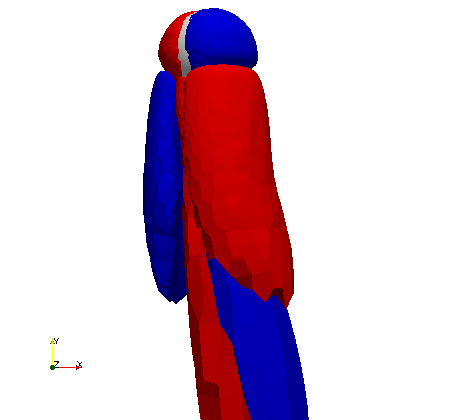}\label{fig:b8va}} $\ $
\subfloat[][\emph{$t_2$}.]
{\includegraphics[width=.15\textwidth]{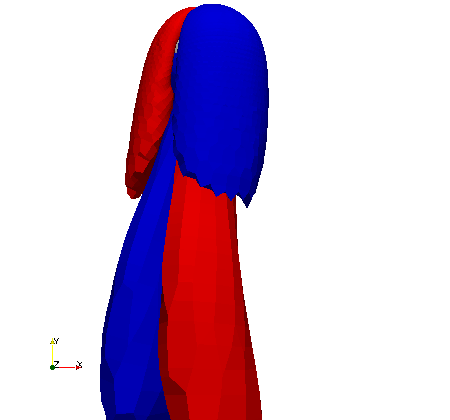}\label{fig:b8vb}} $\ $
\subfloat[][\emph{$t_3$}.]
{\includegraphics[width=.15\textwidth]{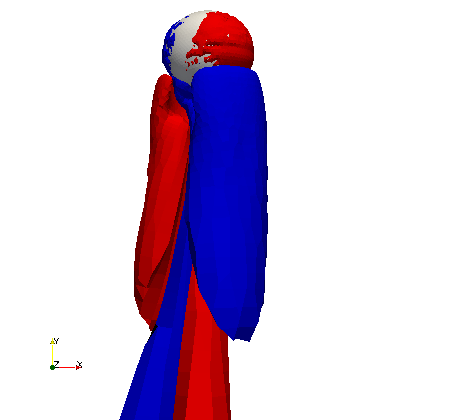}\label{fig:b8vc}} $\ $
\subfloat[][\emph{$t_4$}.]
{\includegraphics[width=.15\textwidth]{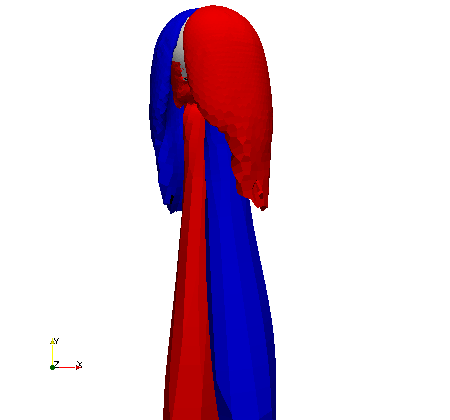}\label{fig:b8vd}} $\ $
\subfloat[][\emph{$t_5$}.]
{\includegraphics[width=.15\textwidth]{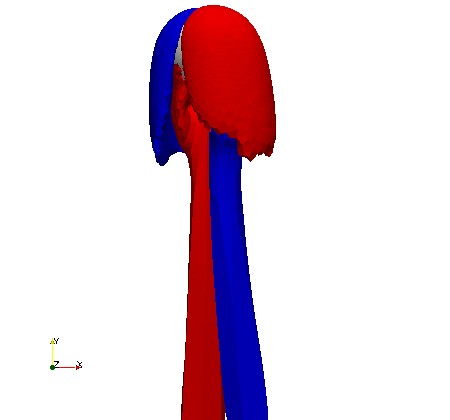}\label{fig:b8ve}} \\
\subfloat[][\emph{$t_6$}.]
{\includegraphics[width=.15\textwidth]{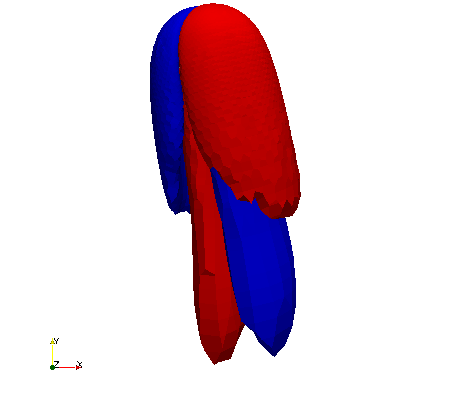}\label{fig:b8vf}} $\ $
\subfloat[][\emph{$t_7$}.]
{\includegraphics[width=.15\textwidth]{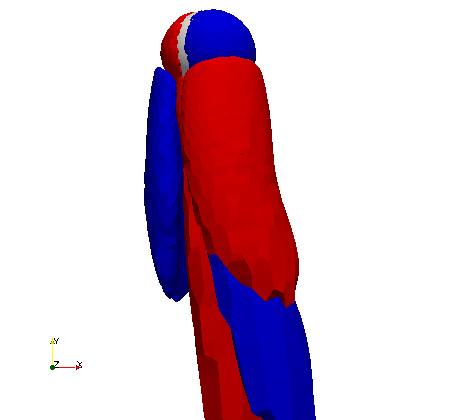}\label{fig:b8vg}}
\caption{Vorticity contour plot ($\omega_y = \pm 10$ 1/s) at different time instances, $c_0 = 5 \cdot 10^{-2}\ \mathrm{mol/m^3}$, $c^{\Sigma}_0 = 5 \% c^{\Sigma}_{eq,3}$.}
\label{fig:b8vort}
\end{figure}

The flow type around the bubble may be characterized by the vorticity ($\boldsymbol{\omega} = \curl \vec{u}$) contour plots in rise direction reported here at various time instances for the three different initial surfactant bulk concentrations; see figures~\ref{fig:b3vort},~\ref{fig:b5vort} and~\ref{fig:b8vort}. Common to all the cases is the strong vorticity production already very close to the interface due to the presence of Marangoni forces. This behaviour related to the surfactant presence is not encountered for path unstable bubbles rising in clean water; see for instance the vorticity distribution in~\cite{mougin2006}~(figures 8 and 9). Moreover, at the end of each period, that is when the bubble completes a full turn (from $t_1$ to $t_5$ in figure~\ref{fig:b3vort} for example), the streamwise vorticity does not vanish.
\FloatBarrier
In the least contaminated case the bubble follows first a helical and then a zig-zag path. This behaviour is confirmed by the vorticity contour plots in figure~\ref{fig:b3vort}. The figures from~\ref{fig:b3va} to~\ref{fig:b3ve} refer to time instances when the bubble path is helical, while the figures from~\ref{fig:b3vf} to~\ref{fig:b3vj} refer to the zig-zag trajectory. As already observed by other authors, e.g. in~\cite{ellingsen2001,mougin2006,canoLozano2016}, along the helical trajectory, the vortical structure is formed by two counter-rotating vortices of opposite sign that produce a bubble inclination in both $x$ and $z$ directions. The two vorticity regions are wrapping around each other without any symmetry plane. On the other hand, when the bubble exhibits a zig-zag trajectory, the inclination changes only in one direction. In this case the wake structure consists of two counter-rotating vortices with a symmetry plane. Common to both trajectories, at each cycle (from one velocity peak to another which corresponds from one side to the other of the path in the $x^{\prime}-y$ view) the two vortices interchange their signs. Due to the high mobility of the interface in the initial stage, the bubble reaches a high terminal velocity and deforms. After the onset of the path instability, the trajectory is helical. With increasing surface contamination, a symmetry between the wake vortices is established and the trajectory changes from helical to zig-zag. Interestingly this happens when the rise velocity is already very close to its quasi-steady value. We, therefore, conclude that not only the pure deceleration but also the indirect influence of the Marangoni forces on the flow pattern around the bubble cause the observed transition.
A similar zig-zag trajectory can be observed for the bubble in figure~\ref{fig:b5vort}. Also in this case two counter-rotating vortices with a symmetry plane are present.
\FloatBarrier
A different behaviour is observed for the most contaminated case; see figure~\ref{fig:b8vort}. The bubble follows a zig-zag trajectory, but the motion is accompanied by a lateral migration. The vortical structure is composed by two counter-rotating vortices with a symmetry plane, but the duration of each half-cycle is not constant any more, as it was for the cases in figures~\ref{fig:b3vort} and~\ref{fig:b5vort}, due to the drift. Considering figure~\ref{fig:b8vort} from $t_1$ to $t_3$, the vorticity production is much higher than from $t_4$ to $t_7$. This means that a bigger portion of fluid around the interface is influenced by bubble motion. Instead, at the sample times $t_5$ and $t_6$ the vorticity production is much less, thus the fluid around the bubble will be less perturbed and the drift towards the left side lasts longer. At $t_7$ the same conditions as in $t_1$ are restored. It seems to be a superimposition of clean case migration and contaminated case oscillation. A possible explanation will be given in section~\ref{subsubsec:5_5_3}.
\FloatBarrier
%

\subsubsection{Forces acting on the interface}
\label{subsubsec:5_5_3}
Several experimental works derived correlations for global lift and drag coefficients of single rising bubbles, e.g.~\cite{tomiyama1998}. In our work, we focus on the local forces acting on the interface and how they influence the integral lift and drag forces. The interfacial jump condition~\eqref{eq:m5} is considered in order to evaluate the forces acting on the interface:
\begin{equation}
\dleftsq p_{\mathrm{tot}}\ \mathrm{\mathbf{I}} - \mathbf{S}^{\mathrm{visc}} \drightsq \cdot \n_{\Sigma} = \sigma \kappa \n_{\Sigma} + \grad_{\Sigma} \sigma,
\label{eq:m5_rep}
\end{equation} 
where $p_{\mathrm{tot}}$ is the total pressure, the sum of dynamic and hydrostatic contributions\footnote{Within the algorithm, equations~\eqref{eq:m1}~to~\eqref{eq:m5} are solved for the modified pressure, or dynamic pressure $p^{\mathrm{dyn}}$ as we will refer to, that is the total pressure minus the hydrostatic contribution,
\begin{equation}
p^{\mathrm{dyn}} = p^{\mathrm{tot}} - p^{\mathrm{hydro}}
\label{eq:n32_a}
\end{equation}
with $p^{\mathrm{hydro}} := \rho \vec{g} \cdot \vec{x}$.
This means that in equation~\eqref{eq:m2} the gravity term disappears and the transmission condition~\eqref{eq:m5} has to be adapted according to the relation~\eqref{eq:n32_a}, too.}.\\
For clarity, we recall that $\vec{f}^{\mathrm{ma}} = \grad_{\Sigma} \sigma$ is the area specific Marangoni force, while $\vec{f}^{\mathrm{ca}} = \sigma \kappa \n_{\Sigma}$ is the area specific capillary pressure force.
Equation~\eqref{eq:m5_rep} at each interface element reads
\begin{equation}
\vec{f}_{B}^{p_{\mathrm{tot}}} - \vec{f}_{A}^{p_{\mathrm{tot}}} - \vec{f}_{B}^{\mathrm{visc}} + \vec{f}_{A}^{\mathrm{visc}} = \vec{f}^{\mathrm{ca}} + \vec{f}^{\mathrm{ma}},
\label{eq:m5_locF}
\end{equation} 
where $\vec{f}^*$ are the area specific forces $\vec{f}^* = \vec{f}^*(\vec{x}^{\Sigma},t)$\footnote{The superscript `*' stands for `$p_{\mathrm{tot}}$', `$\mathrm{visc}$', `$\mathrm{ca}$' or `$\mathrm{ma}$'.}, $A$ represents the liquid phase and $B$ the gas phase. The symbols $\vec{f}^{p^{\mathrm{tot}}}$ and $\vec{f}^{\mathrm{visc}}$ indicate the total pressure and viscous forces, respectively. Comparing the magnitude of the forces between the sides A and B it can be noticed that $\vec{f}_B^*$ is always at least one order of magnitude smaller than the respective force from the A side, thus in the following analysis it will be neglected.

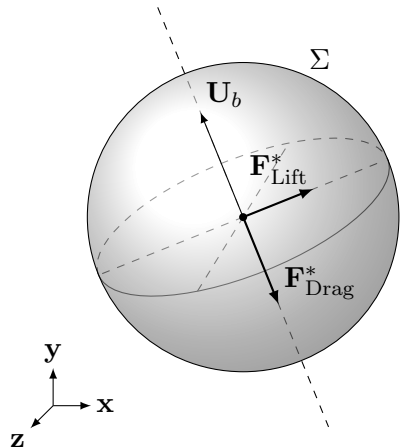
\begin{figure}[ht]
\centering
\begin{tikzpicture}

\def\R{2.05}
\def\angEl{22}

  \draw[dashed] (-0.6,-0.98) -- (0.6,0.98);
    \begin{scope}[rotate around={22:(0,0)}]
    \DrawLatitudeCircle[\R]{0}    
    \draw[dashed] (0,3) -- (0,-3);
    \draw[dashed] (-2,0) -- (2,0);
  \end{scope}
  
  \shade[ball color=white,opacity=0.6] (0,0) circle[x radius=\R, y radius=\R];
  \coordinate[mark coordinate] (sigma) at (0,0);
  \draw (0,0) circle[x radius=\R, y radius=\R];
  \draw (0.8,1.9) node[above right]{$\Sigma$};

  \begin{scope}[rotate around={22:(0,0)}]
    \draw[->] (0,0) -- (0,1.5) node[above right] {$\mathbf{U}_b$};
    \draw[thick,->] (0,0) -- (0,-1.25) node[above right] {$\mathbf{F}^*_{\mathrm{Drag}}$};
    \draw[thick,->] (0,0) -- (1.0,0) node[above left] {$\mathbf{F}^*_{\mathrm{Lift}}$};
  \end{scope}
  
  \draw [->] (-2.5,-2.5) -- (-2.5,-2.0) node[above] {$\mathbf{y}$};
  \draw [->] (-2.5,-2.5) -- (-2.0,-2.5) node[right] {$\mathbf{x}$};
  \draw [->] (-2.5,-2.5) -- (-2.8,-2.8) node[below left] {$\mathbf{z}$};
 
\end{tikzpicture}
\caption{Schematic representation of the lift and drag directions.}
\label{fig:liftDrag}
\end{figure}

\begin{figure}[ht]
\centering
\includegraphics[width=1.0\textwidth]{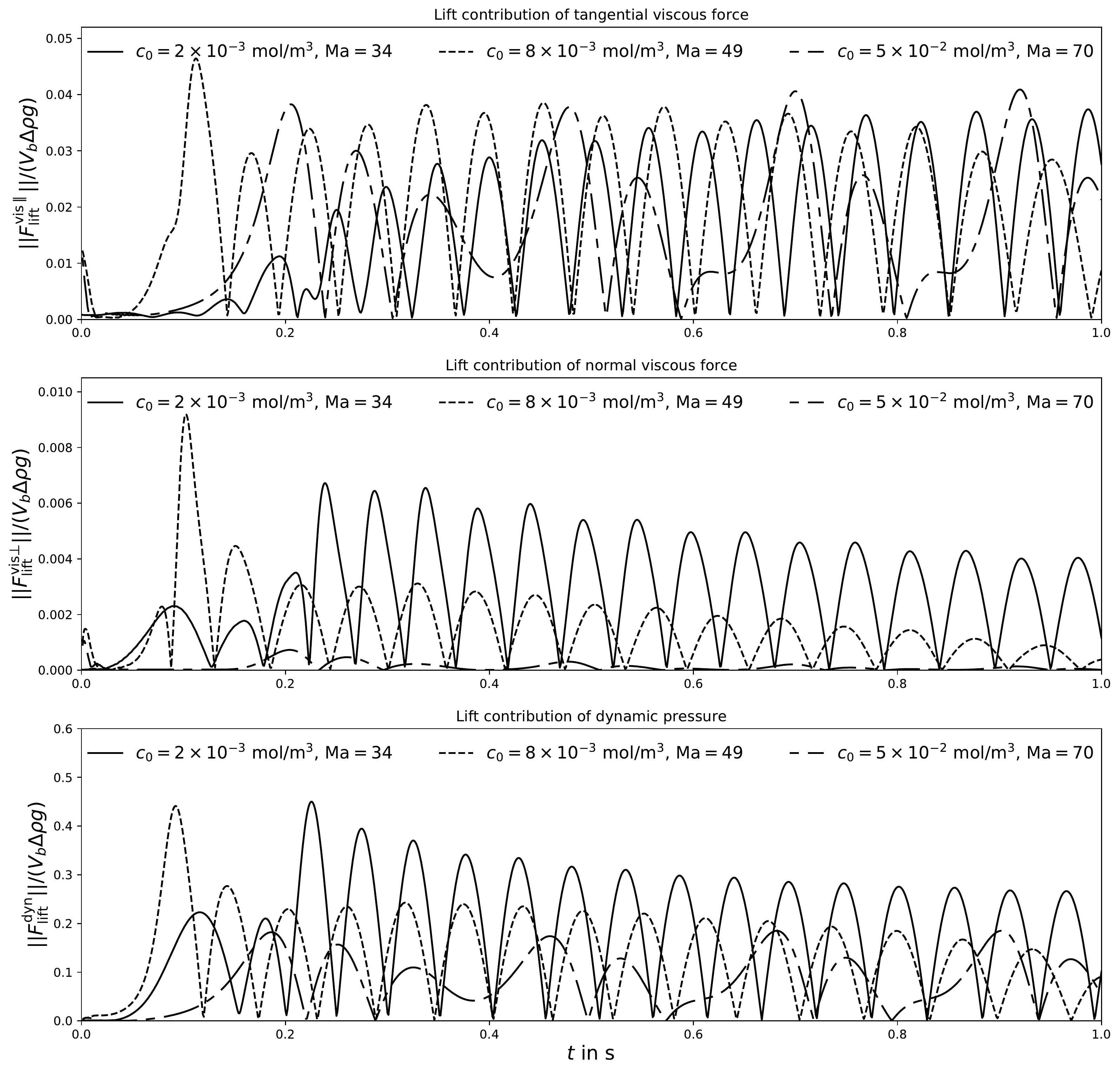}
\caption{Integral lift force contributions, influence of the initial bulk concentration.}
\label{fig:b358_FL}
\end{figure}

\begin{figure}[ht]
\centering
\includegraphics[width=1.0\textwidth]{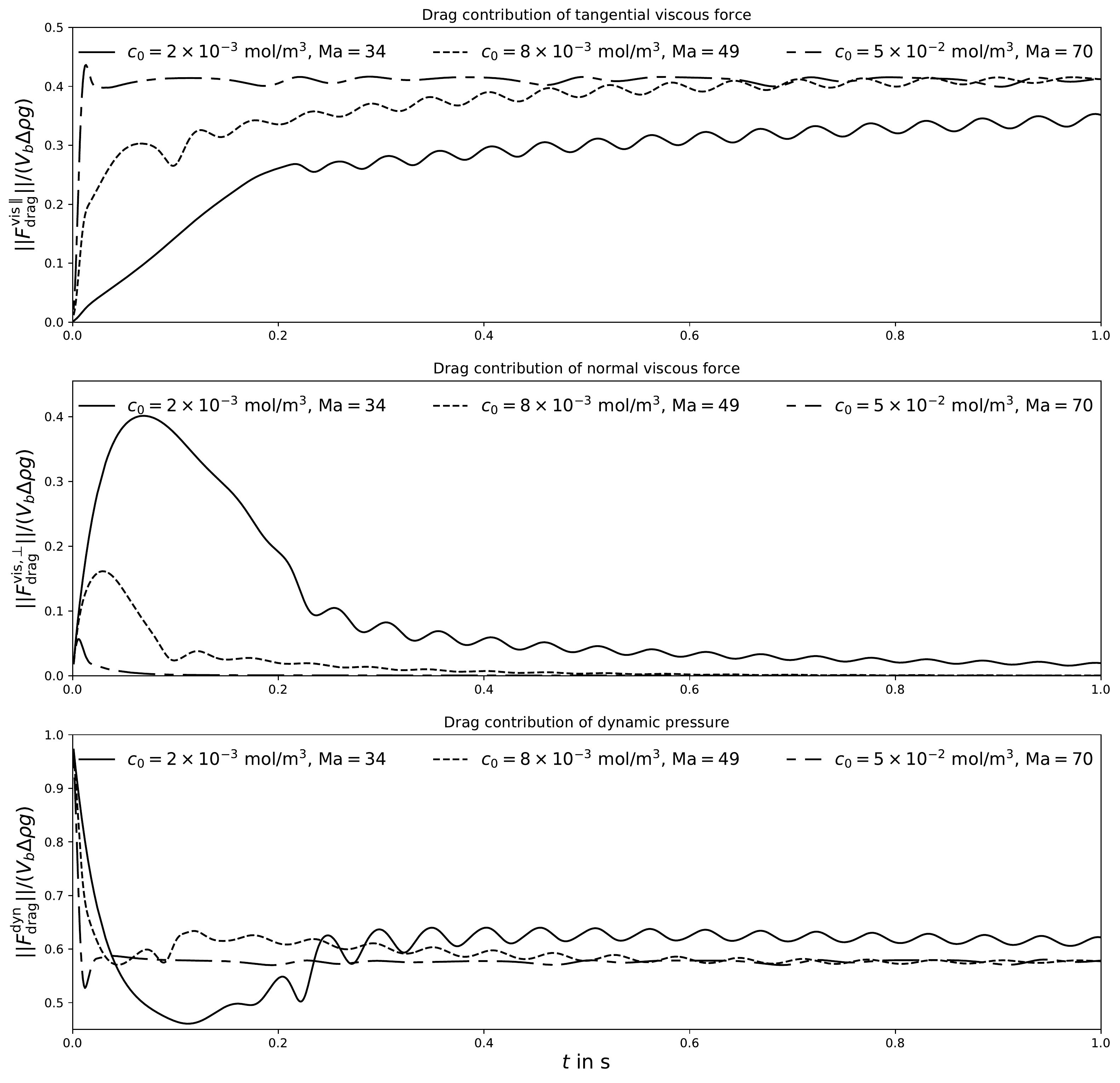}
\caption{Integral drag force contributions, influence of the initial bulk concentration.}
\label{fig:b358_FD}
\end{figure}

The local force balance at the interface~\eqref{eq:m5_locF} is projected in normal and tangential direction to the interface. For the liquid side (A, dropped from here onwards) the two balances read
\begin{equation}
- \vec{f}^{p_{\mathrm{tot}}} + \vec{f}_{\perp}^{\mathrm{visc}} = \vec{f}^{\mathrm{ca}} \quad \mathrm{normal\ to\ }\Sigma,
\label{eq:m5_locF_norm}
\end{equation} 
\begin{equation}
\vec{f}_{\parallel}^{\mathrm{visc}} = \vec{f}^{\mathrm{ma}} \quad \mathrm{tangential\ to\ }\Sigma.
\label{eq:m5_locF_tg}
\end{equation} 
The total pressure force can be further decomposed into the hydrostatic and the dynamic contributions, i.e.
\begin{equation}
\vec{f}^{p_{\mathrm{tot}}} = \vec{f}^{p_{\mathrm{hydro}}} + \vec{f}^{p_{\mathrm{dyn}}}.
\label{eq:pTotDec}
\end{equation} 
Integrating the area specific forces $\vec{f}^*(\vec{x}^{\Sigma},t)$ over the interface, we get the resultant force $\vec{F}^*(t)$ on $\Sigma$ as
\begin{equation}
\vec{F}^*(t) = \int_{\Sigma} \vec{f}^*(\vec{x}^{\Sigma},t) dA.
\label{eq:Fsigma}
\end{equation} 
Thus, the following forces are acting on the bubble surface: the hydrostatic pressure force $\vec{F}^{p_{\mathrm{hydro}}}$, the dynamic pressure force $\vec{F}^{p_{\mathrm{dyn}}}$, normal and tangential viscous forces $\vec{F}_{\perp}^{\mathrm{visc}}$, $\vec{F}_{\parallel}^{\mathrm{visc}}$, the Marangoni force $\vec{F}^{\mathrm{ma}}$, and the capillary pressure force $\vec{F}^{\mathrm{ma}}$. The hydrostatic pressure force is approximately constant over time, so we do not analyse it. As can be observed from equations~\eqref{eq:m5_locF_norm} and~\eqref{eq:m5_locF_tg} the tangential viscous force is balanced by the Marangoni force. Thus we can just consider one of them, say $\vec{F}_{\parallel}^{\mathrm{visc}}$. For the same reason, we drop the capillary pressure force as it is equal in magnitude to the sum of total pressure force and normal viscous force. We are left with three integral forces, $\vec{F}_{\parallel}^{\mathrm{visc}}$, $\vec{F}^{p_{\mathrm{dyn}}}$ and $\vec{F}_{\perp}^{\mathrm{visc}}$, that are decisive for understanding the bubble dynamics. Each force may be written as the sum of contributions parallel and perpendicular to the bubble velocity vector. The parallel component we refer to as drag and the remaining component as lift force:
\begin{equation}
\vec{F}^*(t) = \vec{F}^*_{\mathrm{Lift}} + \vec{F}^*_{\mathrm{Drag}},
\label{eq:LD}
\end{equation} 
as depicted in figure~\ref{fig:liftDrag}. The drag force governs the bubble acceleration/deceleration and the lift force the bubble's change in direction.
\FloatBarrier
Figures~\ref{fig:b358_FL} and~\ref{fig:b358_FD} show the contributions from the three integral forces mentioned above to lift and drag. The different line types correspond to the various initial bulk concentrations. In order to have a common reference, the magnitude of the forces has been made non-dimensional with respect to the buoyancy force.

\begin{figure}[ht]
\centering
\includegraphics[width=0.94\textwidth]{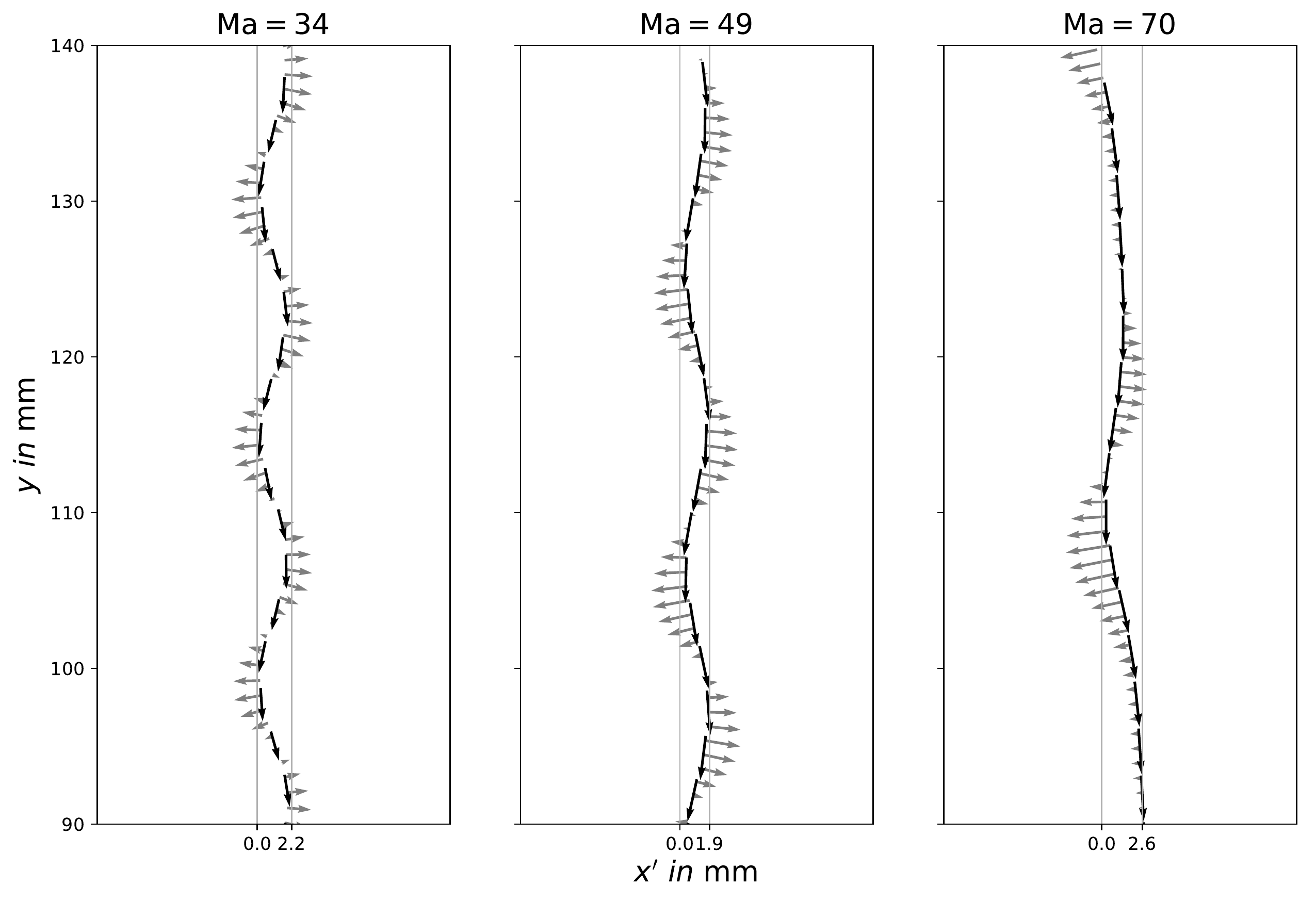}
\caption{Lift (grey) and drag (black) due to tangential viscous forces along the path. Note that the lift force is depicted ten times larger than the drag force.}
\label{fig:b358_FLD_vectors_1}
\end{figure}

\begin{figure}[ht]
\centering
\includegraphics[width=0.94\textwidth]{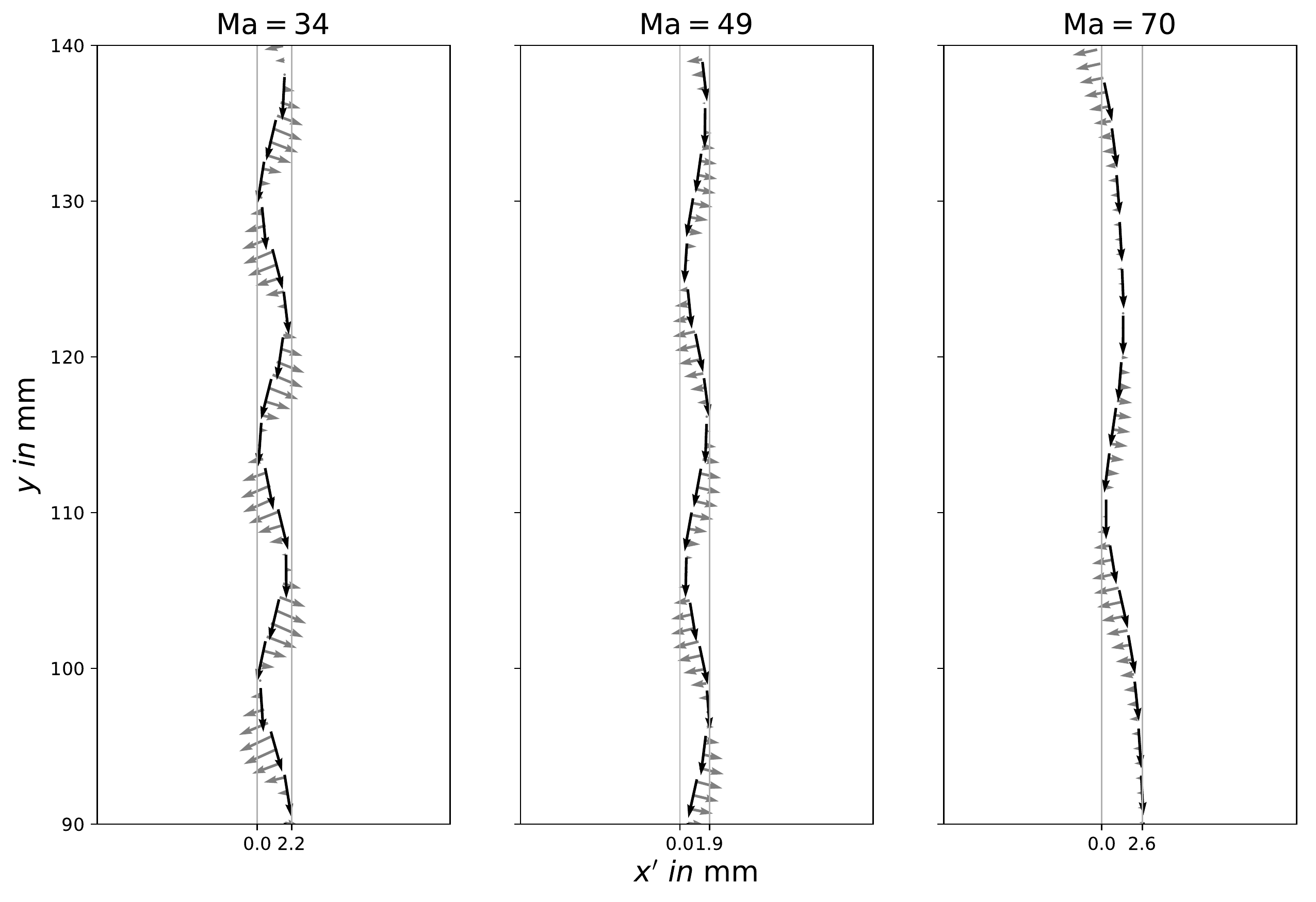}
\caption{Lift (grey) and drag (black) due to dynamic pressure forces along the path. Note that the lift force is depicted two times larger than the drag force.}
\label{fig:b358_FLD_vectors_2}
\end{figure}

As can be noticed from figure~\ref{fig:b358_FL}, the major contribution to the lift force is from the dynamic pressure force (up to 50\% of the buoyancy force). The tangential viscous force contribution to the lift does not exceed 5\%, while the normal viscous force contribution is below 1\%. 
Considering the lift contribution of the dynamic pressure and the bubbles' paths in figure~\ref{fig:b358path}, one can see that a wider trajectory corresponds to a higher lift force (in terms of helical or zig-zag radius); the lower the Marangoni number, the higher the dynamic pressure force and the wider the path. We can see that the lateral motion is mainly driven by the dynamic pressure force. Whether or not the Marangoni forces/tangential viscous forces decrease the lateral motion directly will be clarified in section~\ref{subsec:5_6}. From the plot of the force magnitude, we cannot draw any conclusion on the direction of the bubble motion. For instance, it is not possible to deduce from this plot when the least contaminated bubble ($\operatorname{Ma} = 34$) is changing its trajectory from helical to zig-zag. These aspects will be investigated later in this section; see figures~\ref{fig:b358_FLD_vectors_1} and~\ref{fig:b358_FLD_vectors_2}.
\FloatBarrier
Consider now the force contributions to the drag force, see figure~\ref{fig:b358_FD}. As for the lift, the main contribution comes from the dynamic pressure force, although for the drag, tangential and normal viscous forces cannot be neglected. In the first graph in figure~\ref{fig:b358_FD}, the contribution of the tangential viscous force to the drag is reported. Increasing Marangoni numbers, i.e.~higher surfactant concentrations, lead to a higher drag contribution of $\vec{F}_{\parallel}^{\mathrm{visc}}$. When the bubble reaches the quasi-steady state, after approximately 0.4 s, the tangential viscous force (as the Marangoni force) is still slowly increasing. We believe that this is due to the fact that the equilibrium value of the interfacial concentration has not yet been reached, and thus surfactant is still accumulating on the interface, changing its properties and consequently the Marangoni force. On the other hand, it can be seen from figure~\ref{fig:b358_FD}, that the drag contribution of the normal viscous force decreases with time. At the beginning of the bubble rise, there is a stronger change of the velocity normal to the interface, resulting in higher viscous stresses. In fact, the drag due to viscous forces is the highest for the lowest Marangoni number. For increasing Marangoni numbers, this contribution becomes more and more negligible; see for instance the line corresponding to $\operatorname{Ma} = 70$. 
To conclude the analysis on the drag force, consider the dynamic pressure contribution to it in figure~\ref{fig:b358_FD} (bottom plot). During the initial part of the acceleration phase at the beginning of the rise, the dynamic pressure force contributions reach values comparable to the gravitational force, being the highest for the least contaminated bubble, that is the one with highest rise velocity. After this initial phase, the contribution of dynamic pressure force to the drag drops and oscillates at about 60\% of the buoyancy force. As pointed out in the previous section, all studied surfactant bulk concentrations lead to a similar quasi-steady terminal velocity, even though ad- and desorption are not in equilibrium and the total surface coverage varies significantly. 
The steady state terminal velocity is a consequence of the overall drag force. For higher surfactant bulk concentrations, the viscous drag force increases due to higher surface tension gradients. At the same time, the dynamic pressure force decreases as a result of the decreasing mobility of the interface. These two counteracting effects lead to an approximately constant overall drag force.
\FloatBarrier
In figures~\ref{fig:b358_FLD_vectors_1} and~\ref{fig:b358_FLD_vectors_2} the integral force contributions to the lift and drag from the tangential viscous force and the dynamic pressure force are depicted as vectors along the bubble path. In the two figures, the coordinate $x^{\prime}$ correspond to the direction along which each bubble is translating in a horizontal plane. From these plots, one can clearly deduct how the forces are changing the bubble trajectory. The main contribution to the lift comes from the dynamic pressure; see figure~\ref{fig:b358_FL}. Thus, the deviation from a rectilinear path is mainly caused by the dynamic pressure force and not directly by the tangential viscous force (in response to the Marangoni force). Yet, with increasing contamination, the lateral motion of the bubble decreases, and this effect may be caused by a non-axisymmetric (with respect to the rise velocity vector) distribution of the surfactant on the interface. As can be seen in figure~\ref{fig:b358_FLD_vectors_1}, the Marangoni effect is actually adding to the lift. However, the reduction of the dynamic pressure is much stronger, and consequently, the overall lift is reduced.
Regarding the drag component, the dynamic pressure force is still the dominating contribution, but the tangential viscous force contributes in comparable amounts to the drag.

Even though the dynamic pressure force is the dominating component, locally the flow field is governed by the Marangoni stresses. A study of the local fields is performed in the following section~\ref{subsec:5_6}.
%
%
%

%
\subsection{Local velocity and surface fields under the influence of surfactant}
\label{subsec:5_6}
\begin{figure}[ht]
\centering
\subfloat[][\emph{$t = 0.072$ s}.]
{\includegraphics[width=.54\textwidth]{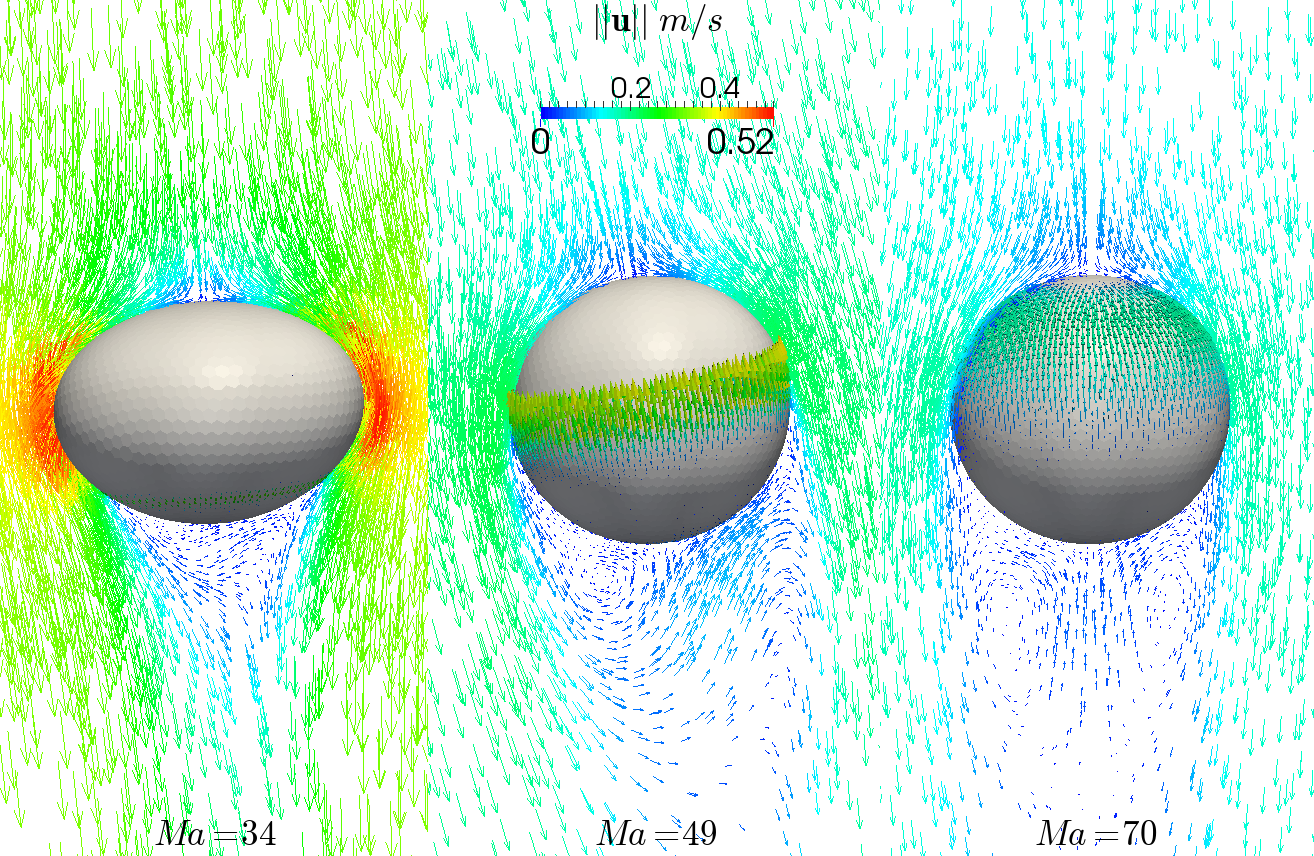}\label{fig:pic24}} \\
\subfloat[][\emph{$t = 0.3$ s}.]
{\includegraphics[width=.54\textwidth]{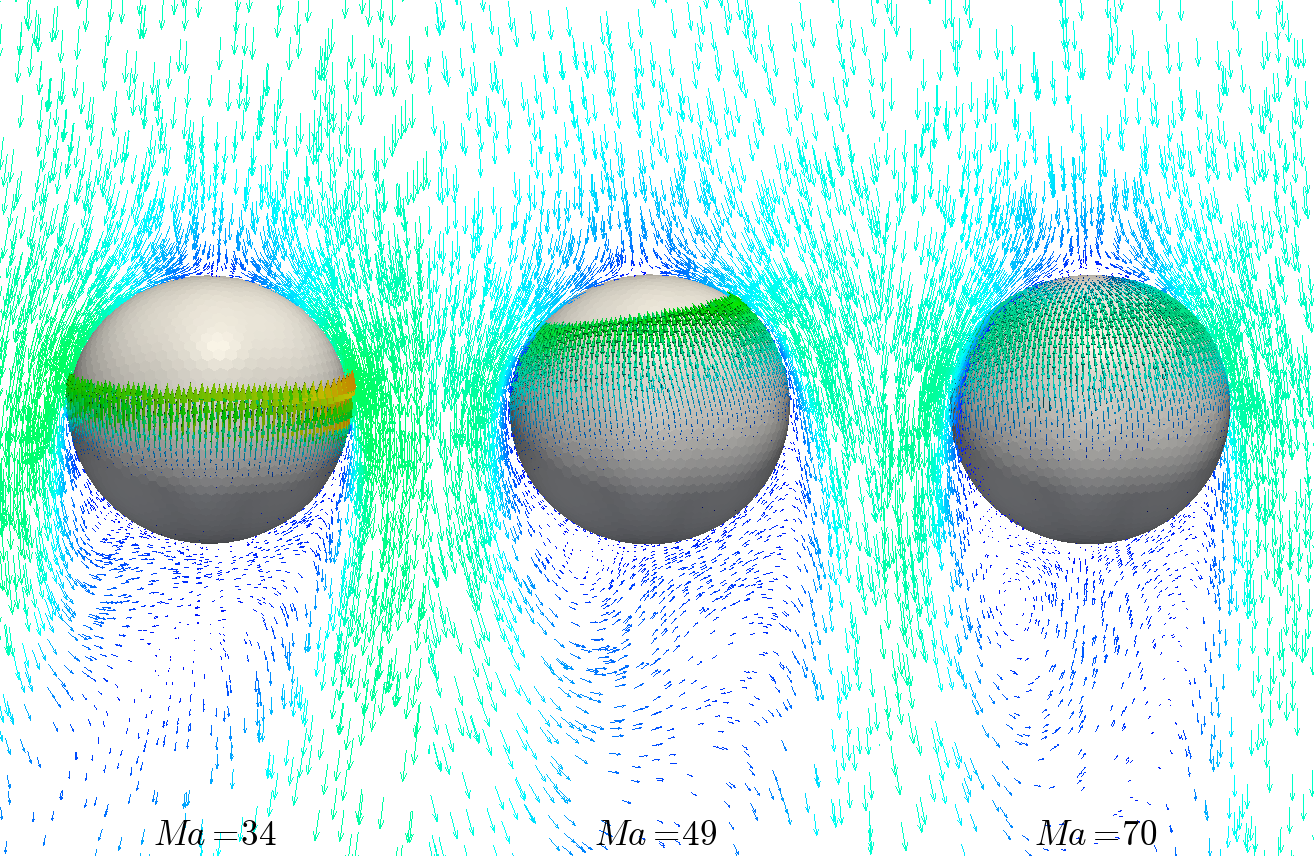}\label{fig:pic100}}  \\
\subfloat[][\emph{$t = 0.9$ s}.]
{\includegraphics[width=.54\textwidth]{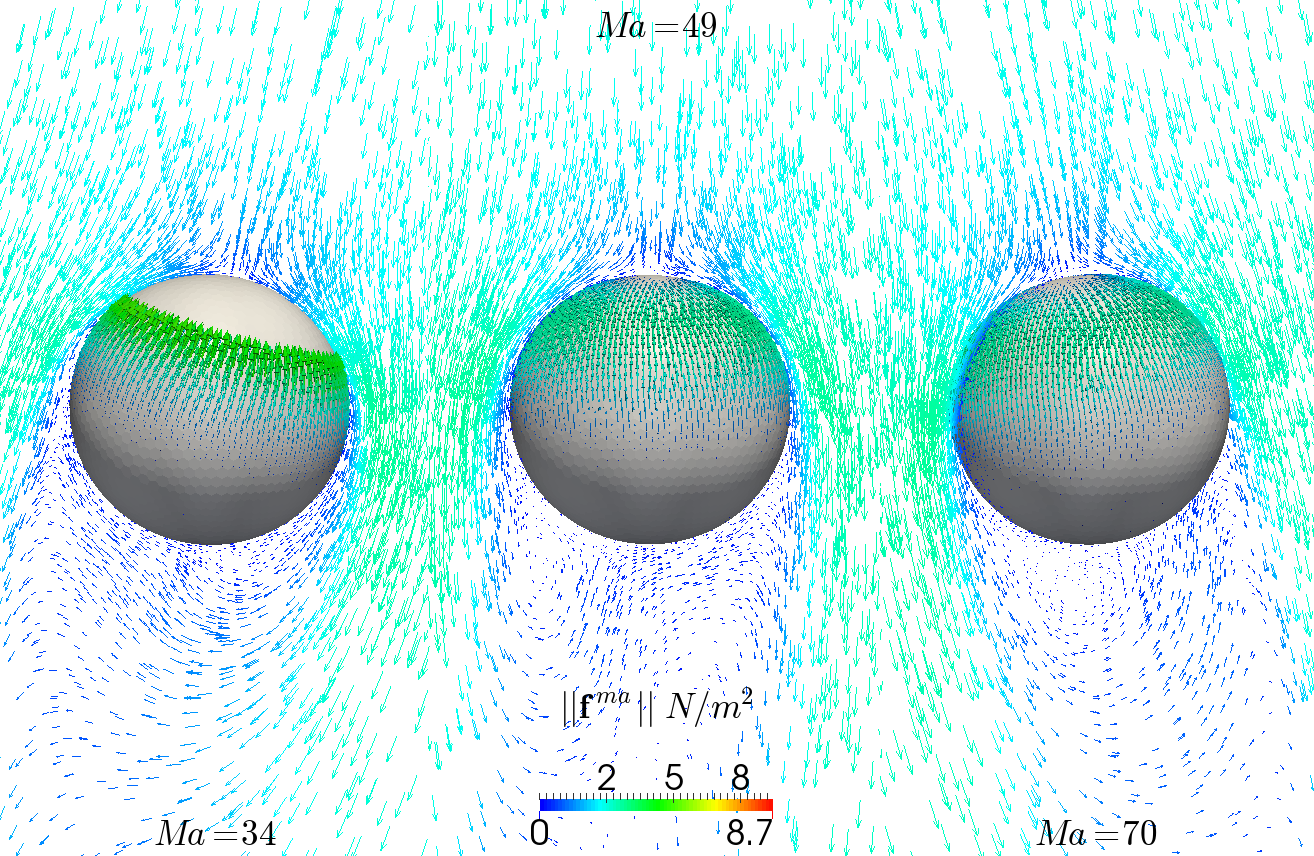}\label{fig:pic300}} 
\caption{Velocity vectors (bulk) and Marangoni forces (interface) at different time instances for $\operatorname{Ma} = 34,\ 49,\ 70$.}
\label{fig:picFlow}
\end{figure}

Figure~\ref{fig:picFlow} shows the velocity field in the liquid phase close to the bubble, while on the bubble surface the local  Marangoni force vectors are depicted for the three initial concentrations at different time instances. At $t = 0.072\ \mathrm{s}$ the bubble rising in the most contaminated solution has already reached a surfactant distribution characteristic of the steady state. In the lower hemisphere, where the surfactant concentration is the highest and uniformly distributed, the Marangoni force is almost zero, while the surface coverage is not yet the equilibrium one. In fact, the surfactant species is still adsorbed, see figure~\ref{fig:totCs358}. For the other two initial bulk concentrations, a longer initial transient stage is visible. The surface coverage is much smaller at the beginning of the rise, while much higher and more confined Marangoni stresses are visible. For the cases on the left and in the middle of figure~\ref{fig:picFlow}, it is clearly visible that the line where the flow detaches corresponds to the region where the Marangoni forces are the highest. As the bubbles are rising, more and more surfactant is adsorbed and the region where the Marangoni stresses are present moves towards the upper hemisphere. The bubble in the middle, at $t = 0.9\ \mathrm{s}$ has reached a similar state as the most contaminated bubble in terms of Marangoni stresses and terminal velocity, even though the surface coverage is approximately 60\% less; see figure~\ref{fig:totCs358}. It is reasonable to predict that the least contaminated bubble, if simulated for a longer time, would reach a similar state as the other two bubbles, but with an even lower surface coverage. 

\begin{figure}[ht]
\centering
\subfloat[][\emph{$t = 0.072$ s}.]
{\includegraphics[width=.8\textwidth]{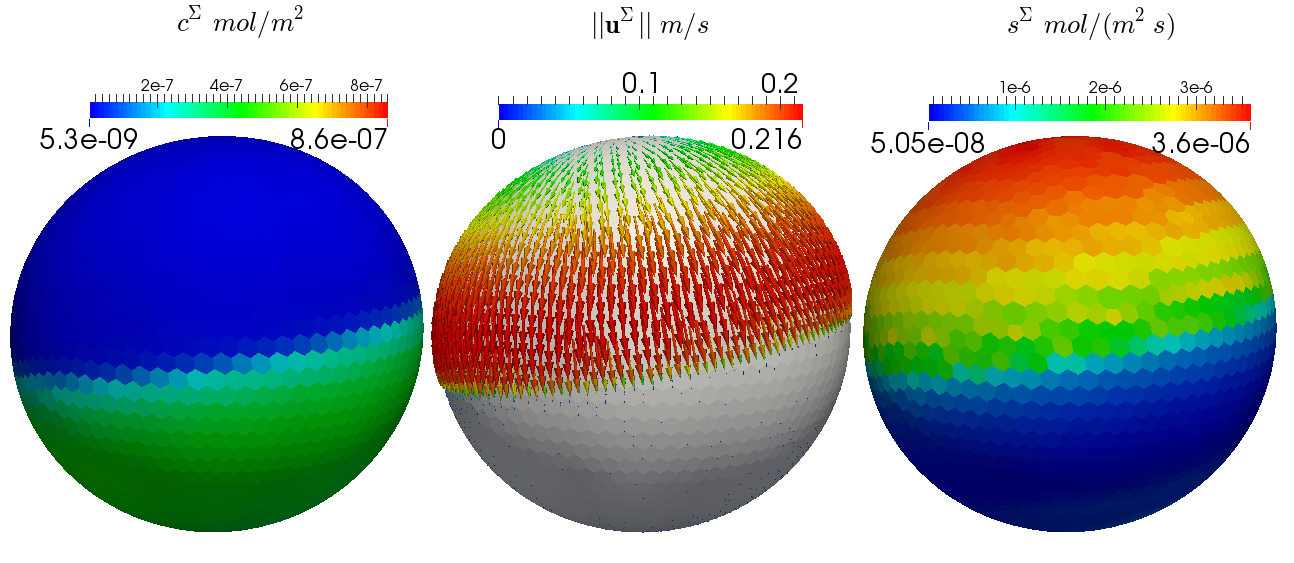}\label{fig:cap24}} \\
\subfloat[][\emph{$t = 0.3$ s}.]
{\includegraphics[width=.8\textwidth]{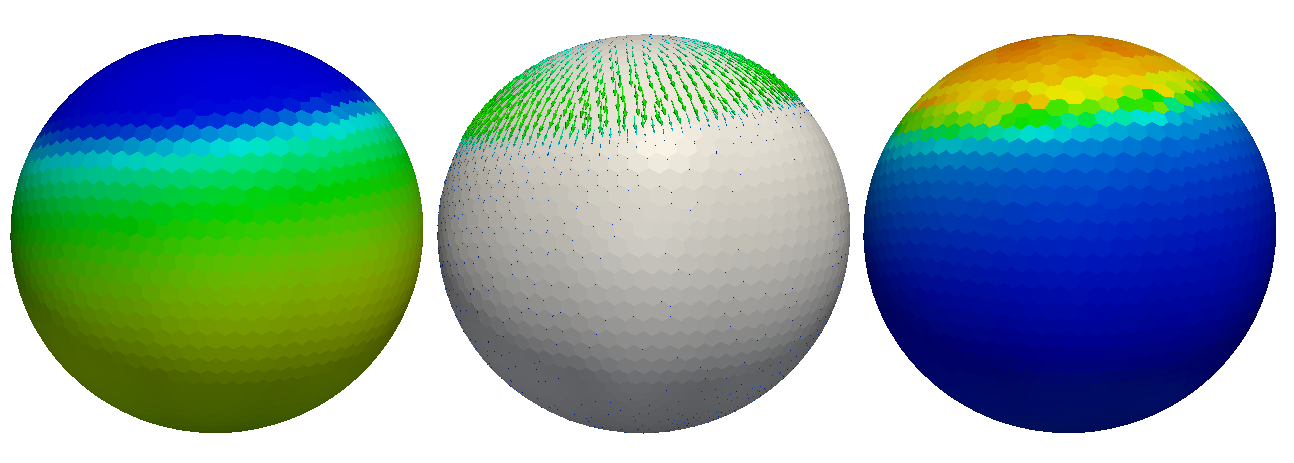}\label{fig:cap100}}  \\
\subfloat[][\emph{$t = 0.9$ s}.]
{\includegraphics[width=.8\textwidth]{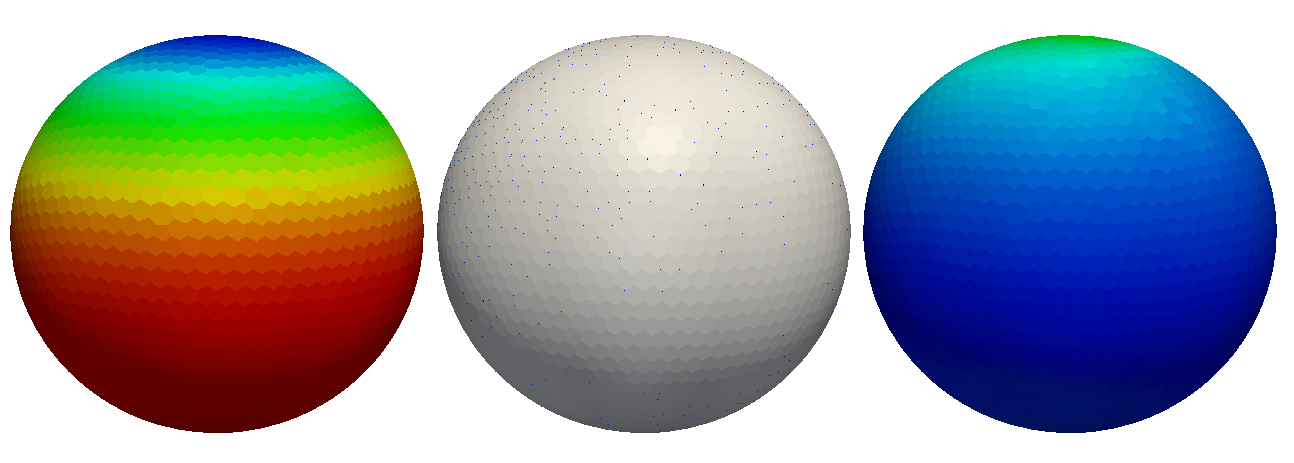}\label{fig:cap300}} 
\caption{From left to right, surfactant distribution on $\Sigma$, interface velocity field and sorption source term at different time instances for the intermediate bulk concentration $\operatorname{Ma} = 49$.}
\label{fig:picCapB5}
\end{figure}

To have a better understanding of the variation of the Marangoni forces and their local distribution, one can consider the adsorption, advection and diffusion processes on the interface; see figures~\ref{fig:picFlow} and~\ref{fig:picCapB5}. Three different stages during the bubble rise can be identified.
\begin{enumerate}
	\item After being released, the bubble undergoes a strong acceleration due to buoyancy force. The surface coverage is low and uniform and, therefore, the interface is fully mobile. A thin concentration boundary layer forms at the interface and the adsorption rates are the highest. The first stage may be very short, depending on the initial surface and bulk concentrations.
	\item Due to the high mobility of the interface, the surfactant is quickly advected to the rear part of the bubble. As a consequence, the surface coverage becomes less uniform and surface tension gradients that are strong enough to locally reduce the tangential interface velocity in the rear part arise. The flow detaches, and vortices are shed. The interface below the detachment ring is almost stagnant, and the adsorption rates are small because the concentration difference with respect to the bulk decreases and no new surfactant is transported there by convection. The front of the bubble is still mobile and the adsorbed surfactant is quickly transported towards the cap. As a consequence, the transition from a very small to a very high contamination happens in a small belt above the ``stagnant cap" zone. Here the highest surface tension gradient and hence Marangoni forces are observed. 
	\item The transition from the second to the third stage happens on a larger time scale than between the first two stages. The convective surfactant transport in the bubble front slowly decreases. This happens, on the one hand, because the bubble decelerates (for small Marangoni numbers), and on the other hand due to the decreasing overall mobility of the interface. The narrow transition zone with high concentration gradients widens and the surfactant distribution in the front becomes approximately linear. Consequently, the resulting Marangoni forces have a smaller magnitude but act almost uniformly on the entire upper hemisphere. The integral tangential viscous force due to the Marangoni stresses is, therefore, higher than in stage two. 
\end{enumerate}
To see a further transition to a fourth stage, a much longer physical time would have to be simulated since also the adsorption steadily decreases. Such an investigation shall be part of future studies.
\FloatBarrier
%

%
%
\section{Conclusion and Outlook}

The focus of the current work is on the dynamics of single bubbles rising in a contaminated solution with surfactant. Within this study, it has been possible to investigate realistic length and time scales thanks to a subgrid-scale model, and the available experimental results for the rising bubble case could be reproduced well.
The necessity of a subgrid-scale model has been proven via specific test cases involving thin species boundary layers. Note that the same methodology that allowed us to simulate realistic surfactant systems can be applied to mass transfer problems to eventually study the effect of surfactant on mass transfer.

We firstly investigated the influence of the initial surface coverage on the rise velocity. In fact, in the experiments there is a certain detachment time including the bubble formation till the release. In this time adsorption mechanisms are already occurring, such that the bubble is pre-contaminated at release. The results show that the initial transient stage is very sensitive to the initial surface concentration. With a parameter study varying the initial surface contamination, we could find the initial surface coverage corresponding to the experiments, a value that was not known a priori. For very high bulk concentrations, we demonstrated that a lower initial surface contamination than the one suggested by the theory (equation~\eqref{eq:estCs0}) was already sufficient to obtain the correct bubble transient velocity. This information is fundamental in view of application cases because from the initial stage depends, for instance, the position of the bubble in a channel or column.

The focus then moved on to study the influence of the initial bulk concentration on the rise velocity and bubble dynamics. From the simulation results, global and local quantities can be evaluated. The bubble path depends both on the initial surface and bulk contaminations. For the least contaminated case, a transition from helical to zig-zag path is observed, as in the experimental work by Tagawa~et~al.~\cite{tagawaTakagi2014}. It has also been found that the quasi-steady state velocity can be reached without an equilibrium of  ad- and desorption. Moreover, the transfer of surfactant in the sub-layer in a steady state regime for the bubble rise velocity is close to the mass transfer at a solid particle. The local vorticity fields have been used to characterize the flow type in the vicinity of the bubble to understand the formation of vortices in the bubble wake.

The forces acting on the bubble surface have been studied considering their contribution to lift and drag forces. The dynamic pressure force, being the major contributor to the lift force, is responsible for the deviation from a rectilinear path. The steady state terminal velocity is a consequence of the overall drag force. In fact, for higher surfactant bulk concentrations, the viscous drag force increases due to higher surface tension gradients. At the same time, the dynamic pressure force decreases due to the reduced mobility of the interface. These two counter-acting effects lead to an approximately constant overall drag force. In other Reynolds regimes, for example for very small bubbles as the one considered in~\cite{takemura2005} that rise along a straight path even if contaminated, these mechanisms could perhaps be different.

From the local distribution of the Marangoni forces, it has been shown that the detachment of the flow from the bubble surface occurs where the Marangoni stresses are the highest. The quasi-steady state situation corresponds to a more uniform distribution of the Marangoni forces on the upper hemisphere of the bubble surface. 
These findings are relevant for deriving simplified models such as an improved stagnant cap model. In fact, one should refer to the quasi-steady state not in terms of ``fully contaminated" surface, but regarding a certain Marangoni stress distribution. The latter depends on the surfactant distribution on the interface and, above a certain threshold, not on the amount of surfactant on $\Sigma$. This implies that at steady state the surface concentration is not necessarily equal to the equilibrium concentration.

Considering the local adsorption, advection and diffusion processes at the interface, three different stages during the bubble rise have been identified. A first stage where the adsorption rates are the highest, a second stage where the transport at the front of the bubble is advection-dominated while in the rear part it is diffusion-dominated, and a third stage with a uniform distribution of the Marangoni stresses in the upper hemisphere of the bubble. A further transition to a fourth stage is foreseeable, but a much longer physical time would have to be simulated since also the adsorption steadily decreases. Such an investigation shall be part of future studies.

\section{Acknowledgements}
We kindly acknowledge the financial support by the German Research Foundation (DFG) within the Priority Program SPP1740 “Reactive Bubbly Flows", Project BO1879/13-2, and the Collaborative Research Center 1194 “Interaction between Transport and Wetting Processes”, Project B02. We also thank Dr.~habil.~Reinhard
Miller and his group at the Max Plank Institute of Colloids and Interfaces for providing the experimental data and the anonymous reviewers for their insightful comments. Calculations for this research were conducted on the Lichtenberg high performance computer of the TU Darmstadt.\\

\noindent A revised form of this contribution subsequent to peer review and editorial inputs by Cambridge University Press has been accepted for publication in the Journal of Fluid Mechanics and can be found at \emph{https://doi.org/10.1017/jfm.2018.723}.

%
\newpage
%
\appendix

\section{Appendix}
\label{sec:app}

\subsection{Algorithm Overview}
\label{subsec:num_3_to_app}
\begin{figure}[ht]
\centering
\includegraphics[width=0.95\textwidth]{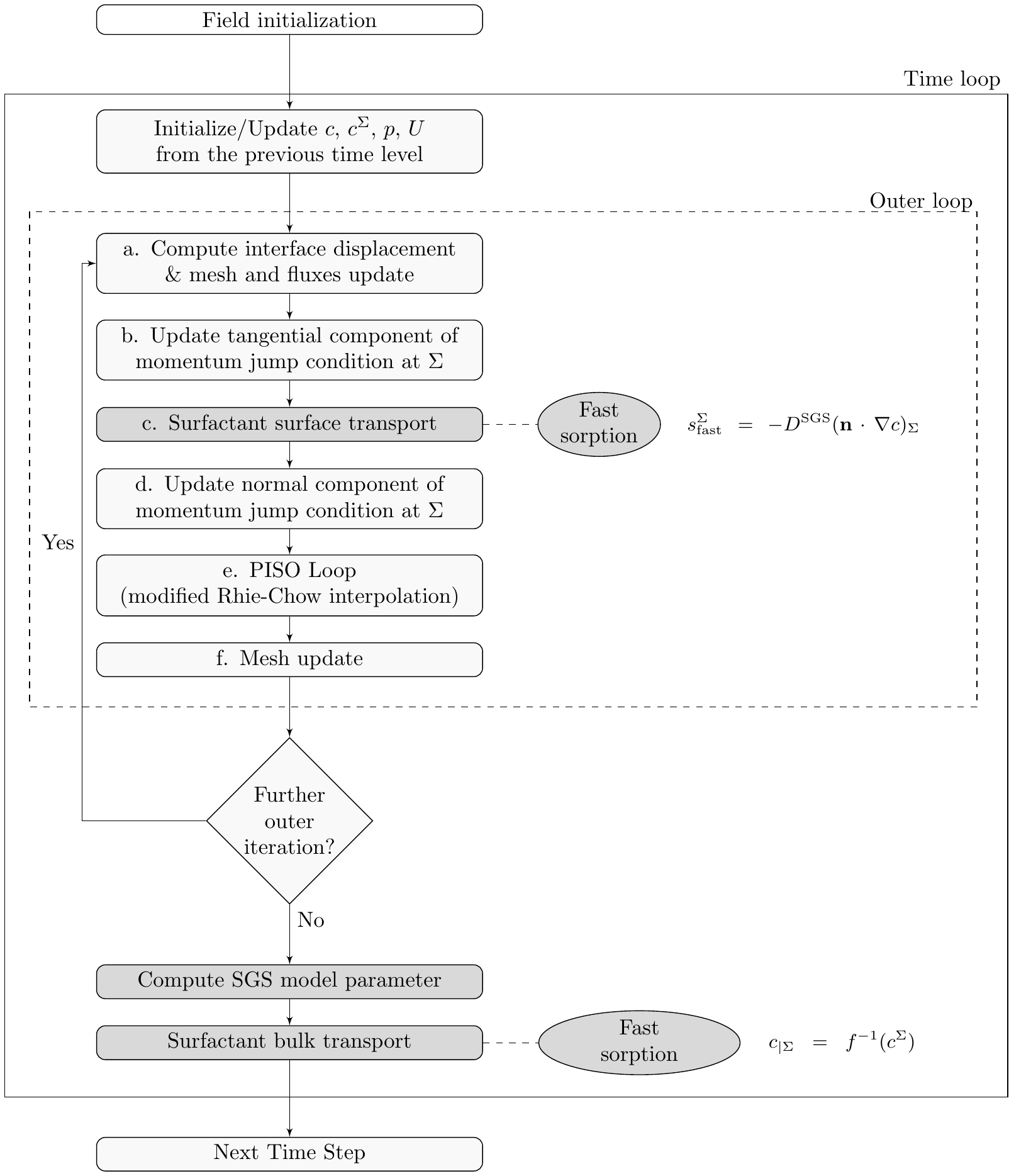}
\caption{Overview of the algorithm to solve the full problem: hydrodynamics with mesh motion, surfactant transport and sorption.}
\label{fig:n5}
\end{figure}

In figure~\ref{fig:n5}, a schematic overview of the numerical solution procedure is depicted.

\FloatBarrier
%

%
\subsection{Implicit SGS model for advection-dominated problems}
\label{subsec:app_0}

\begin{figure}[ht]
\centering
\begin{tikzpicture}
    \coordinate (y) at (0,-5);
    \coordinate (x) at (5,0);
    \draw[<->] (y) node[below] {$y$} -- (0,0) --  (x) node[right] {$x$};

   \coordinate (A) at (0,-1);
   \coordinate (B) at (1.5,-4);
   \coordinate (C) at (4,-4);
   \draw[black,thick] (A) to[out=-80,in=180] (B) -- (C);
   \node[fill=black,circle, minimum size = 0.15cm, inner sep=0pt] at (A) {};
   \draw[] (A) node[left] {$c|_{\scriptscriptstyle \Sigma}$};
   \draw[] (C) node[above] {$c(x,y)$};

   \coordinate (C) at (1.5,-4.5);
   \draw[scale=0.5,domain=0:9,smooth,variable=\y,black,rotate=-90,samples=300,densely dotted,thick] plot(\y,{sqrt(\y)});
   \draw[](C) node[below]{$\delta\left(y\right)$};

   \coordinate (s1) at (1.5,-0.5);
   \coordinate (s2) at (2.5,-0.5);
   \coordinate (s3) at (3.5,-0.5);
   \coordinate (e1) at (1.5,-1.5);
   \coordinate (e2) at (2.5,-1.5);
   \coordinate (e3) at (3.5,-1.5);
   \draw[->,thick] (s1)--(e1);
   \draw[->,thick] (s2)--(e2);
   \draw[->,thick] (s3)--(e3);
   \draw[] (e3) node[below right] {$\mathbf{u}=\left(0,v\right)$};

    myball/.style={shade, ball color=black, circular drop shadow={
    shadow xshift=0pt, shadow yshift=0pt}};
   \coordinate (bc) at (-3,-2.5);
   \shadedraw[shading=ball,ball color=black!20, white] (bc) circle (1);

    \draw[->] (-3,0) to [out=-90,in=150](-2.6,-1.5);
    \draw[->] (-2.6,-1.5) to[out=-30,in=90] (-1.95,-2.5);
    \draw[->] (-1.95,-2.5) to[out=-90,in=30] (-2.6,-3.5);
    \draw[->] (-2.6,-3.5) to[out=210,in=90] (-3,-5.0);

   \coordinate (rad) at (-2,-2.5);
   \node[draw=black, minimum size=0.5cm, circle,thick] at (rad) {};
   \coordinate (up) at (-2,-2.25);
   \coordinate (down) at (-2,-2.75);
   \draw[dashed] (up) -- (0,0);
   \draw[dashed] (down) -- (0,-4.75);

   \coordinate (sig) at (-0.5,-3);
   \draw[inner sep=0.1em] (sig) node[above left] {$\Sigma$};
   \draw[] (sig) -- (0,-3.5);
   \node[] at (-1,-0.5) {$\Omega^-$};
   \node[] at (1,-0.5) {$\Omega^+$};

\end{tikzpicture}
\caption{Simplified 2D model for species transport close to the bubble surface, figure based on~\cite{weiner2017}.}
\label{fig:3}
\end{figure}
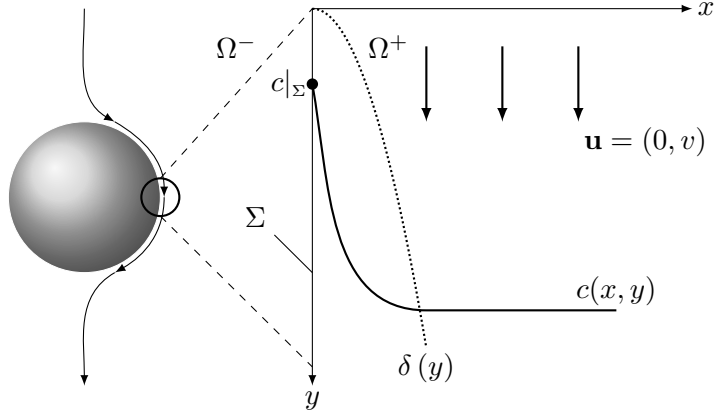

The SGS model for advection-dominated transport is based on a simplified 2D problem formulation of the species convection-diffusion equation~\eqref{eq:m18}. 
Consider the species transport in the vicinity of a bubble surface. Close to the interface $\Sigma$, a situation as sketched in figure~\ref{fig:3} is encountered.
\FloatBarrier
For high P\'{e}clet numbers, constant species concentration in the gas phase (the diffusivity in the gas phase is much higher than the one in the liquid phase) and a fully developed and quasi-stationary boundary layer, equation~\eqref{eq:m18} can be reduced to
\begin{equation}
    v \frac{\partial c}{\partial y} = D \frac{\partial^2 c}{\partial x^2}\quad \mathrm{for}\ x \geq 0\ \mathrm{and}\ y \geq 0 
    \label{eq:n41oo}  
\end{equation}
with the boundary conditions
\begin{equation}
    c(x,y=0) = c_{\infty},\quad c(x \to \infty,y>0) = c_{\infty},\quad c(x=0,y>0) = c|_{\scriptscriptstyle \Sigma}.
    \label{eq:n41ooo}  
\end{equation}
This problem has an analytical solution, describing the species distribution normal to the interface for a given boundary layer thickness $\delta(y)$,
\begin{equation}
    c(x,y) = c|_{\scriptscriptstyle \Sigma} + (c_{\infty} - c|_{\scriptscriptstyle \Sigma})\ \mathrm{erf} \left( \frac{x}{\delta(y)} \right)
    \label{eq:n41oooo}  
\end{equation}
with $\delta(y) = \sqrt{4Dy/v}$. The physical profile derived from the local substitute problem is adopted to compute the fluxes over the faces in the interface cells. The free model parameter $\delta$ computed iteratively to be consistent with the cell centred concentration value. 
The computation of the SGS model parameter is reported in Appendix~\ref{subsec:app_1}.

Consider now the discretized species (surfactant) transport equation in the liquid phase~\eqref{eq:n35}, and reported here in a condensed form,
\begin{equation}
    \frac{ 3 c_{P}^{n} V_{P}^{n} - 4 c_{P}^{o} V_{P}^{o} + c_{P}^{oo} V_{P}^{oo}}{\Delta t} + \sum_f F^{\mathrm{A}}_f = \sum_f F^{\mathrm{D}}_f,
    \label{eq:n41}  
\end{equation}
where $F^{\mathrm{A}}_f = \phi_f c_f^n$ and $F^{\mathrm{D}}_f = D_f \left( \grad c\right)_f^n \cdot \mathbf{S}_f$ are the advective and diffusive species fluxes, respectively. Recall from section~\ref{sec:mathMod} that this equation is completed by the initial condition
\begin{equation}
    c(t=0, \vec{x}) = c_0,\quad \vec{x} \in \Omega^{+}(t=0)
    \label{eq:n42}
\end{equation}
and the Dirichlet boundary condition imposed at the bubble surface $\Sigma(t)$ in case of fast sorption (as outlined in section~\ref{sec:math_2} and equation~\eqref{eq:n37}), i.e.
\begin{equation}
    c(t, \mathbf{x}) = f^{-1} \left(c^{\Sigma}(t)\right),\quad \vec{x} \in \Sigma(t).
    \label{eq:n43}
\end{equation}
When applying the SGS model, the goal is to correctly represent the species distribution around the interface, even if the concentration boundary layer is completely contained in the first cell layer (i.e.~when the DNS cannot resolve the boundary layer). To achieve this, a correction of the \emph{diffusive} and \emph{advective} species fluxes is introduced on the first cell faces normal to $\Sigma$ to counteract the otherwise overestimated numerical fluxes.
\begin{figure}[ht]
\centering
\begin{tikzpicture}[scale=0.7]
\def\R{10.0}
\def\Rr{12.0}
\def\RR{14.0} 
\def\arcI{-100}
\def\arcE{-35}
\def\angA{-45}
\def\angB{-60}
\def\angC{-75}
\def\angD{-90}

\coordinate (A) at ({\R * cos(\arcI)}, {\R * sin(\arcI)});
\coordinate (B) at ({\R * cos(\arcE)}, {\R * sin(\arcE)});
\coordinate (P) at ({\R * cos(\angA)}, {\R * sin(\angA)});
\coordinate (Q) at ({\R * cos(\angB)}, {\R * sin(\angB)});
\coordinate (R) at ({\R * cos(\angC)}, {\R * sin(\angC)});
\coordinate (N) at ({\R * cos(\angD)}, {\R * sin(\angD)});
\coordinate (Pp) at ({\Rr * cos(\angA)}, {\Rr * sin(\angA)});
\coordinate (Qq) at ({\Rr * cos(\angB)}, {\Rr * sin(\angB)});
\coordinate (Rr) at ({\Rr * cos(\angC)}, {\Rr * sin(\angC)});
\coordinate (Nn) at ({\Rr * cos(\angD)}, {\Rr * sin(\angD)});
\coordinate (Nnp) at ({\Rr * cos(\angD-5)}, {\Rr * sin(\angD-5)});
\coordinate (Ppp) at ({\Rr * cos(\angA+5)}, {\Rr * sin(\angA+5)});
\coordinate (PP) at ({(\RR+1) * cos(\angA)}, {(\RR+1) * sin(\angA)});
\coordinate (QQ) at ({(\RR+1) * cos(\angB)}, {(\RR+1) * sin(\angB)});
\coordinate (RR) at ({(\RR+1) * cos(\angC)}, {(\RR+1) * sin(\angC)});
\coordinate (NN) at ({(\RR+1) * cos(\angD)}, {(\RR+1) * sin(\angD)});
\coordinate (NNp) at ({\RR * cos(\angD-5)}, {\RR * sin(\angD-5)});
\coordinate (PPp) at ({\RR * cos(\angA+5)}, {\RR * sin(\angA+5)});
\coordinate (Pfill) at ({\R * cos(\angA+5)}, {\R * sin(\angA+5)});
\coordinate (Nfill) at ({\R * cos(\angD-5)}, {\R * sin(\angD-5)});

\fill[white!60!gray, opacity=1.0] (Pfill) -- ({(\R+0.6) * cos(\angA+6)}, {(\R+0.6) * sin(\angA+6)}) -- ({(\R+1.2) * cos(\angA+4)}, {(\R+1.2) * sin(\angA+4)}) -- (Ppp) arc ((\angA+5):(\angD-5):\Rr)  -- ({(\R+1.2) * cos(\angD-6)}, {(\R+1.2) * sin(\angD-6)}) -- ({(\R+0.6) * cos(\angD-4)}, {(\R+0.6) * sin(\angD-4)}) --  (Nfill) arc ((\angD-5):(\angA+5):\R);

\draw [semithick] (A) arc (\arcI:\arcE:\R);
\draw[dashed] (P) -- (PP);
\draw[dashed] (Q) -- (QQ);
\draw[dashed] (R) -- (RR);
\draw[dashed] (N) -- (NN);
\draw[dashed] (Nnp) arc ((\angD-5):(\angA+5):\Rr);
\draw[dashed] (NNp) arc ((\angD-5):(\angA+5):\RR);
\draw[ultra thick] (R) arc ((\angC):(\angB):\R);
\draw[ultra thick] (Rr) arc ((\angC):(\angB):\Rr);

\draw ({\R * cos(\arcI) + 1.4},{\R * sin(\arcI) + 0.2}) node[above right]{$\tilde{\Omega}^-(t)$};
\draw ({\R * cos(\arcI)},{\R * sin(\arcI) - 0.4}) node[below right]{$\tilde{\Omega}^+(t)$};
\coordinate (LI) at ({\R * cos(\arcE-2)}, {\R * sin(\arcE-2)});
\coordinate (LE) at ({\R * cos(\arcE-2) + 0.5}, {\R * sin(\arcE-2) - 0.5});
\draw (LI) to[out=-2] (LE);
\draw (LE) node[below right]{$\tilde{\Sigma}(t)$};

\coordinate (fisig) at ({\R * cos(\angC+2.5)}, {\R * sin(\angC+2.5)});
\coordinate (fip) at ({\Rr * cos(\angC+2.5)}, {\Rr * sin(\angC+2.5)});
\draw (fisig) node[below]{$f_i^{\Sigma}$};
\draw (fip) node[below right]{$f_i^{\Sigma,o}$};

\coordinate[mark coordinate] (cisig) at ({\R * cos(\angB-7.5)}, {\R * sin(\angB-7.5)});
\coordinate (cisigL) at ({(\R-0.6) * cos(\angB-7.5)}, {(\R-0.6) * sin(\angB-7.5)});
\coordinate[mark coordinate] (cip) at ({(\R+1) * cos(\angB-7.5)}, {(\R+1) * sin(\angB-7.5)});
\draw (cisig) -- (cisigL) node[above]{$c_{f_i^{\Sigma}} = f^{-1} \left(c_i^{\Sigma}\right)$};
\draw (cip) node[below right]{$c_i$};

\coordinate (IO) at ({5 * cos(\angB-7.5))}, {5 * sin(\angB-7.5))});
\def\IOx{5 * cos(\angB-7.5))}
\def\IOy{5 * sin(\angB-7.5))}
\coordinate (IOo) at ({7.5 * cos(\angB-7.5))}, {7.5 * sin(\angB-7.5))});
\def\IOox{7.5 * cos(\angB-7.5))}
\def\IOoy{7.5 * sin(\angB-7.5))}
\fill[white!60!gray, opacity=1.0] (IO) circle[radius=2.5] -- (IO) circle[radius=2.6];
\fill[white] (IO) circle[radius=2.5];
\draw[semithick] (IO) circle[radius=2.5];

\coordinate (aa) at ({7.0 * cos(\angB-11.5))}, {7.0 * sin(\angB-11.5))});
\coordinate (bb) at ({7.0 * cos(\angB-3.6))}, {7.0 * sin(\angB-3.6))});
\coordinate (cc) at ({8 * cos(\angB-11))}, {8 * sin(\angB-11))});
\coordinate (dd) at ({8 * cos(\angB-4))}, {8 * sin(\angB-4))});

\draw[densely dotted, thick] (aa) -- (bb) --(dd) -- (cc) -- (aa);
\draw[densely dotted, thick] ({\IOox - 0.49},{\IOoy - 0.15}) -- (A);
\draw[densely dotted, thick] ({\IOox + 0.45},{\IOoy + 0.24}) -- (B);

\draw (-0.3,-4.5) node[right]{$\tilde{\Omega}^-(t)$};
\draw (-0.8,-5.2) node[left]{$\tilde{\Omega}^+(t)$};
\draw (3.82, -3.0) to[out=-2] (4.15, -2.9);
\draw (4.1, -2.7) node[below right]{$\tilde{\Sigma}(t)$};

\end{tikzpicture} 
\caption{2D sketch for the SGS model with enlarged view of the region near the interface. $\tilde{\Omega}^{\pm}(t),\ \tilde{\Sigma}(t)$ are the discretized counterpart of $\Omega^{\pm}(t),\ \Sigma(t)$.}
\label{fig:4}
\end{figure}
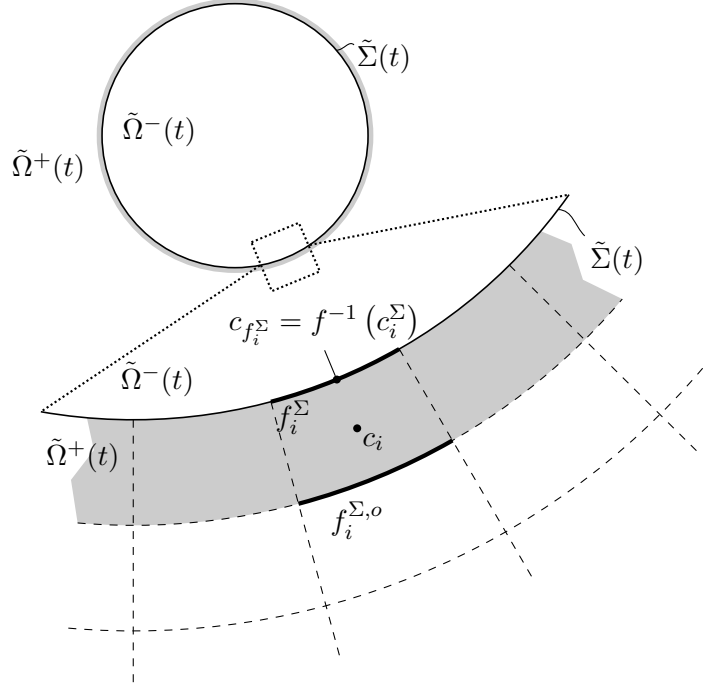
\FloatBarrier
%

\paragraph{\textit{Diffusion}}
\label{par:num_2_3_1}
The diffusive species fluxes $F^{\mathrm{D}}_f$ at the faces belonging to $\Sigma$, $f_i^{\Sigma}$, and at the first cell faces opposite to $\Sigma$, $f_i^{\Sigma,o}$, are considered; see figure~\ref{fig:4} for the notation.

We compute the desired numerical diffusive fluxes at the relevant faces $f_i^{*} = f_i^{\Sigma},\ f_i^{\Sigma,o}$ as
\begin{equation}
    \label{eq:n44}
    F^{\mathrm{D},\mathrm{num}}_{f_i^{*}} = - D_{f_i^{*}} S_{f_i^{*}} \left( \partial_n c \right)^{\mathrm{num}}_{f_i^{*}},
\end{equation} 
where $D_{f_i^{*}}$ is a corrected diffusion coefficient to counteract the numerical effects of the under-resolved species boundary layer. 
To derive an expression for $D_{f_i^{*}}$ we use the diffusive fluxes coming from the SGS modelling
\begin{equation}
    \label{eq:n45}
    F^{\mathrm{D},\mathrm{SGS}}_{f_i^{*}} = - D S_{f_i^{*}} \left( \partial_n c \right)^{\mathrm{SGS}}_{f_i^{*}},
\end{equation}
where $D$ is the molecular diffusivity and $\left( \partial_n c \right)^{\mathrm{SGS}}_{f_i^{*}}$ is provided by the SGS model; see Appendix~\ref{subsec:app_1} for the analytical expression. 
Our goal is to compute $D_{f_i^{*}}$ such that the numerical diffusive fluxes, coming from the standard discretization, equal the SGS-fluxes,
\begin{equation}
    \label{eq:n46}
    F^{\mathrm{D},\mathrm{num}}_{f_i^{*}} \overset{!}{=} F^{\mathrm{D},\mathrm{SGS}}_{f_i^{*}}.
\end{equation} 
Thus, we impose
\begin{equation}
    \label{eq:n47}
    D_{f_i^{*}} \left( \partial_n c \right)^{\mathrm{num}}_{f_i^{*}} \overset{!}{=} D \left( \partial_n c \right)^{\mathrm{SGS}}_{f_i^{*}},
\end{equation} 
to get an expression for the modified diffusion coefficients to be substituted in the discretized transport equation,
\begin{equation}
    \label{eq:n48}
    D_{f_i^{*}} = D \frac{\left( \partial_n c \right)^{\mathrm{SGS}}_{f_i^{*}}}{\left( \partial_n c \right)^{\mathrm{num}}_{f_i^{*}}}.
\end{equation} 
To simplify the notation, below we will address $D_{f_i^{*}}$ as $D^{\mathrm{SGS}}$, where $D^{\mathrm{SGS}}$ contains the modified local values from the SGS model in the required faces. For the other faces the standard molecular diffusivity is kept.
In case the estimated boundary layer thickness is more than 1000 times larger than the first cell width, the SGS correction is not applied to avoid non-physical diffusive fluxes; see Appendix~\ref{subsubsec:app_1_1}.\\


\paragraph{\textit{Advection}}
\label{par:num_2_3_2}
The SGS correction of the advective species fluxes $F^{\mathrm{A}}_f$ is necessary only at the first cell faces opposite to $\Sigma$, $f_i^{\Sigma,o}$, because the velocity normal to the interface in a moving reference frame is zero. Our aim would be to correct directly the concentrations with the prescribed value from the SGS model $c^{\mathrm{SGS}}_{f_i^{\Sigma,o}}$. However, this cannot be done within an implicit framework, thus we correct the convective fluxes to match the prescribed SGS concentration. The numerical fluxes are computed as
\begin{equation}
    \label{eq:n49}
    F^{\mathrm{A},\mathrm{num}}_{f_i^{\Sigma,o}} = c^{\mathrm{num}}_{f_i^{\Sigma,o}} \phi_{f_i^{\Sigma,o}},
\end{equation} 
where $c^{\mathrm{num}}_{f_i^{\Sigma,o}}$ is the concentration value interpolated to the face center and $\phi_{f_i^{\Sigma,o}}$ is a modified advective flux.

The species fluxes computed with the SGS face value are
\begin{equation}
        \label{eq:n50}
        F^{\mathrm{A},\mathrm{SGS}}_{f_i^{\Sigma,o}} = c^{\mathrm{SGS}}_{f_i^{\Sigma,o}} \phi^{\mathrm{num}}_{f_i^{\Sigma,o}}
\end{equation} 
where $c^{\mathrm{SGS}}_{f_i^{\Sigma,o}}$ is provided by the SGS model. Enforcing the SGS fluxes to be equal to the numerical ones
\begin{equation}
        \label{eq:n51}
        F^{\mathrm{A},\mathrm{num}}_{f_i^{\Sigma,o}} \overset{!}{=} F^{\mathrm{A},\mathrm{SGS}}_{f_i^{\Sigma,o}},
    \end{equation}
we get the equality
\begin{equation}
        \label{eq:n52}
        c^{\mathrm{num}}_{f_i^{\Sigma,o}} \phi_{f_i^{\Sigma,o}} \overset{!}{=} c^{\mathrm{SGS}}_{f_i^{\Sigma,o}} \phi^{\mathrm{num}}_{f_i^{\Sigma,o}}
    \end{equation} 
from which we compute the corrected convective fluxes    
    \begin{equation}
        \label{eq:n53}
        \phi_{f_i^{\Sigma,o}} = \phi^{\mathrm{num}}_{f_i^{\Sigma,o}} \frac{c^{\mathrm{SGS}}_{f_i^{\Sigma,o}}}{c^{\mathrm{num}}_{f_i^{\Sigma,o}}}.
    \end{equation}     
Also for the advective term, to simplify the notation, we will address $\phi_{f_i^{\Sigma,o}}$ as $\phi^{\mathrm{SGS}}$, where $\phi^{\mathrm{SGS}}$ contains the modified local values from the SGS model in the required faces. For the other faces the original numerical fluxes are kept. 
Note that $\phi^{\mathrm{SGS}} = \phi^{\mathrm{num}}_{f_i^{\Sigma,o}}\ c^{\mathrm{SGS}}_{f_i^{\Sigma,o}} / c^{\mathrm{num}}_{f_i^{\Sigma,o}}$. Thus, if  $c^{\mathrm{num}}_{f_i^{\Sigma,o}}$ and $c_{f_i^{\Sigma,o}}$ are interpolated with the same scheme, the modification of the advective term at the interested faces translates into enforcing the $c^{\mathrm{SGS}}_{f_i^{\Sigma,o}}$; in fact $\phi^{\mathrm{num}}_{f_i^{\Sigma,o}} \left( c^{\mathrm{SGS}}_{f_i^{\Sigma,o}} / c^{\mathrm{num}}_{f_i^{\Sigma,o}} \right) c_{f_i^{\Sigma,o}} = \phi^{\mathrm{num}}_{f_i^{\Sigma,o}} c^{\mathrm{SGS}}_{f_i^{\Sigma,o}}$. This also assures that our method remains conservative.

The advection correction via the SGS model is applied only if the concentration profile in the first three cell layers close to the interface is monotonic, see Appendix~\ref{subsec:app_2} for more details on exception handling. This condition is fundamental to avoid non-physical (unbounded) concentrations; see~\cite{versteeg1995_2007}.

%
\subsection{Algorithm for the SGS model parameter calculation}
\label{subsec:app_1}

In this section the main steps to compute the SGS model parameter $\delta$ are explained. We adopt an iterative approach, as described in~\cite{weiner2017}, to find the model parameter $\delta$ that fulfils
\begin{equation}
    \bar{\eta}_{\mathrm{C}} = \frac{\bar{c} - c_{|\Sigma}}{c_{\infty} - c_{|\Sigma}} \overset{!}{=} \frac{1}{V} \int_{V} \eta(x/\delta) \mathrm{d}V = \eta_{\mathrm{SGS}},
    \label{eq:eta_bar}
\end{equation}
where $\bar{\eta}_{\mathrm{C}}$ is the volume averaged cell-centred value coming from the finite volume discretization, which has to be equal to the volume average computed with the SGS model. Above, $\eta$ is given as
\begin{equation}
    \eta(x,y) = \frac{c(x,y) - c_{|\Sigma}}{c_{\infty} - c_{|\Sigma}} = \mathrm{erf}(x/\delta(y))
    \label{eq:eta}
\end{equation}
according to equation~\eqref{eq:n41oooo}. The quantity $\bar{c}$ is the average concentration in an interface cell ($c_i$ in figure~\ref{fig:4}), $c_{|\Sigma}$ is the bulk concentration at the interface ($c_{f_i^{\Sigma}}$ in figure~\ref{fig:4}). The iterative solution based on equation~\eqref{eq:eta_bar} requires the evaluation of the volume integral. Here only the main steps from~\cite{weiner2017} are reported. The iterative algorithm is based on the work of Ahn and Shashkov~\cite{ahn2008} and uses a combined Newton-Bisection method to search for $\delta$ which converges very quickly, usually after three iterations. The maximum number of iterations is set to 10.
As initial guess for $\delta_0$ the first two terms of a series expansion for the inverse error function are taken, that is \mbox{$\delta_0 = (l/2)/(0.5 \pi(\eta_c + \pi / 12 \eta_c^3))$}, with $l$ being the first cell thickness. Bounding values for $\delta$ are taken equal to $\delta_{\mathrm{min}} = 1 \cdot 10^{-15}$ and $\delta_{\mathrm{max}} = 10 \delta_0$. The convergence tolerance is set to $\mathrm{tol} = 1 \cdot 10^{-9}$.

In each time step, there is an initialization step for the required parameters. The result of the iterative procedure will be a vector containing all the $\delta$ values (for all the interface cells).
The algorithm is displayed as pseudo-code in Algorithm~\ref{algo:deltaAlgo}. Note that the formula to compute the residual has been corrected with respect to~\cite{weiner2017}. 
%
%
\subsubsection{Exception handling}
\label{subsubsec:app_1_1}
Before the iterative procedure is started a check that the values of $\eta_c$ are between 0 and 1 is done.
If the maximum number of iterations is reached without a converged value for $\delta$ or if the computed $\delta$ is larger than the first cell thickness by a factor of 1000, then $\delta$ is set to -1 and the SGS correction will not be applied at the corresponding face.
\algrenewcommand\algorithmicrequire{\textbf{Data:}}
\algrenewcommand\algorithmicreturn{\textbf{Return:}}
\begin{algorithm}
\begin{algorithmic}
  \Require
  \State{$\eta_c = \left(\bar{c} - c|_{\Sigma}\right)/\left(c_0 - c|_{\Sigma}\right)$}
  \State{$\delta_0 = l/\left[\sqrt{\pi} \left(\eta_c + (\pi/12) \eta_c^3\right)\right]$}
  \State{$\delta_n = \delta_0$}
  \State{$\delta_{\mathrm{min}} = 1 \cdot 10^{-15}$}
  \State{$\delta_{\mathrm{max}} =  10 \delta_0$}
  \State{$\mathrm{tol} = 1 \cdot 10^{-9}$}
  \State{$\eta^{\mathrm{SGS}} = \mathrm{erf}(l/\delta) + (\delta/l) \left[e^{-(l/\delta)^2}-1\right]/\sqrt{\pi}$}
  \State{$\mathrm{res_{min}} = \eta^{\mathrm{SGS}}(\delta_{\mathrm{min}}) - \eta_c$}
  \State{$\mathrm{res_{max}} = \eta^{\mathrm{SGS}}(\delta_{\mathrm{max}}) - \eta_c$}

    \Repeat
      \State{Compute $\eta^{\mathrm{SGS}}(\delta_n)$}
      \State{$\mathrm{res} = \eta^{\mathrm{SGS}}(\delta_n) - \eta_c$}
      \State{$\eta' = \left(e^{-(l/\delta)^2} -1 \right) / (l \sqrt{\pi} )$}
      \State{$\delta_{n+1} = \delta_n - \left( \bar{\eta}^{\mathrm{SGS}}(\delta_n) - \eta_c \right)/\eta'$ }
      \If {$(\delta_{n+1} < \delta_{\mathrm{min}})$ or $(\delta_{n+1} > \delta_{\mathrm{max}})$}
	\If {$\mathrm{res} \cdot \mathrm{res_{max}} > 0$}
	  \State{$\delta_{n+1} = (\delta_{\mathrm{min}} + \delta_n)/2$}
	  \State{$\delta_{\mathrm{max}} = \delta_n$}
	\Else
	  \State{$\delta_{n+1} = (\delta_{\mathrm{max}} + \delta_n)/2$}
	  \State{$\delta_{\mathrm{min}} = \delta_n$}
	\EndIf
      \EndIf
      \State{$\delta_n = \delta_{n+1}$}
    \Until{$\left( \left|\frac{\eta^{\mathrm{SGS}}(\delta_n) - \eta_c}{\eta_c}\right| \le \mathrm{tol} \right)$}\\
$\ $\\
\Return $\delta_n$    
\end{algorithmic}
\caption{Iterative computation of $\delta$ with a Newton-Bisection method.}
\label{algo:deltaAlgo}
\end{algorithm}

%
\subsection{Correction of diffusive and convective fluxes within the SGS modelling}
\label{subsec:app_2}

After the iterative computation of the model parameter $\delta$, the SGS correction is applied to the diffusive and convective fluxes as explained in section~\ref{subsubsec:num_2_3}. The various steps for the flux correction are reported in Algorithm~\ref{algo:fluxCorrAlgo}.

%
\subsubsection{Exception handling}
\label{subsubsec:app_2_1}
The diffusive and convective fluxes are corrected only if the iterative procedure to compute $\delta$ converged, that is $\delta > 0$. Furthermore, a check that the gradient and the concentration close to the interface are non-zero is included, $\left|(\partial_n c)_{f_i^*}^{\mathrm{num}}\right| > 10^{-15}$ and $\left|c^{\mathrm{num}}_{f_i^{\Sigma,o}}\right| > 10^{-15}$. If these checks fail, the standard discretization is used.

An additional exception handling is implemented specifically for the correction of the diffusive fluxes at the second layer of faces $f_i^{\Sigma,o}$. The SGS correction is applied only if the ratio between the SGS gradient and the numerical one is smaller than unity, $\left| (\partial_n c)_{f_i^*}^{\mathrm{SGS}} / (\partial_n c)_{f_i^*}^{\mathrm{num}} \right| < 1$. If the correction factor is larger than one, the SGS model application is not necessary and the diffusivity will not be corrected at the respective face.

The last exception regards the correction of the convective fluxes. The SGS model correction is applied only if the concentration profile  within the first three cells close to the interface is monotonic. If we number the cell centres from the interface outwards as $c_1,\ c_2,\ c_3$, then the SGS correction is applied only if $(c_1 - c_2)(c_2 - c_3) > 0$.
\algrenewcommand{\algorithmiccomment}[1]{\hskip3em$\rightarrow$ #1}
\begin{algorithm}
\begin{algorithmic}
    \ForAll{ (faces $f_i^{\Sigma}$) }
      \If {$(\delta > 0)$}
      $\ $\\
	  \State at $f_i^{\Sigma}$: $(\partial_n c)_{f_i^{\Sigma}}^{\mathrm{SGS}} = \frac{2}{\sqrt{\pi}} \frac{c_0 - c_{f_i^{\Sigma}}}{\delta_i}$
      $\ $\\
	  \State at $f_i^{\Sigma,o}$: $(\partial_n c)_{f_i^{\Sigma,o}}^{\mathrm{SGS}} = \frac{2}{\sqrt{\pi}} \frac{c_0 - c_{f_i^{\Sigma}}}{\delta} e^{-(l_i/\delta_i)^2}$
      $\ $\\
	  \State $\quad \quad \quad \quad c^{\mathrm{SGS}}_{f_i^{\Sigma,o}} = c_{f_i^{\Sigma,o}} + (c_0 - c_{f_i^{\Sigma,o}}) \mathrm{erf}(l_i/\delta_i)$
      $\ $\\
	  \State Diffusion correction: $D_{f_i^*} = D^{\mathrm{mol}} \frac{(\partial_n c)_{f_i^*}^{\mathrm{SGS}}}{(\partial_n c)_{f_i^*}^{\mathrm{num}}}$
      $\ $\\
	  \State Convection correction: $\phi_{f_i^{\Sigma,o}} = \phi^{\mathrm{num}}_{f_i^{\Sigma,o}} \frac{c^{\mathrm{SGS}}_{f_i^{\Sigma,o}}}{c^{\mathrm{num}}_{f_i^{\Sigma,o}}}$
      $\ $\\
      \EndIf
    \EndFor
\end{algorithmic}
\caption{Correction of diffusive and convective fluxes within the SGS model.}
\label{algo:fluxCorrAlgo}
\end{algorithm}
\FloatBarrier

%
\subsection{Validation of the SGS model for species transfer}
\label{subsec:app_3_a}
To validate the solution of the species transfer problem with SGS modelling, three test cases with increasing complexity are presented. The local Sherwood number $\operatorname{Sh_{loc}}$ is used for comparison with the reference solution.
\begin{figure}[ht]
\centering
\begin{tikzpicture}[scale=1.0]
\def\Xo{0.0}
\def\X{15.0}
\def\Yo{0.0} 
\def\Y{3.0}

\coordinate (A) at ({\Xo}, {\Yo});
\coordinate[mark coordinate] (B) at ({\Xo}, {\Y});
\coordinate (C) at ({\X}, {\Y});
\coordinate (D) at ({\X}, {\Yo});
\coordinate (E) at ({\Xo+1}, {\Y});
\coordinate (F) at ({\Xo}, {\Y-1});
\coordinate (G) at ({\X/2}, {\Y/3});
\coordinate (H) at ({\X-2}, {\Y+0.5});
\coordinate (I) at ({\Xo}, {\Y-1.5});
\coordinate (Ii) at ({\Xo+0.3}, {\Y-2});
\coordinate (Iii) at ({\Xo+0.3}, {\Y-2.5});
\coordinate (J) at ({\X}, {\Y-1.5});
\coordinate (Ji) at ({\X-0.3}, {\Y-2});
\coordinate (Jii) at ({\X-0.3}, {\Y-2.5});
\coordinate (K) at ({\X/2}, {\Y-0.1});
\coordinate (L) at ({\Xo}, {\Yo-0.8});
\coordinate (M) at ({\X}, {\Yo-0.8});
\coordinate (N) at ({\X/2}, {\Yo-0.8});

\draw (A) -- (B);
\draw[very thick] (B) -- (C);
\draw (C) -- (D);
\draw (A) -- (D);

\draw[->,very thick] (B) -- (E) node[above] {$y$};
\draw[->,very thick] (B) -- (F) node[left] {$x$};
\draw (G) node{$\Omega_L$};
\draw[dashed] ({\X-2.5}, {\Y}) -- (H) node[above right]{$\Sigma$};
\draw (I) node[right]{Inlet $(x>0, y=0)$:};
\draw (Ii) node[right]{$c = 0$};
\draw (Iii) node[right]{$\vec{u} = (0,v)$};
\draw (J) node[left]{Outlet $(x>0, y=L_y)$:};
\draw (Ji) node[left]{$\partial_y c = 0$};
\draw (Jii) node[left]{$\partial_y \vec{u} = 0$};
\draw (K) node[below]{At $\Sigma$ $(x=0, y)$: $c = 1$, $\partial_y \vec{u} = 0$};

\draw[dashed] (A) -- (L);
\draw[dashed] (D) -- (M);
\draw[<->,dashed] (L) -- (M);
\draw (N) node[above]{$L_y$};

\end{tikzpicture}
\caption{SGS 2D model problem set-up.}
\label{fig:v4}
\end{figure}
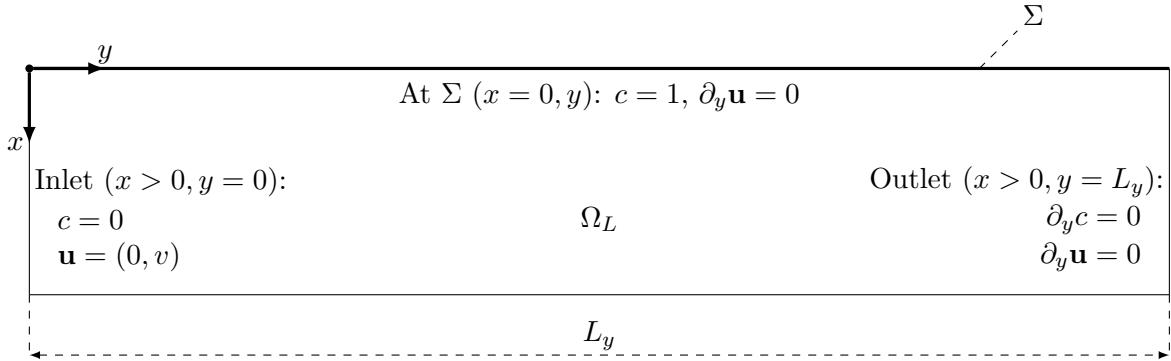
\FloatBarrier

\subsubsection{2D model problem}
\label{subsubsec:app_val_2_1}
This test case refers directly to the simplified problem formulation on which the SGS model is based. The implementation of the SGS model has been validated against the analytical solution taken from~\cite{weiner2017} and reported in Appendix~\ref{subsec:app_0}. The problem set-up under investigation is sketched in figure~\ref{fig:v4}. All the simplifying assumptions of the model problem are fulfilled if the computational domain size is large enough. The distance between the interface and the boundaries in $x$-direction is approximately 50 times the maximum species boundary layer thickness, to ensure that the presence of a finite domain is negligible. The presence of the gas phase is modelled via the boundary condition for the species concentration at $\Sigma$. The boundary and initial conditions can be found in figure~\ref{fig:v4}. Four different mesh resolutions are considered from 5 to 40 $\mu$m. As we are interested in advection-dominated problems, a high P\'{e}clet number of $\operatorname{Pe} = 10^5$ is chosen. The local Sherwood number is computed as 
\begin{equation}
\operatorname{Sh_{loc}}(y_i) = \left( \partial_n c \right)_{f_i^{\Sigma}}\ \frac{L_{y}}{c_{i|\Sigma} - c_{\infty}}
\label{eq:v15}
\end{equation}
with the normal derivative at the interface $\left( \partial_n c \right)_{f_i^{\Sigma}}$ \footnote{Without the SGS model the gradient is computed as $\left( \partial_n c \right)_{f_i^{\Sigma}} = \left(c_{i|\Sigma} - c_{f_i^{\Sigma}}\right)/d_i$, where $d_i$ is the distance between the boundary face center and the boundary cell center, and $c_{f_i^{\Sigma}}$ is the concentration at the interface face; otherwise $\left( \partial_n c \right)^{\mathrm{SGS}}_{f_i^{\Sigma}}$ is used.}, the concentration in the boundary cell center $c_{i|\Sigma}$ and the species concentration far away from the interface $c_{\infty}$.

\begin{figure}[ht]
\centering
\includegraphics[width=1.0\textwidth]{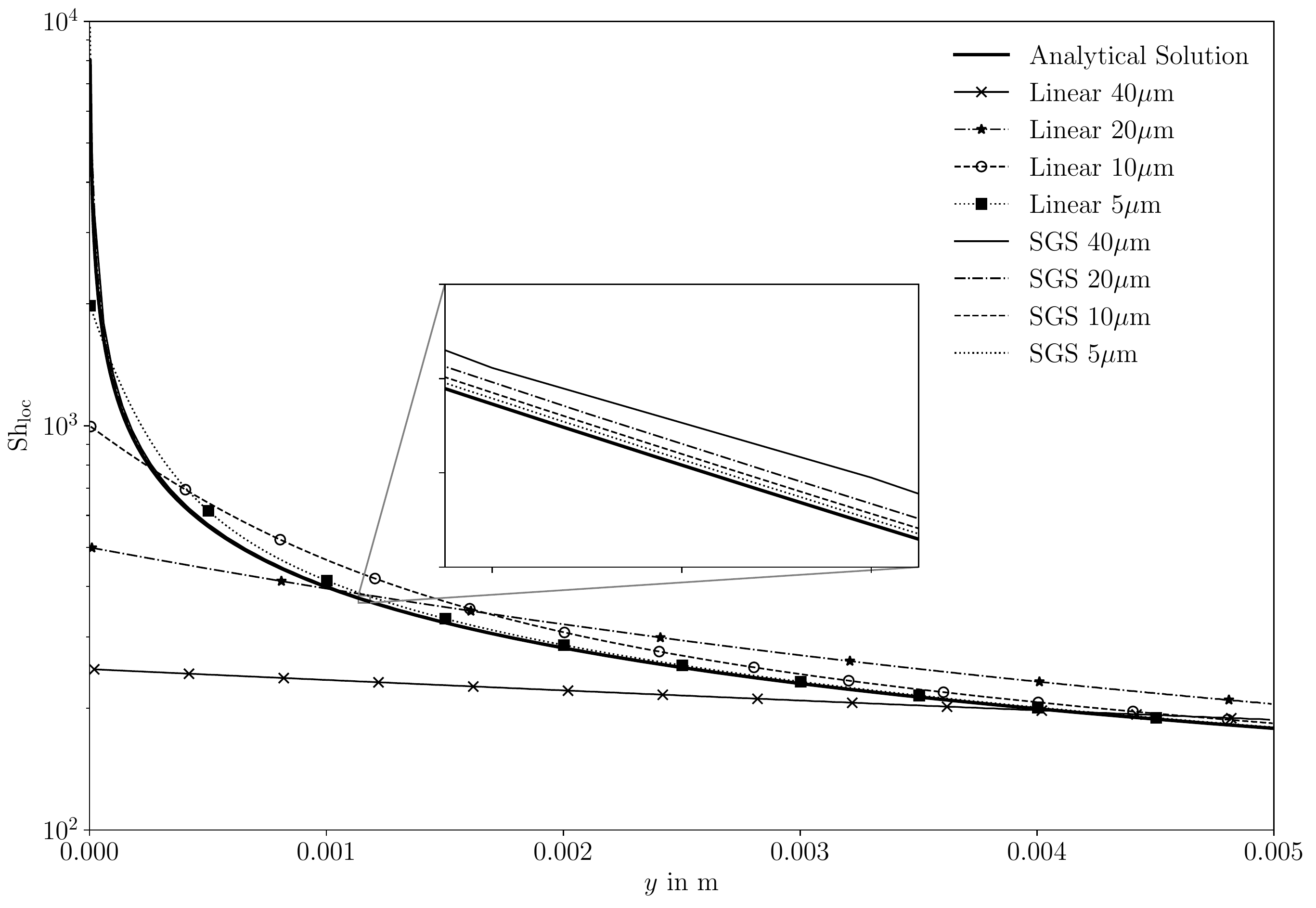}
\caption{Local Sherwood number for the 1D model problem.}
\label{fig:v5}
\end{figure}

Figure~\ref{fig:v5} depicts the comparison between the analytical solution and the numerical results obtained with and without the SGS model. When the problem is solved with linear interpolation, the relatively coarse meshes are not able to predict the solution precisely. The finest mesh (5 $\mu \mathrm{m}$) provides a good approximation of the local Sherwood number except for the region close to the inlet. All the cases where the SGS model is applied are in very good agreement with the reference solution. The enlarged view in figure~\ref{fig:v5} shows also mesh convergence for the SGS model results.
\FloatBarrier

\subsubsection{Spherical bubbles at small Reynolds number}
\label{subsubsec:app_val_2_2}

A spherical bubble at small Reynolds number is considered. For this case a semi-analytical solution of the species transport equation is possible. The velocity field is based on the solution of Satapathy and Smith~\cite{satapathy1960} (spherical particle of radius $r_b$ rising in a larger sphere $R$). On top of this velocity field, the species transport equation can be solved numerically using a very high grid resolution (cell thickness $l \approx 0.06\ \mathrm{\mu m}$ close to the interface). Four different molecular diffusivities are considered corresponding to Schmidt numbers of $\operatorname{Sc} = 10^4,\ 10^5,\ 10^6,\ 10^7$, where $\operatorname{Sc}$ is the ratio between viscous and molecular diffusion $\nu/D$. The bubble radius is $r_{\mathrm{b}} = 1$ mm and the Reynolds number is set to $\operatorname{Re} = 0.56$. The local Sherwood number $\operatorname{Sh_{loc}}(\theta_i)$ is computed as in equation~\eqref{eq:v15}, where $\theta_i$ is the polar angle, i.e.~the angle following a streamline on the bubble surface from the top $(\theta = 0)$ to the bottom $(\theta = \pi)$. The bubble equivalent diameter $d_{eq}$ is taken as reference length.\\

\begin{figure}[ht]
\centering
\includegraphics[width=0.6\textwidth]{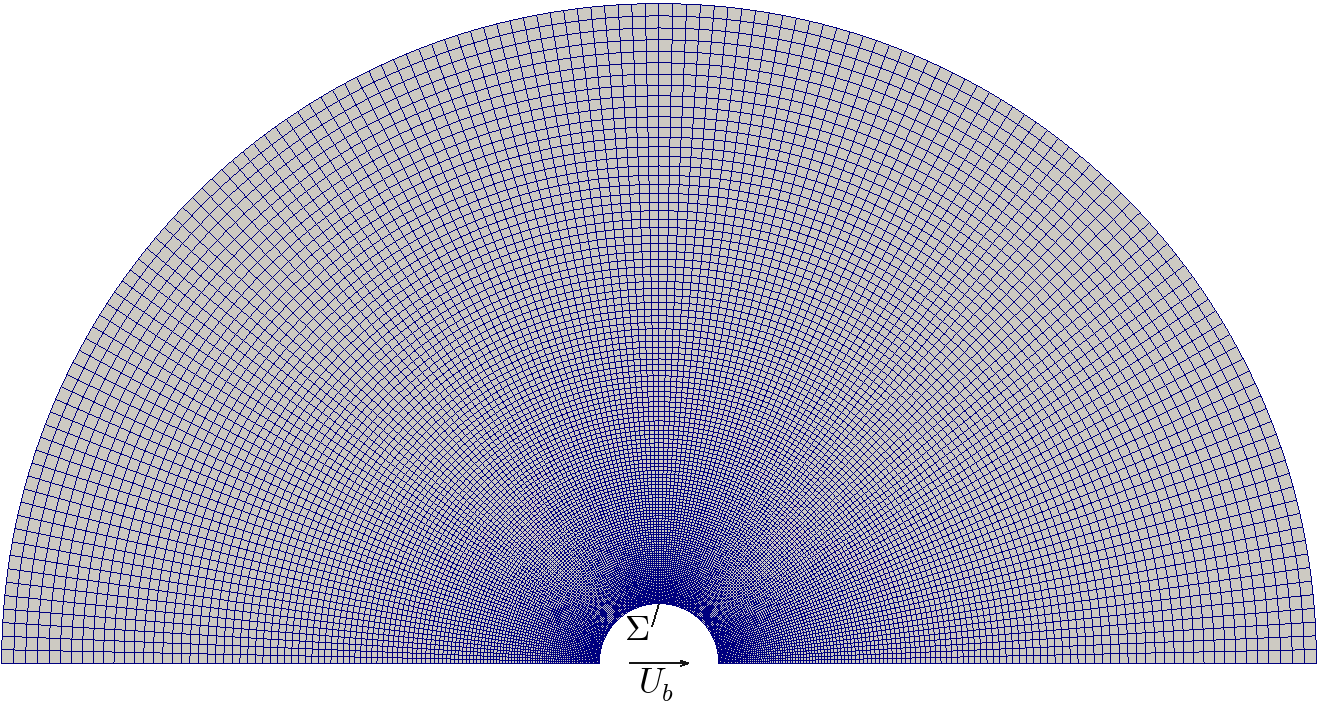}
\caption{Domain used to solve the species transport with the given analytical velocity field.}
\label{fig:v6}
\end{figure}

\paragraph{\emph{Axisymmetric species transfer with given velocity field}}
\label{par:app_val_2_2_1}
The species transport is solved on top of the velocity field provided by the solution of Satapathy and Smith for the different Schmidt numbers. The results obtained with the SGS-model are compared to the mesh independent direct numerical solution. The set-up for this simulations is depicted in figure~\ref{fig:v6}. 
\FloatBarrier
\begin{figure}[ht]
\centering
\includegraphics[angle=-90,width=1.0\textwidth]{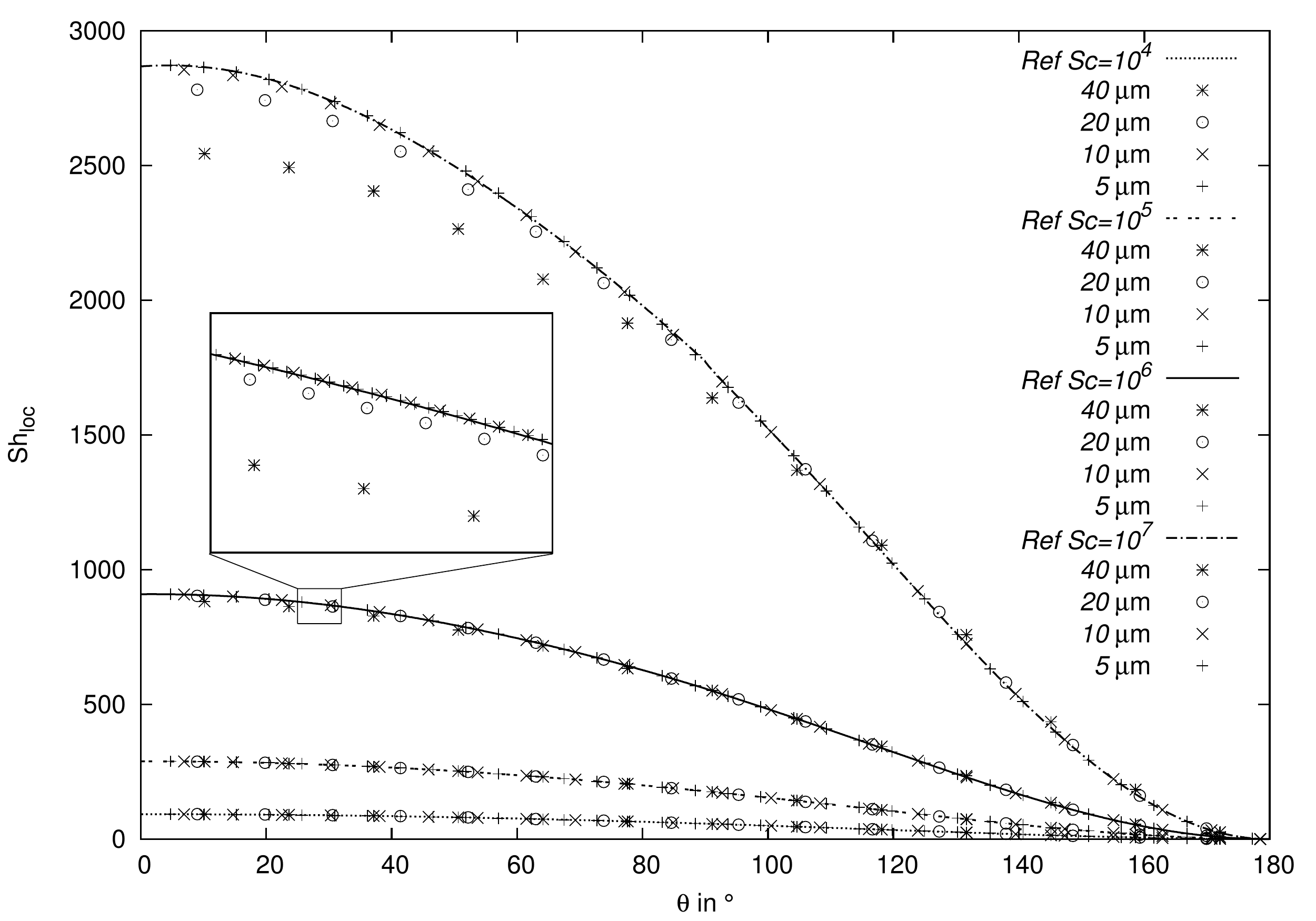}
\caption{Local Sherwood number for the species transfer problem with given Satapathy-Smith velocity profile.}
\label{fig:v7}
\end{figure}
\FloatBarrier
The fluid properties for the liquid side (identified with a $^+$) can be found in table~\ref{tab:tv1}. Four different mesh resolutions are considered with a cell thickness $l$ close to the interface ranging from 5 to 40 $\mu$m. The four different diffusion coefficients are $10^{-8},\ 10^{-9},\ 10^{-10}$ and $10^{-11}$ $\mathrm{m^2/s}$. The species concentration at the interface $\Sigma$ is set to $c|_{\Sigma} = 1\ \mathrm{mol/m^3}$, while the initial bulk concentration in $\Omega$ is set to $c_0 = c_{\infty} = 0$.

In figure~\ref{fig:v7} an overview of the results obtained applying the SGS model compared to the reference solutions is reported. Figure~\ref{fig:v7} shows a very good agreement between the numerical results using the SGS modelling and the respective references for each tested Schmidt number. 
This test case shows that the two coarsest meshes ($l = 40,\ 20\ \mu$m) are not fully capable to properly resolve the species transport for the highest Schmidt number, under-predicting the Sherwood number in the upper part of the bubble. Such behaviour has to be considered in the application case set-up with surfactant transport and sorption, mainly in the choice of the mesh resolution.

\begin{figure}[ht]
\centering
\subfloat[][\emph{Sc = 10$^4$}.]
{\includegraphics[angle=-90,width=.48\textwidth]{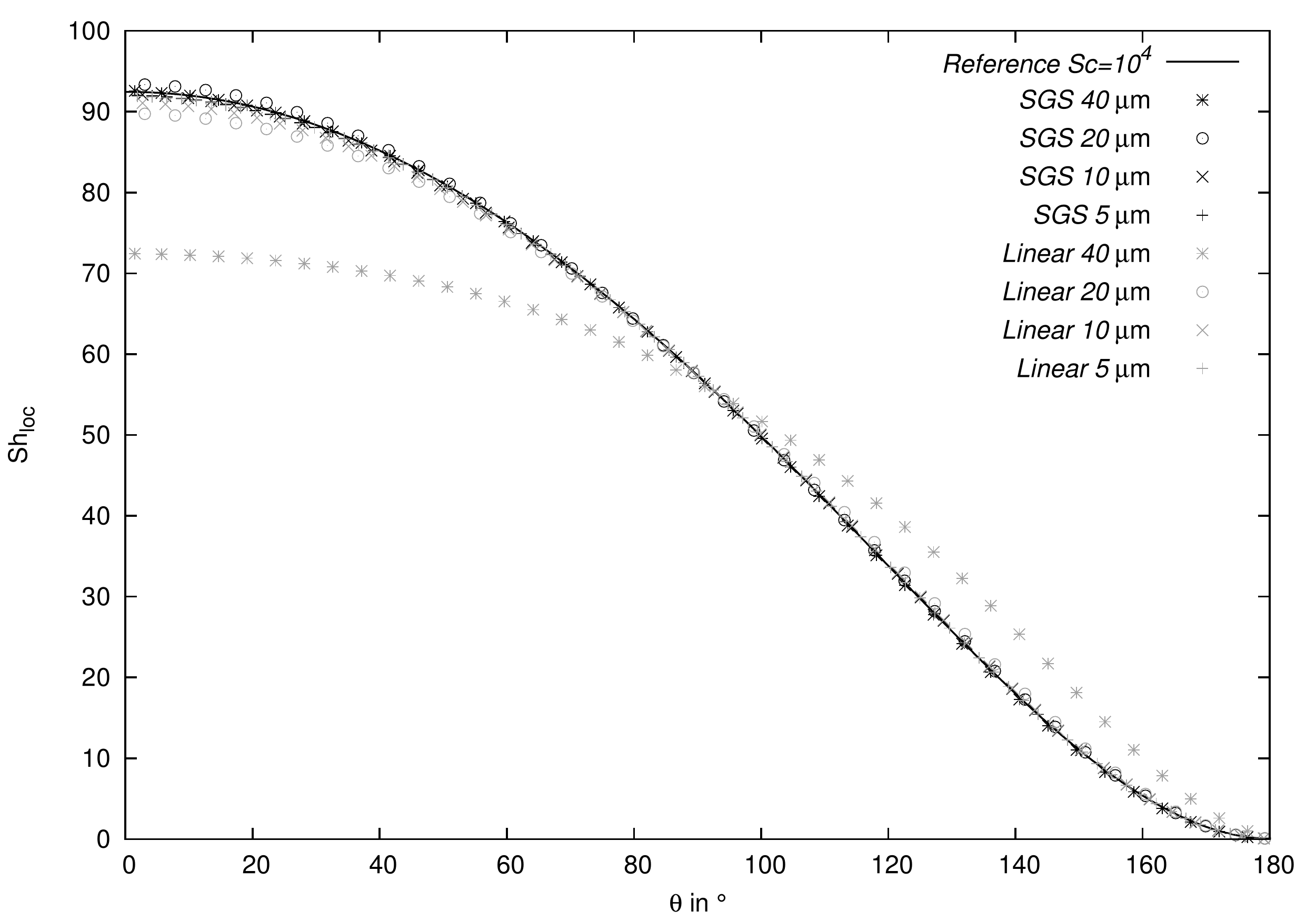}\label{fig:sc4aW}} \quad
\subfloat[][\emph{Sc = 10$^5$}.]
{\includegraphics[angle=-90,width=.48\textwidth]{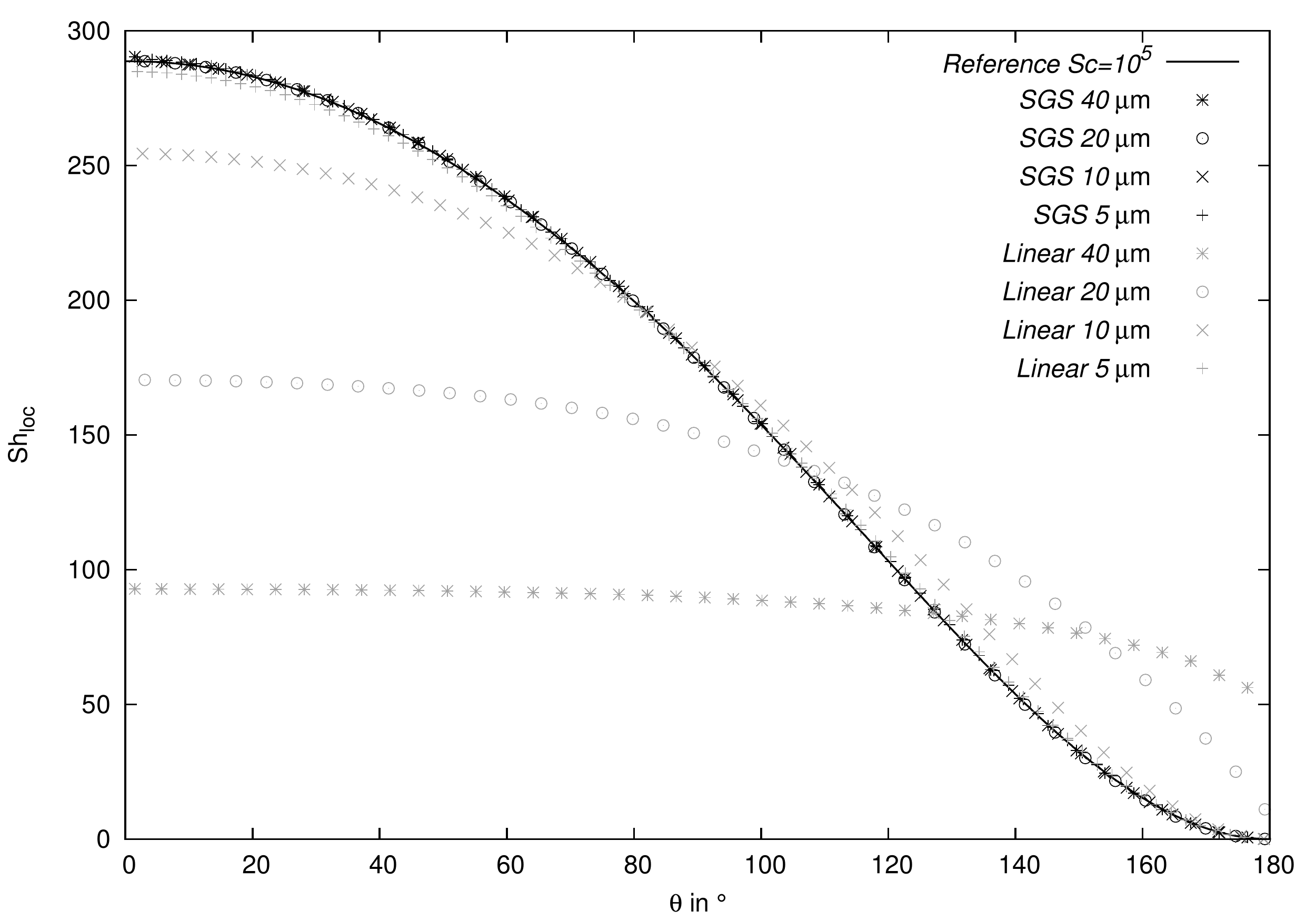}\label{fig:sc5aW}} \\
\subfloat[][\emph{Sc = 10$^6$}.]
{\includegraphics[angle=-90,width=.48\textwidth]{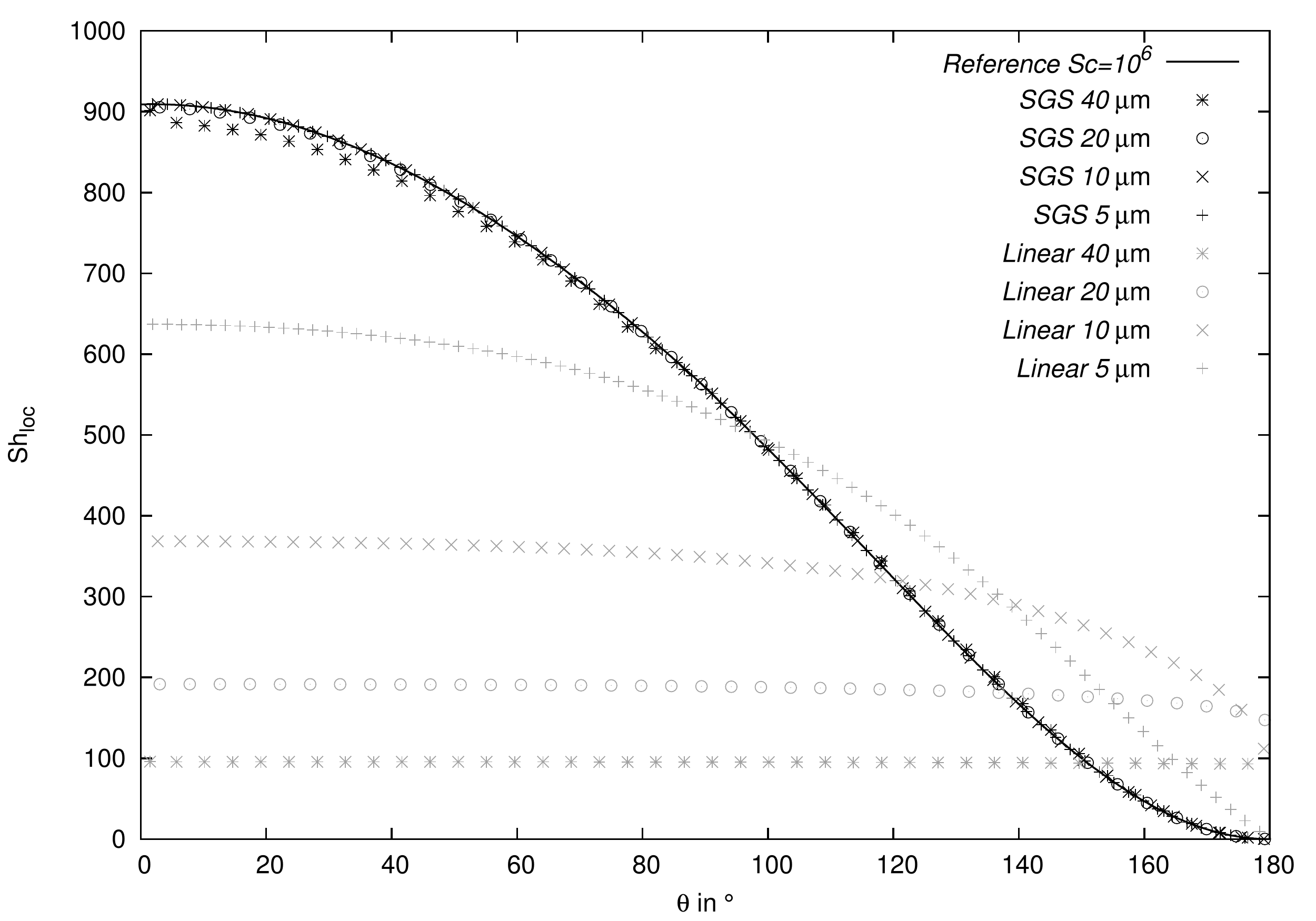}\label{fig:sc6aW}} \quad
\subfloat[][\emph{Sc = 10$^7$}.]
{\includegraphics[angle=-90,width=.48\textwidth]{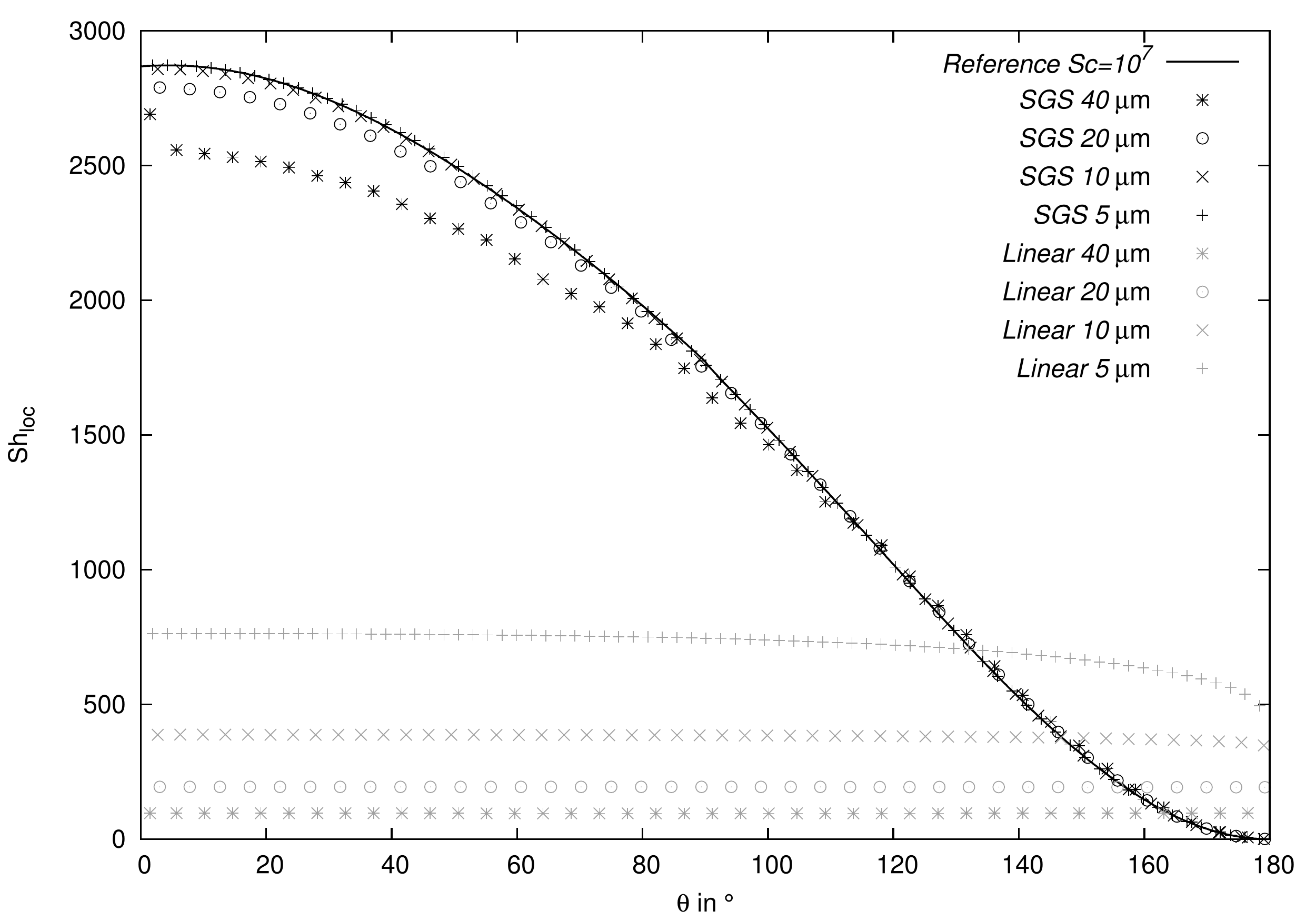}\label{fig:sc7aW}}
\caption{Local Sherwood number for the species transfer problem with given Satapathy-Smith velocity profile. Black symbols: with SGS modelling, grey symbols: linear interpolation.}
\label{fig:scAppWedge}
\end{figure}

For completeness, in figure~\ref{fig:scAppWedge} the comparison between the cases with and without SGS modelling is reported. The results obtained applying the SGS model are coloured in black, while the ones obtained with a linear interpolation method are grey. Already for $\operatorname{Sc} = 10^5$ the standard discretization is inadequate to correctly describe the species transfer close to the interface for the given mesh resolutions. This comparison confirms again that with the SGS model one can save several mesh refinement levels.
\FloatBarrier

\paragraph{\emph{Species transfer with computed velocity field}}
\label{par:val_2_2_2}
The species transport problem from a rising bubble is considered. The full 3D problem, hydrodynamics and species transfer, is solved within the Interface-Tracking framework, see the algorithm in figure~\ref{fig:n5}. 
The case set-up follows the one described in section~\ref{subsec:5_1}. The interface consists of polyhedral faces with an edge length of approximately 50 $\mu$m and a first cell layer thickness of $l = 12\ \mu$m and $l = 25\ \mu$m. The initial shape of the bubble is a sphere of radius $r_b = 1\ \mathrm{mm}$. The bubble is positioned in the center of a spherical domain of radius $10 r_b$. The fact that the interface is deformable is not relevant for the Satapathy-Smith case, because due to the choice of the fluid properties, the bubble does not deform significantly.

\begin{figure}[ht]
\centering
\subfloat[][\emph{Sc = 10$^4$}.]
{\includegraphics[angle=-90,width=.488\textwidth]{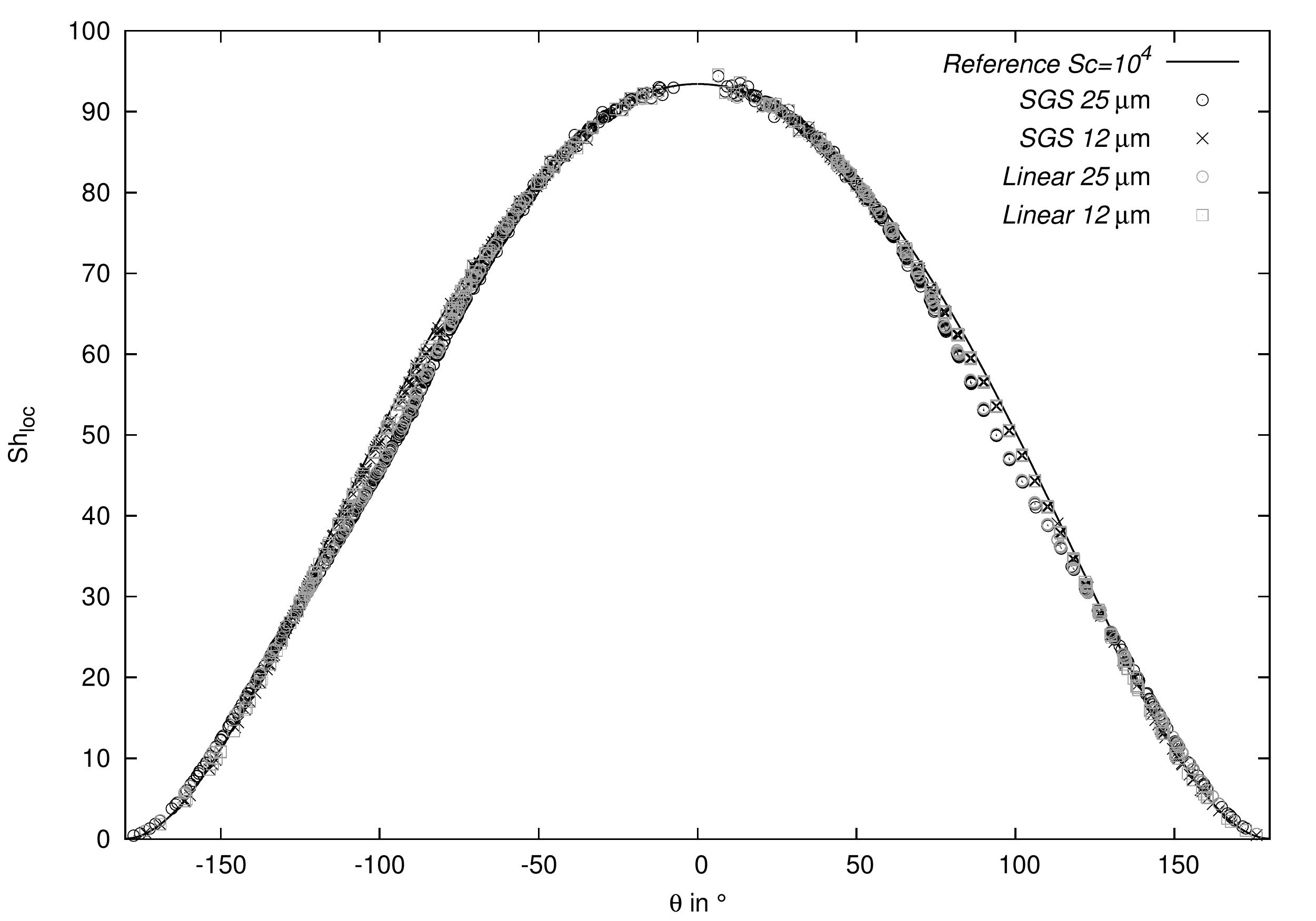}\label{fig:sc4a3D}} \quad
\subfloat[][\emph{Sc = 10$^7$}.]
{\includegraphics[angle=-90,width=.488\textwidth]{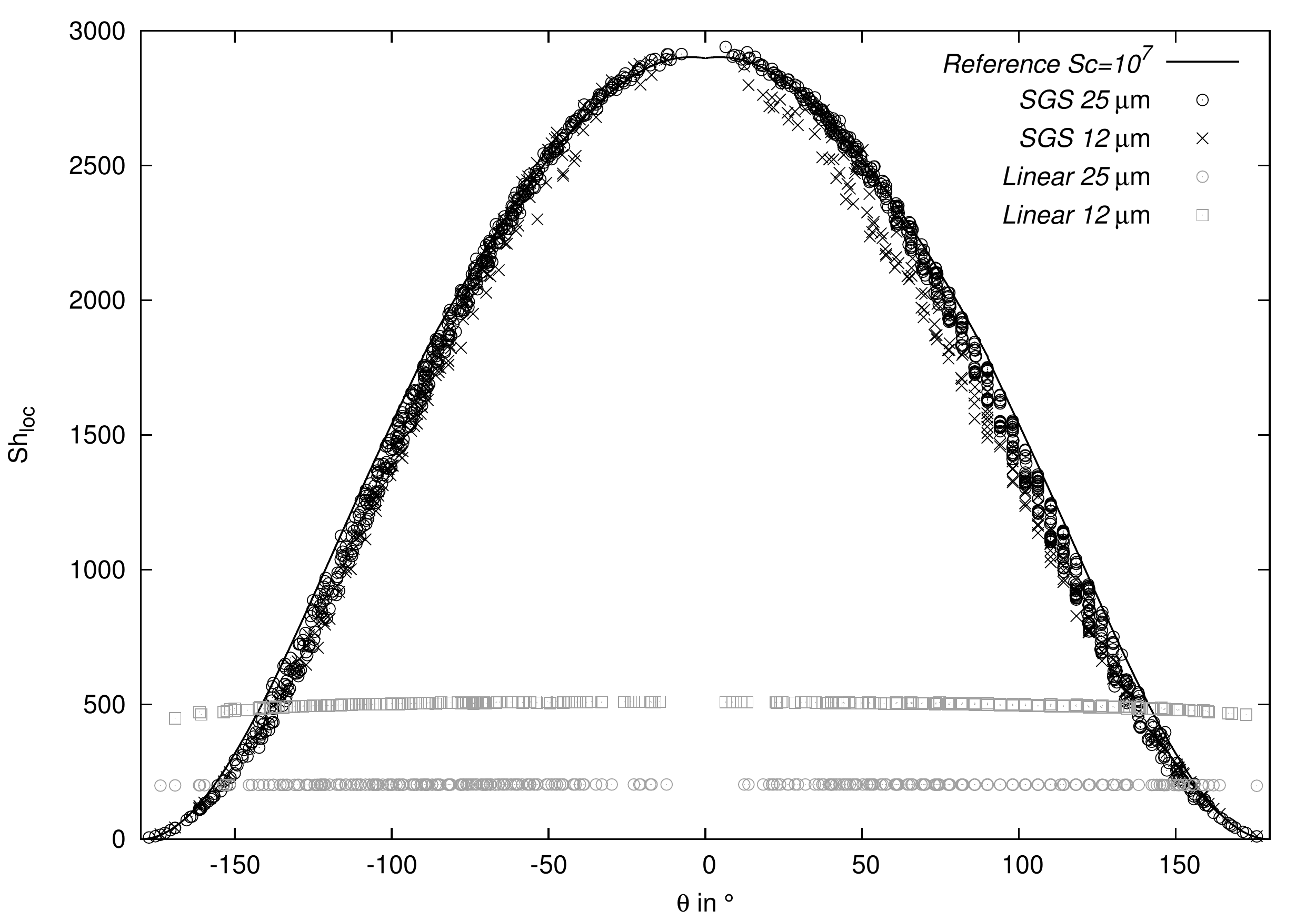}\label{fig:sc7a3D}}
\caption{Local Sherwood numbers for the species transfer problem with Satapathy-Smith set-up. Black symbols: with SGS modelling, grey symbols: linear interpolation.}
\label{fig:3Dsh}
\end{figure}

The initial and boundary conditions for the transferred species are the same as for the semi-analytical solution. The fluids properties are given in Table~\ref{tab:tv1}. For this test case the smallest and the highest Schmidt numbers are considered, i.e.~$\operatorname{Sc} = 10^4,\ 10^7$. As a reference, the semi-analytical solution presented in the former paragraph is used. The calculated velocity profile in the interface-tracking framework slightly differs from the Satapathy-Smith solution (less than 1.2\%, see~\cite{weberPhD2016}~(section 4.1.2 in the reference)), because the latter is based on a Stokes flow. This small difference can have some impact on the concentration profile close to the interface.
\begin{table}[ht]
\caption{Fluid properties for the Satapathy-Smith case.}
\label{tab:tv1}
\centering
\begin{tabular}{ccccccc}
\toprule
$\rho^+\ \mathrm{kg/m^3}$ & $\rho^-\ \mathrm{kg/m^3}$ & $\mu^+\ \mathrm{kg/(ms)}$ & $\mu^-\ \mathrm{kg/(ms)}$ & $\sigma_0\ \mathrm{N/m}$\\
\midrule
$1000$ & $1.1965$ & $0.1$ & $1.8 \cdot 10^{-5}$ & $0.0724$\\
\bottomrule
\end{tabular}
\end{table}
\FloatBarrier
In figure~\ref{fig:3Dsh} the results in terms of Sherwood number for the 3D case are reported. As can be seen from the two graphs there is a good agreement between the reference solution and the numerical one employing the SGS model. As anticipated, the reference solution is computed based on the Satapathy-Smith velocity profile, thus, since we are dealing with highly non-linear functions (species concentration close to $\Sigma$), small deviations in the velocity field could be enough to produce the observed discrepancies in the results. In figure~\ref{fig:3Dsh} also the results without the SGS model are plotted.
For small Schmidt numbers, figure~\ref{fig:sc4a3D}, the standard discretization provides results in good agreement with the reference solution, while the Sherwood numbers resulting from the SGS modelling show a sensitivity to the mesh resolution. On the other hand, for high Schmidt numbers and the given mesh resolution, figure~\ref{fig:sc7a3D}, the standard discretization provides underestimated Sherwood numbers, while the ones obtained with the SGS model are in good agreement with the reference.
\FloatBarrier
%
%
\subsubsection{2D deformable bubbles at higher Reynolds number}
\label{subsubsec:app_val_2_3}
As a final step to validate the SGS model, simulations of a 2D bubble rising in contaminated water are performed with and without SGS modelling for the surfactant transport. This setting aims to demonstrate that the SGS model predicts the surfactant transfer well under dynamic conditions, e.g. when the bubble deforms, accelerates or decelerates, or when the flow detaches and vortices form. As can be seen from figure~\ref{fig:2Dvalid}, the results where the SGS modelling was employed are matching the mesh independent results obtained with standard interpolation.

For these tests, the intermediate initial concentration is used, i.e. $c_0 = 0.008\ \mathrm{mol/m^3}$, with $c^{\Sigma}_{0} = 0\ \mathrm{mol/m^2}$. Different bulk diffusivities, $D = 5\cdot10^{-7},\ 5\cdot10^{-8},\ 5\cdot10^{-9}, 5\cdot10^{-10}\ \mathrm{m^2/s}$, and mesh resolutions, first cell thickness $l_i = 16,\ 12,\ 8,\ 3,\ 1.7,\ 1.2\ \mathrm{\mu m}$, are considered. The changes between $l_i = 1.7\ \mathrm{\mu m}$ and $l_i=1.2 \ \mathrm{\mu m}$ in rise velocity and surfactant transport are always less than 1.15$\%$. Therefore we consider the results on the finest mesh employing standard discretization as mesh independent and use them as reference solution (lines in the plots). The results for mesh resolutions with first cell thickness equal to 16, 8, 3 and 1.2 $\mathrm{\mu m}$ are selected for the plots below. The results for higher surfactant bulk diffusivities are not depicted in figure~\ref{fig:2Dvalid} because they look qualitatively similar. In fact, if the mesh resolution is sufficient, then all the results lay on the reference curve.

\begin{figure}[ht]
\centering
\includegraphics[width=0.8\textwidth]{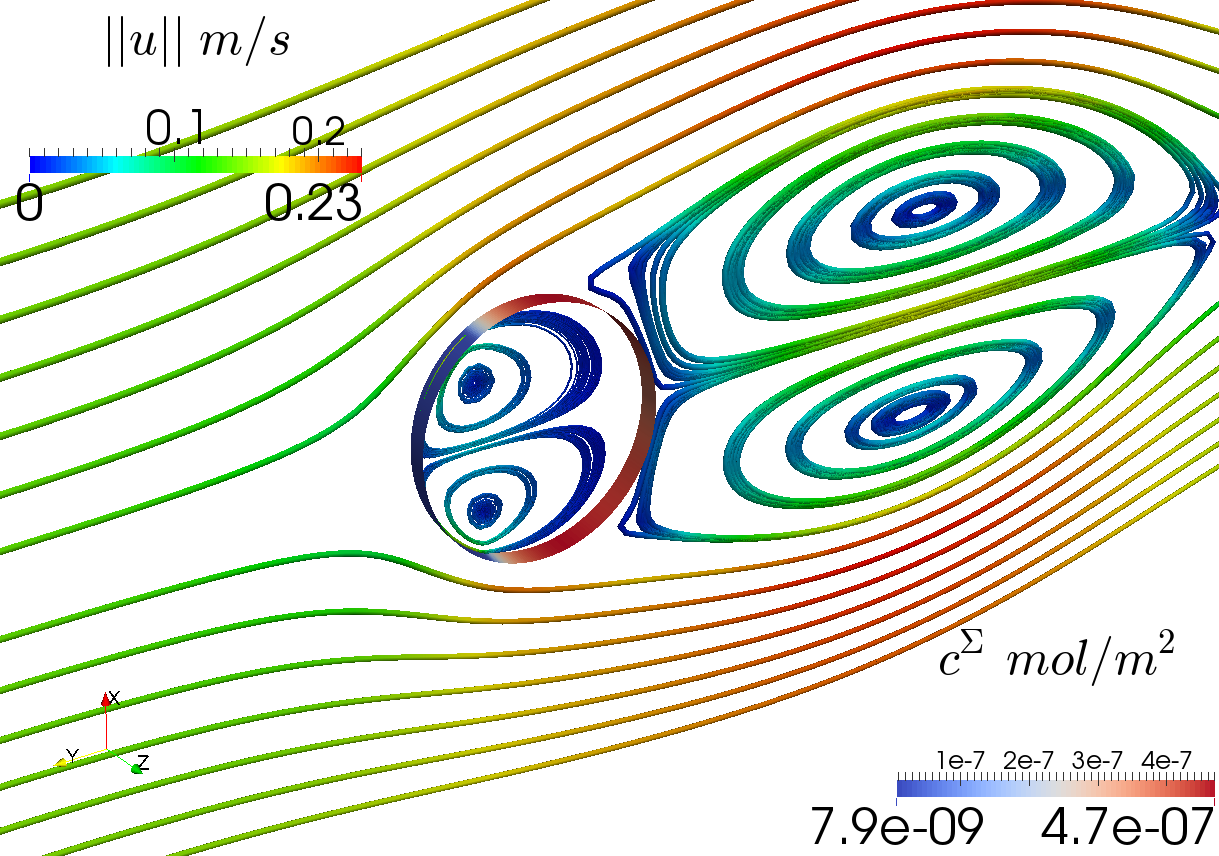}
\caption{Flow field around and inside the rising bubble. The bubble surface is coloured by the surfactant concentration; $t = 0.2$ s.}
\label{fig:2Dfields}
\end{figure}

\begin{figure}[ht]
\centering
\subfloat[][\emph{Rise velocity, $D = 5 \cdot 10^{-10}\ mol/m^2$}.]
{\includegraphics[width=.62\textwidth]{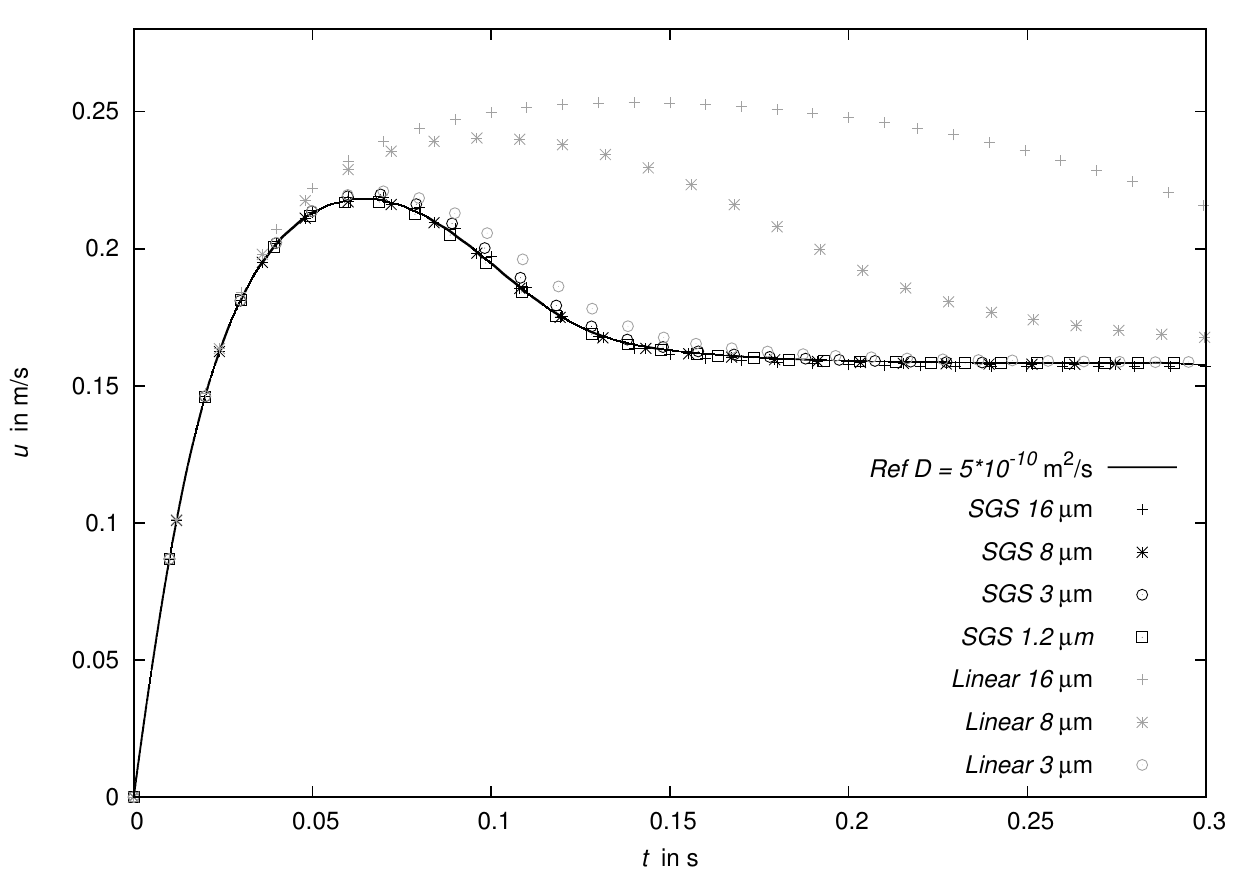}\label{fig:2Dvel}} \\
\subfloat[][\emph{Total moles on the interface}.]
{\includegraphics[width=.62\textwidth]{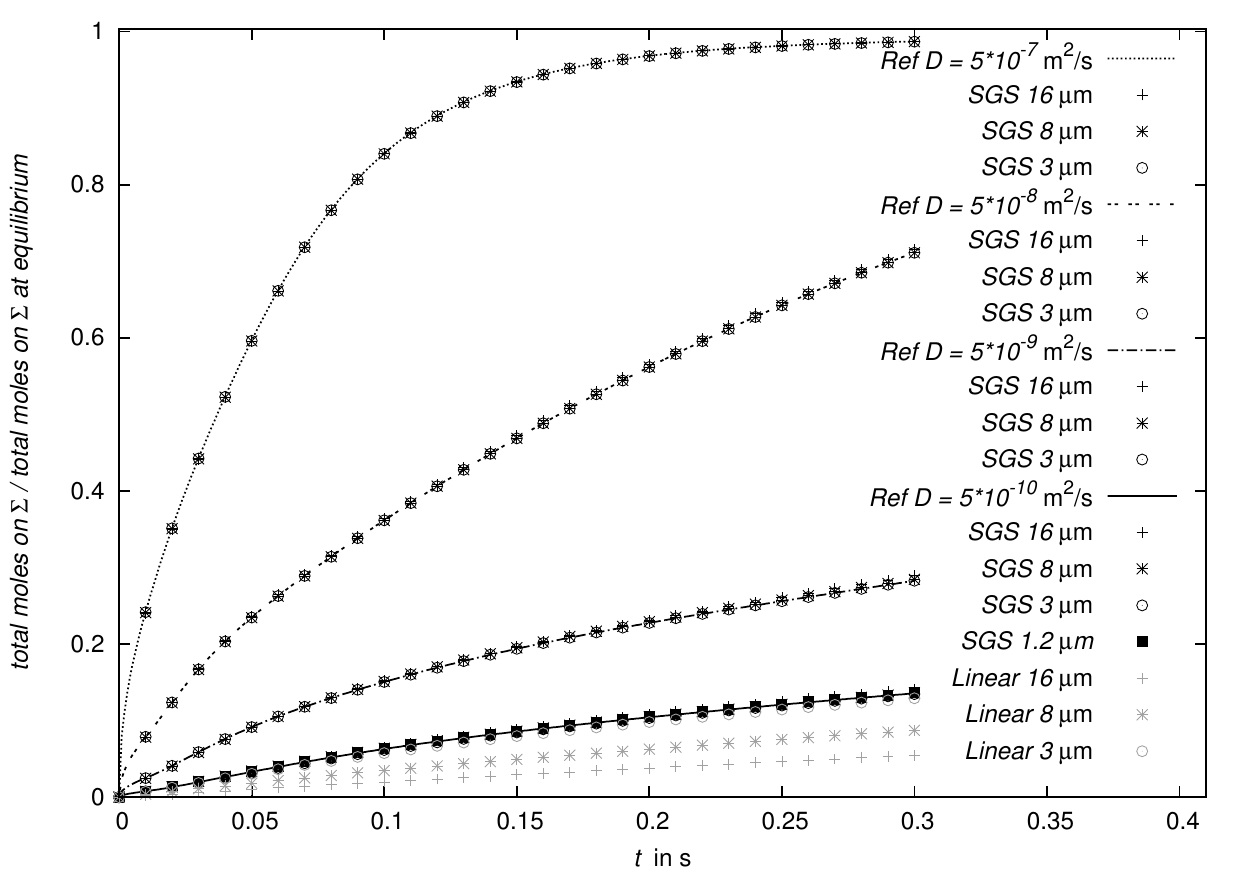}\label{fig:2Dmoles}} \quad
\subfloat[][\emph{Local Sherwood number at t = 0.2 s}.]
{\includegraphics[width=.62\textwidth]{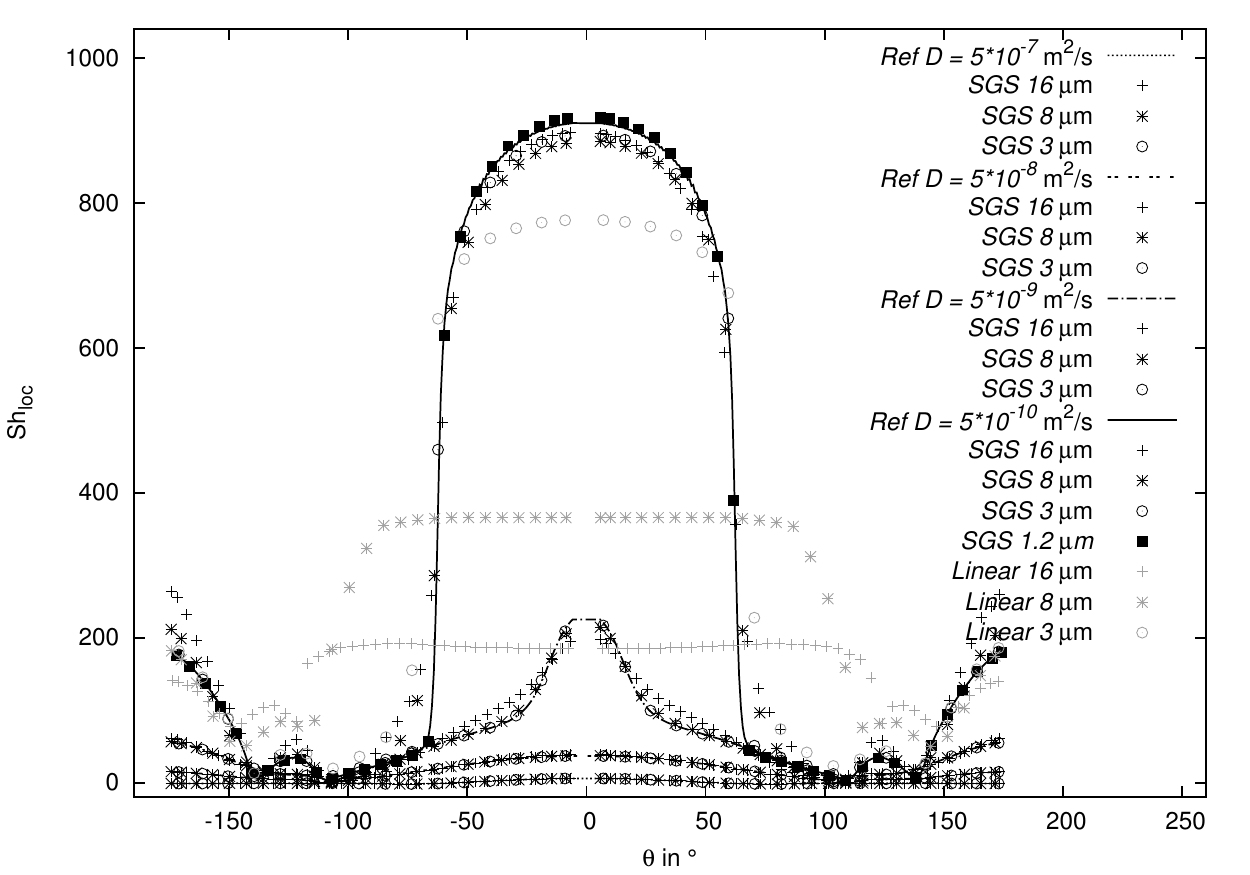}\label{fig:2DlocSh}}
\caption{Simulation results for a 2D bubble rising in contaminated water; comparison between cases with and without SGS modelling. Black symbols: with SGS modelling, grey symbols: linear interpolation.}
\label{fig:2Dvalid}
\end{figure}

Figure~\ref{fig:2Dvel} shows that the rise velocities obtained applying the SGS modelling are all in agreement with the reference. On the other hand, the results obtained with standard interpolation follow a very different trend. Only the 3 $\mu$m mesh gets close to the reference. Not only the rise velocities are in good agreement with the reference, but also the total amount of surfactant on the interface, as shown in figure~\ref{fig:2Dmoles} for different diffusion coefficients. As can be seen from the graph, the results obtained with the SGS modelling are all laying on the reference curves, while for $D = 5 \cdot 10^{-10}\ \mathrm{m^2/s}$, only the 3 $\mu$m case with standard interpolation tends to the correct result. So far we considered global quantities for comparison. Further confirmation that the SGS modelling is performing well and corresponding to the standard interpolation results is given by the local Sherwood numbers for the different diffusivities at $t = 0.2\ \mathrm{s}$, see figure~\ref{fig:2DlocSh}. Here it can be seen that all the cases where the SGS model has been used deliver a very good approximation of the local Sherwood number. Moreover, the shape of the local Sherwood number profile reflects the flow field around the bubble, see figure~\ref{fig:2Dfields}.
\FloatBarrier
\FloatBarrier
%

%
\subsection{Computation of the forces acting on the interface}
\label{subsec:app_4}

The starting point for the derivation of the expression to compute the different forces acting on the interface is the interfacial jump condition~\eqref{eq:m5}, namely
\begin{equation}
\dleftsq p_{\mathrm{tot}}\ \mathrm{\mathbf{I}} - \mathbf{S}^{\mathrm{visc}} \drightsq \cdot \n_{\Sigma} = \sigma \kappa \n_{\Sigma} + \grad_{\Sigma} \sigma.
\label{eq:m5_app}
\end{equation} 
Equation~\eqref{eq:m5_app} can be decomposed in normal and tangential component to the interface, in order to better understand the meaning of each force acting on the bubble.
With some notions of tensor calculus and knowing the definition of the surface operators, the projection of the jump condition in direction normal ($\n_{\Sigma}$) to the interface reads
\begin{equation}
\dleftsq p_{\mathrm{tot}} \drightsq \n_{\Sigma} + 2 \dleftsq \mu \drightsq (\grad_{\Sigma} \cdot \vec{v}) \n_{\Sigma} = \sigma \kappa \n_{\Sigma},
\label{eq:m5_app_n}
\end{equation} 
while in the direction tangential ($\vec{t}_{\Sigma}$) to $\Sigma$ we obtain
\begin{equation}
- \dleftsq \mu \n \cdot \grad \vec{v} \drightsq  - \dleftsq \mu (\grad_{\Sigma} \vec{v}) \cdot \n \drightsq - \dleftsq \mu \drightsq \n_{\Sigma} (\grad_{\Sigma} \cdot \vec{v}) = \grad_{\Sigma} \sigma.
\label{eq:m5_app_t}
\end{equation} 
As before, if we indicate with A the liquid side and with B the gas side, we can specify all the terms in the jump brackets\footnote{Note that the interface normal is $\n_{\Sigma}$ with $\n_A = \n_{\Sigma}$, while $\n_B = - \n_{\Sigma}$.} as follows,
\begin{equation}
p_{\mathrm{tot},B} \n_{\Sigma} - p_{\mathrm{tot},A} \n_{\Sigma} + 2 \mu_B (\grad_{\Sigma} \cdot \vec{v}) \n_{\Sigma} - 2 \mu_A (\grad_{\Sigma} \cdot \vec{v}) \n_{\Sigma} = \sigma \kappa \n_{\Sigma},
\label{eq:m5_app_nAB}
\end{equation} 
for the normal direction, and
\begin{align}
- \mu_B \left[ (\n \cdot \grad \vec{v})_B - (\grad_{\Sigma} \vec{v})_B \cdot \n_{\Sigma} + \n_{\Sigma} (\grad_{\Sigma} \cdot \vec{v}) \right]& \\
+ \mu_A \left[ (\n \cdot \grad \vec{v})_A  + (\grad_{\Sigma} \vec{v})_A \cdot \n_{\Sigma}  + \n_{\Sigma} (\grad_{\Sigma} \cdot \vec{v}) \right]& = \grad_{\Sigma} \sigma,
\label{eq:m5_app_tAB}
\end{align} 
for the tangential direction.
Each term in equations~\eqref{eq:m5_app_nAB} and~\eqref{eq:m5_app_tAB}, when multiplied by the face area will give a force contribution.
\begin{itemize}
    \item Marangoni force
      \begin{itemize}
          \item area specific force at face $i\ \in \Sigma$
	    \begin{equation}
	    \vec{f}^{\mathrm{ma}}_{i} = \grad_{\Sigma} \sigma_i
	    \label{eq:fm}
	    \end{equation} 
          \item resultant force on $\Sigma$
	    \begin{equation}
	    \vec{F}^{\mathrm{ma}} = \sum_{i}^{N_f} \vec{f}^{\mathrm{ma}}_{i}\ S_{f_i}
	    \label{eq:Fmt}
	    \end{equation} 
	    where $N_f$ is the number of faces on the interface and $S_{f_i}$ the face area.
      \end{itemize}           
    \item Capillary pressure force
      \begin{itemize}
          \item area specific force at face $i\ \in \Sigma$
	    \begin{equation}
	    \vec{f}^{\mathrm{ca}}_{i} = \sigma_i k_i\ \n_{\Sigma_i}
	    \label{eq:fc}
	    \end{equation} 
          \item resultant force on $\Sigma$
	    \begin{equation}
	    \vec{F}^{\mathrm{ca}} = \sum_{i}^{N_f} \vec{f}^{\mathrm{ca}}_{i}\ S_{f_i}
	    \label{eq:Fct}
	    \end{equation} 
      \end{itemize}      
    \item Total pressure force jump
      \begin{itemize}
          \item area specific force at face $i\ \in \Sigma$
	    \begin{equation}
	    \vec{f}^{p_{\mathrm{tot}}}_{i} = (p_{\mathrm{tot},B_i} - p_{\mathrm{tot},A_i})\ \n_{\Sigma_i}
	    \label{eq:fp}
	    \end{equation} 
          \item resultant force on $\Sigma$
	    \begin{equation}
	    \vec{F}^{p_{\mathrm{tot}}} = \sum_{i}^{N_f} \vec{f}^{p_{\mathrm{tot}}}_{i}\ S_{f_i}
	    \label{eq:Fpt}
	    \end{equation} 
      \end{itemize}
    \item Dynamic pressure force jump
      \begin{itemize}
          \item area specific force at face $i\ \in \Sigma$
	    \begin{equation}
	    \vec{f}^{p_{\mathrm{dyn}}}_{i} = (p_{\mathrm{dyn},B_i} - p_{\mathrm{dyn},A_i})\ \n_{\Sigma_i}
	    \label{eq:fPdyn}
	    \end{equation} 
	  where the dynamic pressure is computed as $p_{\mathrm{dyn}} = p_{\mathrm{tot}} - p_{\mathrm{hydro}}$, with the hydrostatic pressure $p_{\mathrm{hydro}} = \rho\ \vec{g} \cdot \vec{x}_{f_i}$
          \item resultant force on $\Sigma$
	    \begin{equation}
	    \vec{F}^{p_{\mathrm{dyn}}} = \sum_{i}^{N_f} \vec{f}^{p_{\mathrm{dyn}}}_{i}\ S_{f_i}
	    \label{eq:FPdynt}
	    \end{equation} 
      \end{itemize}     
    \item Normal viscous force
      \begin{itemize}
          \item area specific forces at face $i\ \in \Sigma$
	    \begin{align}
	    \vec{f}^{\mathrm{visc}}_{\perp,B_i} &= 2 \mu_B (\grad_{\Sigma} \cdot \vec{v})_i\ \n_{\Sigma_i} \label{eq:fv_nB}\\
	    \vec{f}^{\mathrm{visc}}_{\perp,A_i} &= - 2 \mu_A (\grad_{\Sigma} \cdot \vec{v})_i\ \n_{\Sigma_i} \label{eq:fv_nA}\\
	    \vec{f}^{\mathrm{visc}}_{\perp,i} &= \vec{f}^{\mathrm{visc}}_{\perp,B_i} + \vec{f}^{\mathrm{visc}}_{\perp,A_i} \label{eq:fv_nJ}
	    \end{align} 
          \item resultant forces on $\Sigma$
	    \begin{align}
	    \vec{F}^{\mathrm{visc}}_{\perp,B} &= \sum_{i}^{N_f} \vec{f}^{\mathrm{visc}}_{\perp,B_i}\ S_{f_i} \label{eq:fv_nBt}\\
	    \vec{F}^{\mathrm{visc}}_{\perp,A} &= \sum_{i}^{N_f} \vec{f}^{\mathrm{visc}}_{\perp,A_i}\ S_{f_i} \label{eq:fv_nAt}\\
	    \vec{F}^{\mathrm{visc}}_{\perp} &= \sum_{i}^{N_f} \vec{f}^{\mathrm{visc}}_{\perp,i}\ S_{f_i} \label{eq:fv_nJt}
	    \end{align} 
      \end{itemize}
    \item Tangential viscous force
      \begin{itemize}
          \item area specific forces at face $i\ \in \Sigma$
	    \begin{align}
	    \vec{f}^{\mathrm{visc}}_{\parallel,B_i} &= \mu_B \left[ - (\n \cdot \grad \vec{v})_{B_i} + (\grad_{\Sigma} \vec{v})_{B_i} \cdot \n_{\Sigma_i} - \n_{\Sigma_i} (\grad_{\Sigma} \cdot \vec{v})_i \right] \label{eq:fv_tB}\\
	    \vec{f}^{\mathrm{visc}}_{\parallel,A_i} &= \mu_A \left[ (\n \cdot \grad \vec{v})_{A_i}  + (\grad_{\Sigma} \vec{v})_{A_i} \cdot \n_{\Sigma_i}  + \n_{\Sigma_i} (\grad_{\Sigma} \cdot \vec{v})_i \right] \label{eq:fv_tA}\\
	    \vec{f}^{\mathrm{visc}}_{\parallel,i} &= \vec{f}^{\mathrm{visc}}_{\parallel,B_i} + \vec{f}^{\mathrm{visc}}_{\parallel,A_i} \label{eq:fv_tJ}
	    \end{align} 
          \item resultant forces on $\Sigma$
	    \begin{align}
	    \vec{F}^{\mathrm{visc}}_{\parallel,B} &= \sum_{i}^{N_f} \vec{f}^{\mathrm{visc}}_{\parallel,B_i}\ S_{f_i} \label{eq:fv_tBt}\\
	    \vec{F}^{\mathrm{visc}}_{\parallel,A} &= \sum_{i}^{N_f} \vec{f}^{\mathrm{visc}}_{\parallel,A_i}\ S_{f_i} \label{eq:fv_tAt}\\
	    \vec{F}^{\mathrm{visc}}_{\parallel} &= \sum_{i}^{N_f} \vec{f}^{\mathrm{visc}}_{\parallel,i}\ S_{f_i}. \label{eq:fv_tJt}
	    \end{align} 
      \end{itemize}
\end{itemize}
If we write the jump condition in terms of global forces then we obtain the following expression
\begin{equation}
\vec{F}^{p_{\mathrm{tot}}} + \vec{F}^{\mathrm{visc}} = \vec{F}^{\mathrm{ca}} + \vec{F}^{\mathrm{ma}}
\label{eq:eqSigma}
\end{equation} 
that can serve as a check of the fulfilment of the jump condition at the interface at the end of the simulation.

%
%

\bibliographystyle{plain}
\bibliography{bibliography}

\end{document}